\documentclass[iop,revtex4,onecolumn]{emulateapj}
\shorttitle{CONTINUITY EQUATION AND ABUNDANCE MATCHING}
\shortauthors{R. AVERSA ET AL.}

\begin{document}

\title{Black Hole and Galaxy Coevolution\\ from Continuity Equation and Abundance Matching}
\author{R. Aversa\altaffilmark{1,4}, A. Lapi\altaffilmark{1,2,3,4},
G. de Zotti\altaffilmark{1,5}, F. Shankar\altaffilmark{6}, L.
Danese\altaffilmark{1,3,4}} \altaffiltext{1}{SISSA, Via Bonomea 265, 34136
Trieste, Italy} \altaffiltext{2}{Dip. Fisica, Univ. `Tor Vergata',Via Ricerca
Scientifica 1, 00133 Roma, Italy} \altaffiltext{3}{INAF-Osservatorio
Astronomico di Trieste, Via Tiepolo 11, 34131, Trieste, Italy}
\altaffiltext{4}{INFN-Sezione di Trieste, via Valerio 2, 34127, Trieste,
Italy}\altaffiltext{5}{INAF-Osservatorio Astronomico di Padova, Vicolo
Osservatorio 5, 35122 Padova, Italy}\altaffiltext{6}{School of Physics and
Astronomy, University of Southampton, Southampton SO17 1BJ, UK}

\begin{abstract}
We investigate the coevolution of galaxies and hosted supermassive black
holes throughout the history of the Universe by a statistical approach
based on the continuity equation and the abundance matching technique.
Specifically, we present analytical solutions of the continuity equation
without source term to reconstruct the supermassive black hole (BH) mass
function from the AGN luminosity functions. Such an approach includes
physically-motivated AGN lightcurves tested on independent datasets, which
describe the evolution of the Eddington ratio and radiative efficiency from
slim- to thin-disc conditions. We nicely reproduce the local estimates of
the BH mass function, the AGN duty cycle as a function of mass and
redshift, along with the Eddington ratio function and the fraction of
galaxies with given stellar mass hosting an AGN with given Eddington ratio.
We exploit the same approach to reconstruct the observed stellar mass
function at different redshift from the UV and far-IR luminosity functions
associated to star formation in galaxies. These results imply that the
buildup of stars and BHs in galaxies occurs via \emph{in-situ} processes,
with dry mergers playing a marginal role at least for stellar masses $\la
3\times 10^{11}\, M_\odot$ and BH masses $\la 10^9\, M_\odot$, where the
statistical data are more secure and less biased by systematic errors. In
addition, we develop an improved abundance matching technique to link the
stellar and BH content of galaxies to the gravitationally dominant dark
matter component. The resulting relationships constitute a testbed for
galaxy evolution models, highlighting the complementary role of stellar and
AGN feedback in the star formation process. In addition, they may be
operationally implemented in numerical simulations to populate dark matter
halos or to gauge subgrid physics. Moreover, they may be exploited to
investigate the galaxy/AGN clustering as a function of redshift, mass
and/or luminosity. In fact, the clustering properties of BHs and galaxies
are found to be in full agreement with current observations, so further
validating our results from the continuity equation. Finally, our analysis
highlights that (i) the fraction of AGNs observed in slim-disc regime,
where anyway most of the BH mass is accreted, increases with redshift; (ii)
already at $z\ga 6$ substantial dust amount must have formed over
timescales $\la 10^8$ yr in strongly starforming galaxies, making these
sources well within the reach of \textsl{ALMA} surveys in (sub-)millimeter
bands.
\end{abstract}


\setcounter{footnote}{0}

\section{Introduction}\label{sec|intro}

Kinematic and photometric observations of the very central regions in local,
massive early-type galaxies strongly support the almost ubiquitous presence
of black holes (BHs) with masses $M_{{\rm BH}}\ga 10^6\, M_{\odot}$ (Dressler
1989; Kormendy \& Richstone 1995; Magorrian et al. 1998; for a recent review
Kormendy \& Ho 2013). Their formation and evolution is a major problem in
astrophysics and physical cosmology.

The correlations between the central BH mass and galaxy properties such as
the mass in old stars (Kormendy \& Richstone 1995; Magorrian et al. 1998;
Marconi \& Hunt 2003; H\"{a}ring \& Rix 2004; McLure \& Dunlop 2004;
Ferrarese \& Ford 2005; Graham 2007; Sani et al. 2011; Beifiori et al. 2012;
McConnell \& Ma 2013; Kormendy \& Ho 2013), the velocity dispersion
(Ferrarese \& Merritt 2000; Gebhardt et al. 2000; Tremaine et al. 2002;
G\"{u}ltekin et al. 2009; McConnell \& Ma 2013; Kormendy \& Ho 2013; Ho \&
Kim 2014), and the inner light distribution (Graham et al. 2001; Lauer et al.
2007; Graham \& Driver 2007; Kormendy \& Bender 2009) impose strong ties
between the formation and evolution of the BH and that of the old stellar
population in the host galaxy (Silk \& Rees 1998; Fabian 1999; King 2005; for
a recent review see King 2014).

A central role in this evolution is played by the way dark matter (DM) halos
and associated baryons assemble. So far it has been quite popular, e.g. in
most semi-analytic models, to elicit merging as the leading process; as to
the baryons, `wet' and `dry' mergers or a mixture of the two kinds have been
often implemented (for a recent review see Somerville \& Dav\'e 2015). On the
other hand, detailed analyses of DM halo assembly indicate a two-stage
process: an early fast collapse during which the central regions reach
rapidly a dynamical quasi-equilibrium, followed by a slow accretion that
mainly affects the halo outskirts (e.g., Zhao et al. 2003; Wang et al. 2011;
Lapi \& Cavaliere 2011). Thus one is led to consider the rapid starformation
episodes in the central regions during the fast collapse as the leading
processes in galaxy formation (e.g., Lapi et al. 2011, 2014; Cai et al.
2013). Plainly, the main difference between merging and fast collapse models
relates to the amount of stars formed \emph{in-situ} (e.g., Moster et al.
2013).

While $N-$body simulations of DM halo formation and evolution are nowadays
quite robust (though details of their results are not yet fully understood),
the outcomes of hydrodynamical simulations including star formation and
central BH accretion are found to feature large variance (Scannapieco et al.
2012; Frenk \& White 2012). This is expected since most of the relevant
processes involving baryons such as cooling, gravitational instabilities,
angular momentum dissipation, star formation and supermassive BH accretion
occur on spatial and temporal scales well below the current resolution.

On the other hand, observations of active galactic nuclei (AGNs) and galaxies
at different stages of their evolution have spectacularly increased in the
last decade at many wavelengths. In particular, the AGN luminosity function
is rather well assessed up to $z\sim 6$ though with different uncertainties
in the X-ray (Ueda et al. 2014; Fiore et al. 2012; Buchner et al. 2015; Aird
et al. 2010, 2015), UV/optical (Richards et al. 2006; Croom et al. 2009;
Masters et al. 2012; Ross et al. 2012; Fan et al. 2006; Jiang et al. 2009;
Willott et al. 2010a), and IR bands (Richards et al. 2006; Fu et al. 2010;
Assef et al. 2011; Ross et al. 2012); these allow to infer the BH accretion
rate functions at various redshifts. In addition, luminosity functions of
galaxies are now available up to $z\sim 10$ in the ultraviolet (UV; Bouwens
et al. 2015; Finkelstein et al. 2014, Weisz et al. 2014; Cucciati et al.
2012; Oesch et al. 2010; Reddy \& Steidel 2009; Wyder et al. 2005) and up to
$z\sim 4$ in the far-infrared (FIR) band (Lapi et al. 2011; Gruppioni et al.
2013; Magnelli et al. 2013); these allow to infer the star formation rate
(SFR) function at various redshifts.

As for galaxies selected by their mid- and far-IR emission, the distribution
function of the luminosity associated to the formation of massive stars shows
that at $z\la 4$ the number density of galaxies endowed with star formation
rates $\dot M_\star\ga 10^2\, M_{\odot}$ yr$^{-1}$ is $N(\log \dot
M_\star)\ga 10^{-3}$ Mpc$^{-3}$. The density is still significant, $N(\log
\dot M_\star)\ga 10^{-5}$ Mpc$^{-3}$, for $\dot M_\star\approx 10^3\,
M_{\odot}$ yr$^{-1}$. On the other hand, the UV selection elicits galaxies
forming stars at much lower rates $\dot M_\star\la 30\, M_{\odot}$ yr$^{-1}$
up to $z\la 10$. The complementarity between the two selections is ascribed
to the increasing amount of dust in galaxies with larger SFRs (Steidel et al.
2001; Mao et al. 2007; Bouwens et al. 2013, 2015; Fan et al. 2014; Cai et al.
2014; Heinis et al. 2014). From deep, high resolution surveys with
\textsl{ALMA} at (sub-)mm wavelengths there have been hints of possible
source blending at fluxes $S_{870\,\mu{\rm m}}\ga 10$ mJy (Karim et al. 2013;
Ono et al. 2014; Simpson et al. 2015b). On the other hand, observations at
high spatial resolution of sub-mm selected, high redshift galaxies with the
\textsl{SMA} and follow-ups at radio wavelengths with the \textsl{VLA} show
that $z\la 6$ galaxies exhibiting $\dot M_\star\approx$ a few $10^3\,
M_{\odot}$ yr$^{-1}$ have a number density $N\sim 10^{-6}$ Mpc$^{-3}$ (Barger
et al. 2012, 2014), fully in agreement with the results of Lapi et al. (2011)
and Gruppioni et al. (2013) based on \textsl{Herschel} (single dish) surveys.

Studies on individual galaxies show that several sub-mm galaxies at high
redshift exhibit $\dot M_\star\ga 10^3\, M_{\odot}$ yr$^{-1}$ concentrated on
scales $\la 10$ kpc (e.g.,  Finkelstein et al. 2014; Neri et al. 2014; Rawle
et al. 2014; Riechers et al. 2014; Ikarashi et al. 2014; Simpson et al.
2015a; Scoville et al. 2014). Size ranging from a few to several kpc of
typical high$-z$ strongly star forming galaxies has been confirmed by
observations of many gravitational lensed objects (e.g., Negrello et al.
2014). In addition, high spatial resolution observations around optically
selected quasars put in evidence that a non negligible fraction of host
galaxies exhibits $\dot M_\star\ga 10^3\, M_\odot$ yr$^{-1}$ (Omont et al.
2001, 2003; Carilli et al. 2001; Priddey et al. 2003; Wang et al. 2008;
Bonfield et al. 2011; Mor et al. 2012).

The clustering properties of luminous sub-mm selected galaxies (Webb et al.
2003; Blain et al. 2004; Weiss et al. 2009; Hickox et al. 2012; Bianchini et
al. 2015) indicate that they are hosted by large halos with masses $M_{\rm
H}\ga$ several $10^{12}\, M_{\odot}$ and that the star formation timescale is
around $\sim 0.5-1$ Gyr.

The statistics on the presence of AGNs along the various stages of galaxy
assembling casts light on the possible reciprocal influence between star
formation and BH accretion (for a recent review, see Heckman \& Best 2014 and
references therein), although the fine interpretation of the data is still
debated. On one side, some authors suggest that star formation and BH
accretion are strongly coupled via feedback processes, while others support
the view that the two processes are only loosely related and that the final
relationships among BH mass and galaxy properties are built up along the
entire Hubble time with a relevant role of dry merging processes.

Most recently, Lapi et al. (2014) have shown that the wealth of data at $z\ga
1$ strongly support the view that galaxies with final stellar mass
$M_{\star}\ga 10^{11}\, M_{\odot}$ proceed with their star formation at an
almost constant rate over $\sim 0.5-1$ Gyr, within a dusty interstellar
medium (ISM). At the same time several physical mechanisms related to the
star formation, such as gravitational instabilities in bars or dynamical
friction among clouds of starforming gas or radiation drag (Norman \&
Scoville 1988; Shlosman et al. 1989, 1990; Shlosman \& Noguchi 1993;
Hernquist \& Mihos 1995; Noguchi 1999; Umemura 2001; Kawakatu \& Umemura
2002; Kawakatu et al. 2003; Thompson et al. 2005; Bournaud et al. 2007, 2011;
Hopkins \& Quataert 2010, 2011), can make a fraction of the ISM loose angular
momentum and flow into a reservoir around the seed BH. The accretion from the
reservoir to the BH can be as large as $30-50$ times the Eddington rate,
leading to slim-disc conditions (Abramowicz et al. 1988; Watarai et al. 2001;
Blandford \& Begelman 2004; Li 2012; Begelman 2012; Madau et al. 2014;
Volonteri \& Silk 2014), with an Eddington ratio $\lambda \la 4$ and an
average radiative efficiency $\epsilon\la 0.1$. This results in an
exponential increase of the BH mass and of the AGN luminosity, with an
$e-$folding timescale $\tau_{\rm ef}$ ranging from a few to several $10^7$
years. Eventually, the AGNs at its maximum power can effectively transfer
energy and momentum to the ISM, removing a large portion of it from the
central regions and so quenching the star formation in the host. The
reservoir around the BH is no more fed by additional gas, so that even the
accretion and the nuclear activity come to an end.

More in general, we can implement lightcurves for the luminosity associated
to the star formation and to the BH accretion in a continuity equation
approach. In the context of quasar statistics, the continuity equation has
been introduced by Cavaliere et al. (1971) to explore the optical quasar
luminosity evolution and its possible relation with the radiosource
evolution. Soltan (1982) and Choksi \& Turner (1992) exploited the
mass-energy conservation to derive an estimate of the present mass density in
inactive BHs. The extension and the derivation of the BH mass function has
been pioneered by Small \& Blandford (1992), who first attempted to connect
the present day BH mass function to the AGNs luminosity evolution. A
simplified version in terms of mass-energy conservation has been used by
Salucci et al. (1999), who have shown that the distribution of the mass
accreted onto central BHs during the AGNs activity well matches the mass
function of local inactive BHs. A detailed discussion of the continuity
equation in the quasar and central supermassive BH context has been presented
by Yu \& Lu (2004, 2008). In the last decade the continuity equation has been
widely used, though the lightcurve of the AGN, one of the fundamental
ingredients, was largely based on assumptions (e.g., Marconi et al. 2004;
Shankar et al. 2009, 2013; Merloni \& Heinz 2008; Wang et al. 2009; Cao
2010). Results on the BH mass function through the continuity equation have
been reviewed by Kelly \& Merloni (2012) and Shankar (2013).

We will also implement the continuity equation for the stellar content of
galaxies. This has become possible because the UV surveys for Lyman Break
Galaxies, the wide surveys \textsl{HerMES} and \textsl{H-ATLAS} obtained with
the \textsl{Herschel} space observatory made possible to reconstruct the star
formation rate function in the Universe out to $z\la 6$ for SFRs $\dot
M_\star \sim 10-1000\, M_{\odot}$ yr$^{-1}$. Therefore, we can exploit the
continuity equation, in an analogous manner as routinely done for the AGNs;
the BH mass is replaced by the mass in stars and the bolometric luminosity
due to accretion is replaced by the luminosity generated by the formation of
young, massive stars.

As for the stellar mass function, it is inferred by exploiting the observed
luminosity function in the wavelength range of the SED dominated by the
emission from older, less massive stars. The passage from stellar luminosity
to mass is plagued by several problems, which result in uncertainties of
order of a factor of $2$, increasing for young, dusty galaxies (e.g.,
Cappellari et al. 2013; Conroy et al. 2014). Therefore the mass estimate is
more robust for galaxies with quite low star formation and/or in passive
evolution. All in all, the stellar mass function of galaxies is much easier
to estimate, and hence much better known, than the BH mass function,
particularly at high redshift. Reliable stellar mass functions are available
both for local and redshift up to $z\sim 6$ galaxy samples (e.g., Bernardi et
al. 2013; Maraston et al. 2013; Ilbert et al. 2013; Santini et al. 2012;
Stark et al. 2009; Gonz\'alez et al. 2011; Duncan et al. 2014). The
comparison between the observed stellar mass function and the results from
the continuity equation sheds light on the relative contribution of dry
merging and of in-situ star formation. In the present paper we will solve the
continuity equation for AGN and for the stellar component after inserting the
corresponding lightcurves derived from the data analysis of Lapi et al.
(2011, 2014, see above).

Once the stellar and the BH mass functions at different redshifts are known,
these can also be compared with the abundance of DM halos to obtain
interesting relationships between halo mass and galaxy/BH properties. Such a
technique, dubbed abundance matching, has been exploited by several authors
(e.g., Vale \& Ostriker 2004; Shankar et al. 2006; Moster et al. 2010, 2013;
Behroozi et al. 2013; Behroozi \& Silk 2015). In this paper, the technique is
refined and used in connection with the outcomes of the continuity equation
to tackle the following open issues in galaxy formation and evolution:

\begin{itemize}

\item Is the BH mass function reflecting the past AGN activity? What was
    the role of merging in shaping it? (see Sect.~2.1 and Appendix B)

\item How does the BH duty cycle evolve? What can we infer on the radiative
    efficiency and on the Eddington ratio of active BHs? (see Sect.~2.1.4)

\item Is there any correlation between the central BH mass and the halo
    mass and how does it evolve with time? (see Sect.~3.1.1)

\item Which is the relationship between the AGN bolometric luminosity and
    the host halo mass? Can we use this relationship with the duty cycle to
    produce large simulated AGN catalogs? (see Sect.~3.1.1)

\item Which are the bias properties of AGNs? Do they strongly depend on
    luminosity and redshift? (see Sect.~3.1.1)

\item Can the evolution of the stellar mass function be derived through the
    continuity equation as in the case of the BH mass function, by
    replacing the accretion rate with the SFR? Does dry merging play a
    major role in shaping the stellar mass function of galaxies? What is
    the role of the dust in the star formation process within galaxies?
    (see Sect.~2.2, App. B and C)

\item Which is the relationship between the SFR, the stellar mass of the
    galaxies, and the mass of the host halo? Does the star formation
    efficiency (i.e., the fraction of baryons going into stars) evolve with
    cosmic time? (see Sect.~3.1.2)

\item Which is the relationship between the bolometric luminosity of
    galaxies due to star formation and the host halo mass? Can we use this
    relationship with the stellar duty cycle to produce large simulated
    catalogs of star forming galaxies? (see Sect.~3.1.2)

\item Which are the bias properties of star forming and passively evolving
    galaxies? (see Sect.~3.1.2)

\item How does the specific star formation rate evolve with redshift and
    stellar mass? (see Sect.~3.1.3)

\item Which is the relationship between the BH mass and the stellar mass at
    the end of the star formation and BH mass accretion epoch? Does it
    evolve with time? (see Sect.~3.1.4)

\item How and to what extent can we extrapolate the relationships for both
    galaxies and hosted AGNs to higher, yet unexplored, redshift? (see
    Sects.~3 and 4)

\end{itemize}

To answer these questions we have organized the paper as follows. In
Sect.~\ref{sec|continuity} we present the statistical data concerning AGN and
starforming galaxies, introduce and motivate the corresponding lightcurves,
and solve the continuity equation to derive the BH and stellar mass functions
at different redshifts. In Sect.~\ref{sec|abundance} we exploit the abundance
matching technique to infer relationships among the properties of the BH,
stellar, and dark matter component in galaxies. In
Sect.~\ref{sec|conclusions} we discuss and summarize our findings.

Throughout this work we adopt the standard flat concordance cosmology (Planck
Collaboration 2014, 2015) with round parameter values: matter density
$\Omega_M= 0.3$, baryon density $\Omega_b=0.05$, Hubble constant $H_0 = 100\,
h$ km s$^{-1}$ Mpc$^{-1}$ with $h = 0.7$, and mass variance $\sigma_8 = 0.8$
on a scale of $8\, h^{-1}$ Mpc. Stellar masses and luminosities (or SFRs) of
galaxies are evaluated assuming the Chabrier's (2003) IMF.

\section{Continuity equation}\label{sec|continuity}

Given an evolving population of astrophysical sources, we aim at linking the
luminosity function $N(L,t)$ tracing a generic form of baryonic accretion
(like that leading to the growth of the central BH or the stellar content in
the host galaxies) to the corresponding final mass function $N(M,t)$. To this
purpose, we exploit the standard continuity equation approach (e.g., Small \&
Blandford 1992; Yu \& Lu 2004), in the integral formulation
\begin{equation}\label{eq|continuity}
N(L,t) = \int_0^\infty{\rm d}M\, \left[\partial_t
N(M,t)-S(M,t)\right]\, \sum_i{{\rm d}\tau_i\over{\rm d}L}(L|M,t)~;
\end{equation}
here $\tau$ is the time elapsed since the triggering of the activity (the
internal clock, different from the cosmological time $t$), and ${\rm
d}\tau/{\rm d} L$ is the time spent by the object with final mass $M$ in the
luminosity range $[L, L+{\rm d}L]$ given a lightcurve $L(\tau|M,t)$; the sum
allows for multiple solution $\tau_i$ of the equation $L=L(\tau|M,t)$. In
addition, $S(M,t)$ is a source term due to `dry' merging (i.e., adding the
whole mass content in stars or BHs of merging objects without contributing
significantly to star formation or BH accretion). In solving
Eq.~(\ref{eq|continuity}) we shall set the latter to zero, and investigate
the impact of dry merging in the dedicated App.~B. Note that by integrating
Eq.~(\ref{eq|continuity}) in ${\rm d}t$ from $0$ to the present time $t_0$,
one recovers Eq.~(18) of Yu \& Lu (2004).

If the timescales $\tau_i$ (that encase the mass-to-energy conversion
efficiency) are constant in redshift and luminosity, then a generalized
Soltan (1982) argument concerning the equivalence between the integrated
luminosity density and the local, final mass density can be straightforwardly
recovered from Eq.~(\ref{eq|continuity}) without source term by multiplying
by $L$ and integrating it over $L$ and $t$
\begin{equation}
\int_0^{t_0}{\rm d}t\,\int_0^\infty{\rm d}L\,L\,N(L,t) = \int_0^\infty{\rm d}M\,
N(M,t)\,\int_0^\infty{\rm d}L\, L\,\sum_i{{\rm d}\tau_i\over
{\rm d}L} = {\rm const}\times \int_0^\infty{\rm d}M\, M\, N(M,t)~,
\end{equation}
where the last equivalence holds since $\sum_i\int{\rm d}\tau_i\, L\equiv$
const $\times M$. Specifically, for the BH population the constant is equal
to $\epsilon\, c^2/(1-\epsilon)$ in terms of an average radiative efficiency
$\epsilon\sim 0.1$. We shall see that an analogous expression holds for the
stellar component in galaxies.

More in general, Eq.~(\ref{eq|continuity}) constitutes an
integro-differential equation in the unknown function $N(M,t)$, that can be
solved once the input luminosity function $N(L,t)$ and the lightcurve
$L(\tau|M,t)$ have been specified. Specifically, we shall use it to derive
the mass function of the supermassive BH and stellar component in galaxies
throughout the history of the Universe. Remarkably, we shall see that
Eq.~(\ref{eq|continuity}) can be solved in closed analytic form under quite
general assumptions on the lightcurve.

\subsubsection{Connection with standard approaches for
BHs}\label{sec|std_continuity}

It is useful to show the connection of Eq.~(\ref{eq|continuity}) with the
standard, differential form of the continuity equation for the evolution of
the BH mass function as pioneered by Small \& Blandford (1992) and then used
in diverse contexts by many authors (e.g., Salucci et al. 1999; Yu \&
Tremaine 2002; Marconi et al. 2004; Shankar et al. 2004, 2009; Merloni \&
Heinz 2008; Wang et al. 2009; Cao 2010). Following Small \& Blandford (1992),
BHs are assumed to grow in a single accretion episode, emitting at a constant
fraction $\lambda\equiv L_{\rm AGN}/L_{\rm Edd}$ of their Eddington
luminosity $L_{\rm Edd}\equiv M_{{\rm BH}} c^2/t_{\rm Edd}\approx 1.38\times
10^{38}\, M_{{\rm BH}}/M_\odot$ erg s$^{-1}$ in terms of the Eddington time
$t_{\rm Edd}\approx 4\times 10^8$ yr. The resulting lightcurve can be written
as
\begin{equation}
L_{\rm AGN}(\tau|M_{{\rm BH}},t)={\lambda\, M_{{\rm BH}}\, c^2\over t_{\rm
Edd}}\, e^{(\tau-\tau_{\rm life})/\tau_{\rm ef}}~~~~\tau\leq \tau_{\rm life};
\end{equation}
here $M_{{\rm BH}}$ is the final BH mass, $\tau_{\rm life}=\int{\rm d}\tau$
is the total duration of the luminous accretion phase, and $\tau_{\rm
ef}=\epsilon\, t_{\rm Edd}/\lambda\,(1-\epsilon)$ is the $e-$folding time in
terms of the mass-energy conversion efficiency $\epsilon$. Then one has
\begin{equation}
{{\rm d}\tau\over {\rm d}L_{\rm AGN}} = {\tau_{\rm ef}\over L_{\rm AGN}}\,
\Theta_{H}\left[L_{\rm AGN}\leq L_{\rm AGN}(M_{{\rm BH}})\right]~,
\end{equation}
where the Heaviside step function $\Theta_{H}(\cdot)$ specifies that a BH
with final mass $M_{{\rm BH}}$ cannot have shone at luminosity exceeding
$L_{\rm AGN}(M_{{\rm BH}})\equiv \lambda\, M_{{\rm BH}} c^2/t_{\rm Edd}$.
Equivalently, only BHs with final masses exceeding $M_{{\rm BH}}(L_{\rm
AGN})\equiv L\,t_{\rm Edd}/\lambda\, c^2$ can have attained a luminosity
$L_{\rm AGN}$ and so can contribute to the integral on the right hand side of
Eq.~(\ref{eq|continuity}). Hence such an equation can be written as
\begin{equation}
L_{\rm AGN}\,N(L_{\rm AGN},t) = \int_{M_{{\rm BH}}(L_{\rm AGN})}^\infty{\rm d}M_{{\rm BH}}\, [\partial_t
N(M_{{\rm BH}},t)-S(M_{{\rm BH}},t)]\, \tau_{\rm ef}~.
\end{equation}
Differentiating both sides with respect to $L$ and rearranging terms yields
\begin{equation}
\partial_t N(M_{{\rm BH}},t) + {1\over \tau_{\rm ef}}\,
\partial_{M_{{\rm BH}}}[L_{\rm AGN}\, N(L_{\rm AGN},t)]_{|L_{\rm AGN}(M_{{\rm BH}})}=S(M_{{\rm BH}},t)~.
\end{equation}
Now one can formally write that
\begin{equation}
N(L_{\rm AGN},t) = N(M_{{\rm BH}},t)\, {{\rm d}M_{{\rm BH}}\over {\rm d}L_{\rm AGN}}\,
\delta_{\rm AGN}(M_{{\rm BH}},t)~~~~~~~~~~\langle L_{\rm AGN}\rangle = \delta_{\rm AGN}(M_{{\rm BH}},t)\, L_{\rm AGN}
\end{equation}
in terms of the BH duty cycle $\delta_{\rm AGN}(M_{{\rm BH}},t)\equiv
\tau_{\rm life}(M_{{\rm BH}},t)/t\la 1$. Since by definition $\langle L_{\rm
AGN}\rangle = \epsilon\,\langle\dot M_{{\rm BH}}\rangle\,c^2/(1-\epsilon)$,
one finally obtains the continuity equation in the form
\begin{equation}
\partial_t N(M_{{\rm BH}},t) + \partial_{M_{\rm BH}}[\langle
\dot M_{{\rm BH}}\rangle\,N(M_{{\rm BH}},t)] = S(M_{{\rm BH}},t)~;
\end{equation}
the underlying rationale is that, although individual BHs turn on and off,
the evolution of the BH population depends only on the mean accretion rate
$\langle\dot M_{{\rm BH}}\rangle$.

\subsection{The BH mass function}\label{sec|BH_MF}

We now solve Eq.~(\ref{eq|continuity}) to compute the BH mass function at
different redshifts.

\subsubsection{BHs: luminosity function}\label{sec|AGN_LF}

Our basic input is constituted by the bolometric AGN luminosity functions,
that we build up as follows. We start from the AGN luminosity functions at
different redshifts observed in the optical band by Richards et al. (2006),
Croom et al. (2009), Masters et al. (2012), Ross et al. (2012), Fan et al.
(2006), Jiang et al. (2009), Willott et al. (2010a), and in the hard X-ray
band by Ueda et al. (2014), Fiore et al. (2012), Buchner et al. (2015), Aird
et al. (2015).

Then we convert the optical and X-ray luminosities to bolometric ones by
using the Hopkins et al. (2007) corrections\footnote{Most of the optical data
are given in terms of magnitude $M_{1450}$ at $1450$ {\AA}. First, we convert
them to $B-$band ($4400$ {\AA}) using the relation $M_B=M_{1450}-0.48$ (Fan
et al. 2001), then we pass to $B-$band luminosities in solar units $\log
L_B/L_{B,\odot}=-0.4\,(M_B-5.48)$, and finally we go to bolometric
luminosities in solar units after $L_{\rm AGN}/L_\odot = k_B\,
L_B/L_{B,\odot}\times L_{B,\odot}/L_\odot$. For this last step we recall that
the $B-$band luminosity of the Sun $L_{B,\odot}\approx 2.13\times 10^{33}$
erg s$^{-1}\approx L_\odot/2$ is about half its bolometric one
$L_{\odot}\approx 3.9\times 10^{33}$ erg s$^{-1}$. In some other instances
the original data are expressed in terms of a $z=2$ $K$-corrected $i-$band
magnitude $M_i(z=2)$. We adopt the relation with the $1450$ magnitude
$M_{1450}=M_i(z=2)+1.486$ (Richards et al. 2006) and then convert to
bolometric as above.}. Note that in the literature several optical and X-ray
bolometric corrections have been proposed (see Marconi et al. 2004; Hopkins
et al. 2007; Shen et al. 2011; Lusso et al. 2012; Runnoe et al. 2012). Those
by Marconi et al. (2004) and Lusso et al. (2012) are somewhat smaller by $\la
40\%$ in the optical and by $\la 30\%$ in the hard X-ray band with respect to
Hopkins et al. (2007). In fact, since bolometric corrections are
intrinsically uncertain by a factor $\sim 2$ (e.g., Vasudevan \& Fabian 2007;
Lusso et al. 2012; Hao et al. 2014), these systematic differences between
various determinations are not relevant. We shall show in
Sect.~\ref{sec|BH_results} that our results on the BH mass function are
marginally affected by bolometric corrections. In addition, we correct the
number density for the fraction of obscured (including Compton thick) objects
as prescribed by Hopkins et al. (2007) for the optical data and according to
Ueda et al. (2014; see also Ueda et al. 2003) for the hard X-ray data. We
stress that both the bolometric and the obscuration correction are rather
uncertain, with the former affecting the luminosity function mostly at the
bright end, and the latter mostly at the faint end.

Given the non-homogeneous nature and the diverse systematics affecting the
datasets exploited to build up the bolometric luminosity functions, a formal
minimum $\chi^2-$fit is not warranted. We have instead worked out an analytic
expression providing a sensible rendition of the data in the relevant range
of luminosity and redshift. For this purpose, we use a \emph{modified}
Schechter function with evolving characteristic luminosity and slopes. The
luminosity function in logarithmic bins $N(\log L_{\rm AGN})=N(L_{\rm AGN})\,
L_{\rm AGN}\, \ln(10)$ writes
\begin{equation}\label{eq|AGN_LF}
N(\log L_{\rm AGN},z) = \Phi(z)\, \left[L_{\rm AGN}\over
L_c(z)\right]^{1-\alpha(z)}\,\exp{\left\{-\left[L_{\rm AGN}\over
L_c(z)\right]^{\omega(z)}\right\}}
\end{equation}
The normalization $\log\Phi(z)$, the characteristic luminosity $\log L_c(z)$,
and the characteristic slopes $\alpha(z)$ and $\omega(z)$ evolve with
redshift according to the same parametrization
\begin{equation}
p(z) = p_0 + k_{p1}\, \chi + k_{p2}\, \chi^2 + k_{p3}\, \chi^3
\end{equation}
with $\chi= \log [(1+z)/(1+z_0)]$ and $z_0=0.1$. The parameter values are
reported in Table~1. The functional form adopted here is similar to the
widely-used double powerlaw shape (e.g., Ueda et al. 2014; Aird et al. 2015),
but with a smoother transition between the faint and bright end slopes; all
in all, it provides a data representation of comparable quality. In fact, we
stress that the results of the continuity equation approach are insensitive
to the specific parameterization adopted for the luminosity function and its
evolution, provided that the quality in the rendition of the data be similar
to ours. For example, in Sect.~\ref{sec|BH_results} we shall show explicitly
that our results on the BH mass function are marginally affected when using a
double powerlaw shape in place of a modified Schechter to represent the AGN
luminosity functions.

\begin{figure}
\epsscale{1.}\plotone{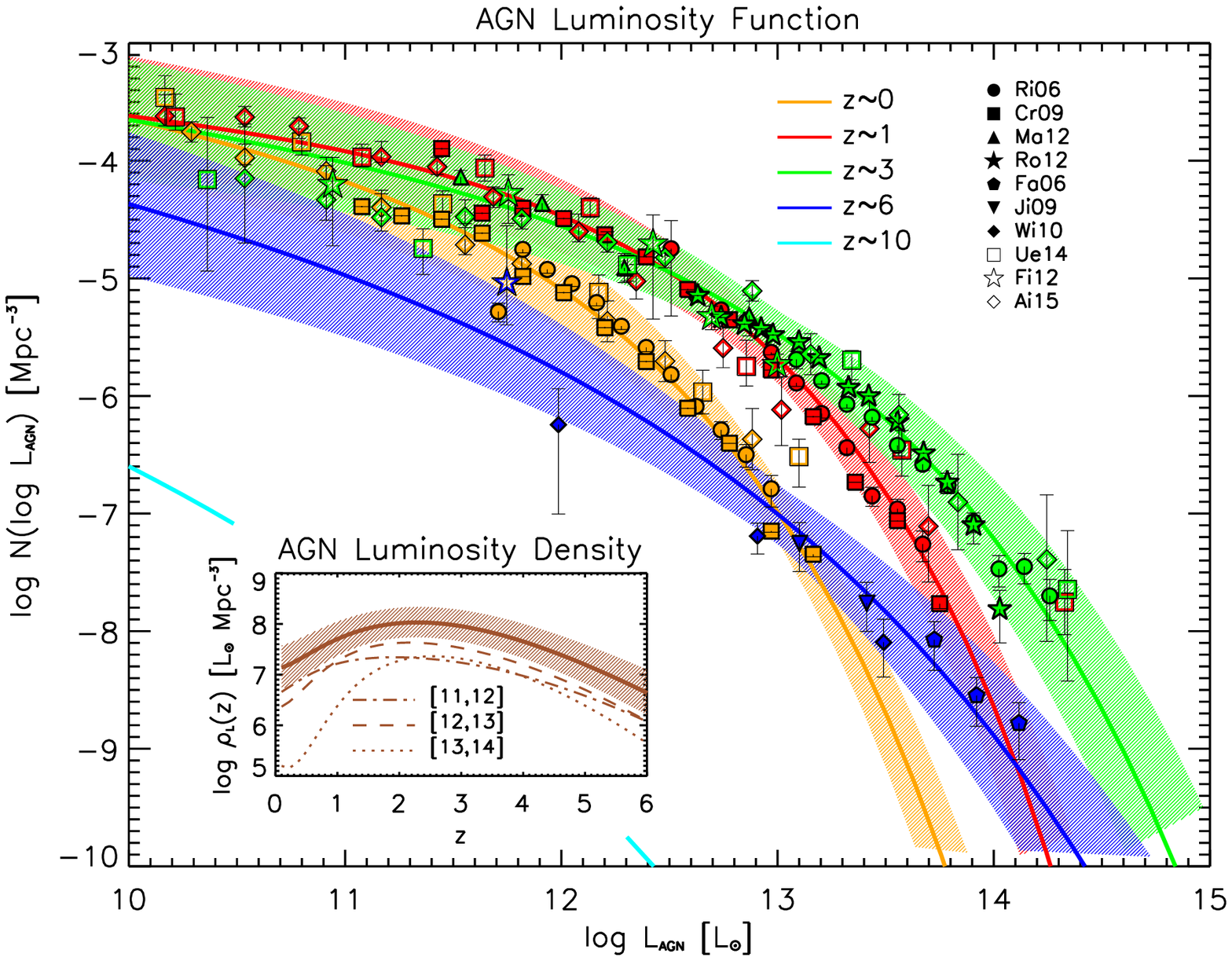} \caption{The bolometric AGN luminosity
function $N(\log L_{\rm AGN})$ at redshift $z=0$ (orange), $1$ (red), $3$
(green), and $6$ (blue). Optical data are from Richards et al. (2006; filled
circles), Croom et al. (2009; filled squares), Masters et al. (2012; filled
triangles), Ross et al. (2012; filled stars), Fan et al. (2006; filled pentagons),
Jiang et al. (2009; filled reversed triangles), Willott et al. (2010a; filled
diamonds); X-ray data are from Ueda et al. (2014; open squares), Fiore et al.
(2012; open stars), and Aird et al. (2015, open diamonds). The optical and
X-ray luminosities have been converted to bolometric by using the Hopkins et
al. (2007; see their Fig.~1) corrections, while the number densities have been
corrected for the presence of obscured AGNs according to Ueda et al. (2003,
2014). The solid lines illustrate the analytic rendition of the
luminosity functions as described in Sect.~\ref{sec|AGN_LF}, and the hatched
areas represent the associated uncertainty; the cyan line is the
extrapolation to $z=10$ plotted for illustration. The inset shows the AGN
luminosity density as a function of redshift, for the overall luminosity
range probed by the data (solid line with hatched area), and for AGN
bolometric luminosity $\log L_{\rm AGN}/L_\odot$ in the ranges $[11,12]$
(dot-dashed line), $[12,13]$ (dashed line), $[13,14]$ (dotted
line).}\label{fig|AGN_LF}
\end{figure}

In Fig.~\ref{fig|AGN_LF} we illustrate the bolometric AGN luminosity function
at various redshifts, including both our data collection and our analytic
parameterization of Eq.~(\ref{eq|AGN_LF}), with an estimate of the associated
$1\sigma$ uncertainty; the $z=10$ extrapolation is also shown for
illustration. In the inset we plot the evolution with redshift of the AGN
luminosity density, computed as
\begin{equation}
\rho_{L_{\rm AGN}}(z)=\int{\rm d}\log L_{\rm AGN}\, N(\log L_{\rm AGN},z)\,
L_{\rm AGN}~
\end{equation}
and the contribution to the total by specific luminosity ranges.

\subsubsection{BHs: lightcurve}\label{sec|AGNlightcurve}

As a further input to the continuity equation, we adopt the following
lightcurve (Yu \& Lu 2004)
\begin{eqnarray}\label{eq|AGNlightcurve}
L_{\rm AGN}(\tau|M_{{\rm BH}},t)\nonumber &=& {\lambda_0\,M_{{\rm BH}, \rm P}\, c^2/
t_{\rm Edd}}\,e^{(\tau -\tau_{\rm P})/\tau_{\rm ef}} ~~~~ 0 \le \tau
\le\tau_{\rm P}\\
\nonumber \\
&=& {\lambda_0\,M_{{\rm BH}, \rm P}\, c^2/ t_{\rm Edd}}\, e^{-(\tau
-\tau_{\rm P})/\tau_{\rm D}} ~~~~ \tau_{\rm P}\le \tau
\le\tau_{\rm P}+\zeta\tau_{\rm D} \\
\nonumber\\
\nonumber &=& 0 ~~~~~~~~~~~~~~~~~~~~~~~~~~~~~~~~~~~\tau > \tau_{\rm
P}+\zeta\tau_{\rm D}
\end{eqnarray}
This includes two phases: an early one up to a peak time $\tau_{\rm P}$ when
the BH grows exponentially with a timescale $\tau_{\rm ef}$ to a mass
$M_{{\rm BH}, \rm P}$ and emits with an Eddington ratio $\lambda_0$ until it
reaches a peak luminosity $\lambda_0\, M_{{\rm BH}, \rm P}\,c^2/t_{\rm Edd}$;
a late phase when the luminosity declines exponentially on a timescale
$\tau_{\rm D}$ up to a time $\tau_{\rm P}+\zeta\tau_{\rm D}$ when it shuts
off. With $\lambda_0$, $\epsilon_0$, we denote the average Eddington ratio
and radiative efficiency during the early, ascending phase. The $e-$folding
time associated to them is $\tau_{\rm ef}=\epsilon_0\, t_{\rm
Edd}/\lambda_0\,(1-\epsilon_0)$.

The lightcurve in Eq.~(\ref{eq|AGNlightcurve}) has been set in Lapi et al.
(2014) in order to comply with the constraints imposed by large observational
datasets concerning:
\begin{itemize}

\item the fraction of X-ray detected AGNs in FIR/K-band selected host
    galaxies (e.g., Alexander et al. 2005; Mullaney et al. 2012a; Wang et
    al. 2013a; Johnson et al. 2013);

\item the fraction of FIR detected galaxies in X-ray AGNs (e.g., Page et
    al. 2012; Mullaney et al. 2012b; Rosario et al. 2012) and optically
    selected quasars (e.g., Mor et al. 2012; Wang et al. 2013b; Willott et
    al. 2015);

\item related statistics via stacking of undetected sources (e.g.,
    Basu-Zych et al. 2013).

\end{itemize}
The same authors also physically interpreted the lightcurve according to a
specific BH-galaxy coevolution scenario.

In a nutshell, the scenario envisages that the early growth of the BH occurs
in an interstellar medium rich in gas and strongly dust-enshrouded (Lapi et
al. 2014; also Chen et al. 2015). The BH accretes in a demand-limited fashion
with values of Eddington ratios $\lambda$ appreciably greater than unity,
though the radiative efficiency $\epsilon$ may keep to low values because
slim-disc conditions develop. Since the BH mass is still small, the nuclear
luminosity, though appreciably super-Eddington, is much lower than that of
the starforming host galaxy, and the whole system behaves as a sub-mm bright
galaxy with an X-ray nucleus. On the other hand, close to the peak of the AGN
lightcurve, the BH mass has grown to large values, and the nuclear emission
becomes comparable or even overwhelms that of the surrounding galaxy. Strong
winds from the nucleus remove gas and dust from the ambient medium stopping
the star formation in the host, while the whole system shines as an optical
quasar. If residual gas mass is still present in the central regions, it can
be accreted in a supply-driven fashion so originating the declining part of
the lightcurve; this phase corresponds to the onset of the standard thin disk
accretion, which yields the observed SEDs of UV/optically-selected type-1
AGNs (Elvis et al. 1994; Hao et al. 2014). Actually, the data concerning the
fraction of starforming galaxies in optically-selected quasar samples suggest
such a descending phase to be present only for luminous objects, while in
low-luminosity ones tiny residual mass is present and the AGN fades more
drastically after the peak. When the accreting gas mass ends, the BH becomes
silent, while the stellar populations in the galaxy evolve passively. For the
most massive objects, the outcome will be a local elliptical-type galaxy with
a central supermassive BH relic.

All in all, we set the fiducial values of the parameters describing the BH
lightcurve on the basis of the Lapi et al. (2014) analysis. We shall discuss
the effects of varying them in Sect.~\ref{sec|BH_results}. Specifically, we
fiducially adopt $\tau_{\rm D}=3\, \tau_{\rm ef}$ and $\zeta\approx 3$ for
luminous AGNs with peak luminosity $L\ga 10^{13}\, L_\odot$, while $\tau_{\rm
D}=0$, i.e., the declining phase is almost absent for low-luminosity objects.
To interpolate continuously between these behaviors, we use a standard ${\rm
erfc}$-function smoothing
\begin{equation}\label{eq|tau_d}
{\tau_{\rm D}\over \tau_{\rm ef}} = 3\, \left[1-{1\over 2}\,{\rm
erfc}\left({1\over 2}\,\log{L_{\rm AGN}\over 10^{13} L_\odot}\right)\right]
\end{equation}
which is illustrated in Fig.~\ref{fig|parameters} (bottom panel). We note
that our results will be insensitive to the detailed shape of the smoothing
function. The value of $\zeta=3$ is fiducially adopted, since after a time
$\zeta\, \tau_{\rm D}$ after the peak the BH mass has almost saturated to its
final value. Results are unaffected by modest variation of this parameter.

We also fiducially assume that the Eddington ratio $\lambda_0$ of the
ascending phase depends on the cosmic time $t$ (or redshift $z$) after
\begin{equation}\label{eq|lambda_z}
\lambda_0(z)= 4\,\left[1-{1\over 2}\,{\rm erfc}\left({z-2\over 3}\right)\right]~,
\end{equation}
as illustrated in Fig.~\ref{fig|parameters} (top panel). As shown by Lapi et
al. (2006, 2014), such moderately super-Eddington values at high redshift
$z\ga 4$ are necessary to explain the bright end of the quasar luminosity
function (see also Shankar et al. 2009, 2013). During the demand-limited,
ascending phase of the lightcurve, $\lambda_0$ exceeds the characteristic
value $\lambda_{\rm thin}\approx 0.3$ for the onset of a slim accretion disc
(Laor \& Netzer 1989). On the other hand, during the declining phase of the
lightcurve, the Eddington ratio declines almost exponentially, so that after
the characteristic time $\tau_{\rm thin}\approx \tau_{\rm D}\,\log
\lambda_0/\lambda_{\rm thin}$ the transition to a thin accretion disc takes
place. At high redshift where $\lambda_0\approx 4$, the thin-disc regime sets
in only after a time $\tau_{\rm thin}\approx 2.5\, \tau_{\rm D}$ after the
peak, while at low redshift where $\lambda_0\la 1$ it sets in about
$\tau_{\rm thin}\approx 1.2\,\tau_{\rm D}$ after the peak. We notice that
statistically the fraction of slim discs should increase toward high-$z$, as
suggested by the data analysis of Netzer \& Trakhtenbrot (2014), paving the
way for their use as standard candles for cosmological studies (Wang et al.
2013c). The time-averaged value of $\lambda$ during the declining phase is,
to a good approximation, given by $\langle \lambda\rangle\simeq \lambda_0\,
(1-e^{-\zeta})/\zeta\approx \lambda_0/\zeta$, while the time-averaged value
during the thin disk regime $\langle \lambda\rangle\simeq (\lambda_{\rm
thin}-\lambda_0\, e^{-\zeta})/(\zeta-\log \lambda_0/\lambda_{\rm thin})$
ranges from $0.1$ at $z\la 1$ to $0.3$ at $z\ga 3$. We will see that such
values $\langle\lambda\rangle$ averaged over the Eddington distribution
associated to the adopted lightcurve reproduce well the observational
determinations (Vestergaard \& Osmer 2009; Kelly \& Shen 2013).

As to the radiative efficiency, we take into account the results of several
numerical simulations and analytic works (Abramowicz et al. 1988; Mineshige
et al. 2000; Watarai et al. 2001; Blandford \& Begelman 2004; Li 2012;
Begelman 2012; Madau et al. 2014), that indicate a simple prescription to
relate the efficiency $\epsilon$ and the Eddington ratio $\lambda$ in both
slim and thin disc conditions
\begin{equation}\label{eq|epslambda}
\epsilon = {\epsilon_{\rm thin}\over 2}\, {\lambda\over e^{\lambda/2}-1}~;
\end{equation}
here $\epsilon_{\rm thin}$ is the efficiency during the thin disc phase,
which may range from $0.057$ for a non-rotating to $0.32$ for a maximally
rotating Kerr BH (Thorne 1974). We will adopt $\epsilon_{\rm thin}=0.1$ as
our fiducial value (see Davis \& Laor 2011). In conditions of mildly
super-Eddington accretion with $\lambda\ga$ a few the radiative efficiency
$\epsilon\la 0.3\, \epsilon_{\rm thin}$ applies, while in sub-Eddington
accretion regime with $\lambda\la 1$ it quickly approaches the thin disc
value $\epsilon=\epsilon_{\rm thin}$. We also take into account that along
the declining portion of the lightcurve $\epsilon$ increases given the almost
exponential decrease of $\lambda$. The time averaged values
$\langle\epsilon\rangle$ of the efficiency during the declining phase and
during the thin disc regime are illustrated in Fig.~\ref{fig|parameters}. We
expect that the redshift dependence of the average efficiency is negligible
during the thin disc regime; this is in qualitative agreement with the
findings by Wu et al. (2013) based on spectral fitting in individual type-1
quasars (see also Davis \& Laor 2011 for a low-$z$ determination), and by Cao
(2010) based on continuity equation analysis. However, we caution the reader
that the determination of the radiative efficiency is plagued by several
systematic uncertainties and selection effects (see discussion by Raimundo et
al. 2012). Large samples of AGNs with multiwavelength SEDs and BH masses are
crucial in fully addressing the issue.

The final BH mass $M_{\rm BH}$ is easily linked to the mass at the peak
$M_{{\rm BH},\rm P} $ appearing in Eq.~(\ref{eq|AGNlightcurve}). One has
\begin{equation}\label{eq|BH_mass}
M_{{\rm BH}} = \int_0^{\tau_{\rm P}+\zeta\,\tau_{\rm D}}\, d\tau'
{1-\epsilon\over \epsilon\, c^2}\, L_{\rm AGN}(\tau') = M_{{\rm BH},\rm
P}\,\left[1+f_\epsilon\,{\tau_{\rm D}\over \tau_{\rm
ef}}\,(1-e^{-\zeta})\right]~.
\end{equation}
The correction factor $f_\epsilon$ takes into account the modest change of
the quantity $(1-\epsilon)/\epsilon$ along the declining phase. We have
checked that $f_\epsilon\approx 0.8$ for any reasonable value of
$\epsilon_{\rm thin}$. Notice that at high redshift where $\lambda_0\approx
4$, the fraction of BH mass accumulated in thin-disc conditions is only $5\%$
of the total, while it can be as large as $20\%$ at low redshift where
$\lambda_0\approx 1$. This is relevant since most of the BH mass estimates at
high-$z$ are based on single-epoch method, that probes the UV/optical bright
phase.

\begin{figure}
\epsscale{0.5}\plotone{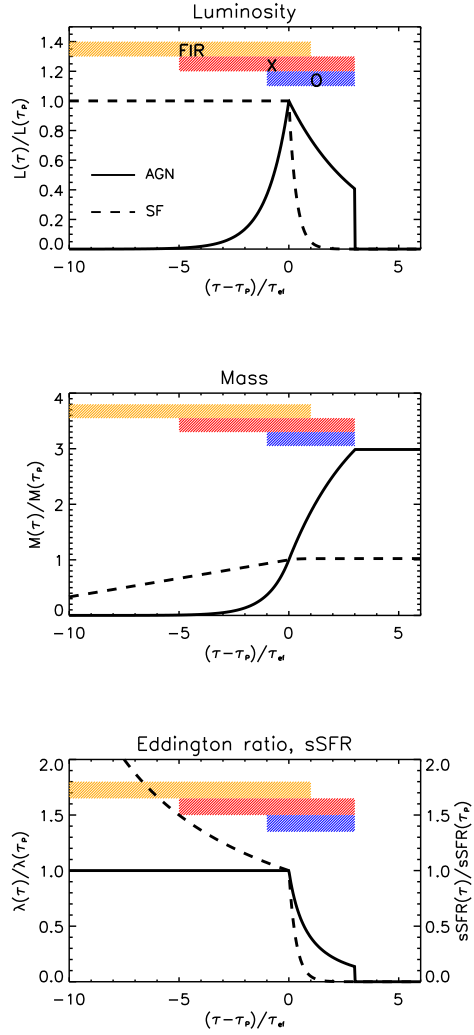}\caption{Time evolution of the
luminosity (top panel), mass (middle panel), and Eddington ratio/specific SFR
(bottom panel) normalized to the value at time of the AGN luminosity peak.
Solid lines refer to AGN-related, and dashed lines to star-formation related
quantities. The orange area highlights the stage when the galaxy
is starforming and appears as a FIR bright source (orange); the red and blue
areas highlight the stages when the BH is detectable as an X-ray AGN and as
an optical quasar, respectively. See main text below
Eq.~(\ref{eq|AGNlightcurve}) for details.}\label{fig|lightcurves}
\end{figure}

\begin{figure}
\epsscale{0.7}\plotone{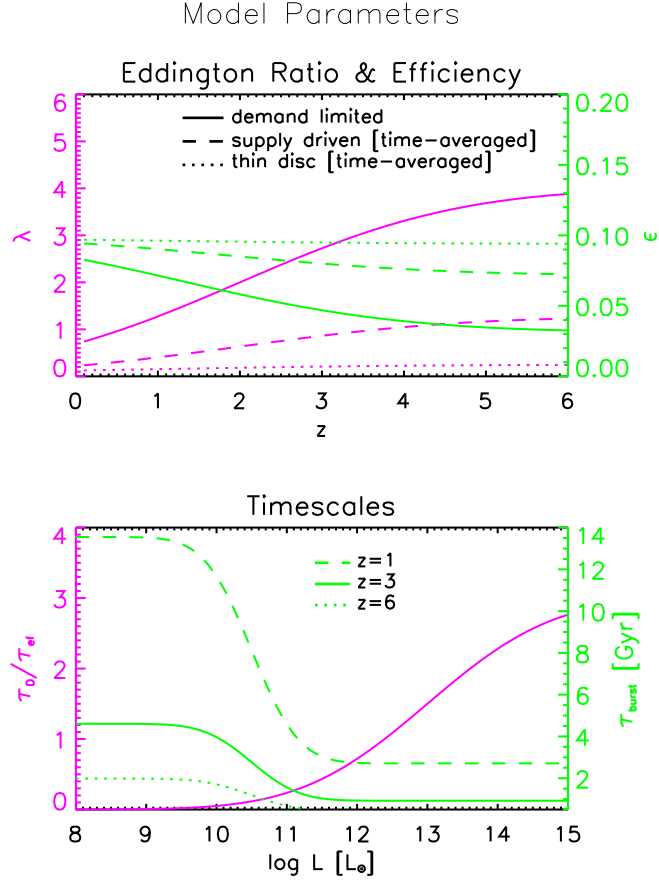}\caption{Top panel: The adopted
Eddington ratio (magenta lines) and radiative efficiency (green line) as a
function of redshift. The values in the ascending, demand-limited phase
(solid lines) and the time-averaged values during the descending,
supply-driven phase (dashed lines) and during the thin-disc regime (dotted
lines) are also shown. Bottom panel: the characteristic timescale $\tau_{\rm
D}/\tau_{\rm ef}$ of the AGN descending phase (magenta line) and the duration
$\tau_{\rm burst}$ of the stellar burst (green lines) at redshift $z=1$
(dashed), $3$ (solid), and $6$ (dotted) as a function of the peak AGN and of
the SFR luminosity, respectively.}\label{fig|parameters}
\end{figure}

The evolution with the internal time $\tau$ of the AGN luminosity, mass, and
Eddington ratio are sketched in Fig.~\ref{fig|lightcurves}. We also
schematically indicate with colors the stages (according to the framework
described below Eq.~\ref{eq|AGNlightcurve}) when the galaxy is detectable as
a FIR-bright source, and the nucleus is detectable as an X-ray AGN and as an
optical quasar.

\subsubsection{BHs: solution}\label{sec|BH_solution}

Given the lightcurve in Eq.~(\ref{eq|AGNlightcurve}), the fraction of the
time spent by the BH per luminosity bin reads
\begin{equation}\label{eq|dtaudL}
\sum_i{{\rm d}\tau_i\over {\rm d}L_{\rm AGN}} = {\tau_{\rm ef}+\tau_{\rm D}\over L_{\rm AGN}}\,
\Theta_{H}\left[L_{\rm AGN}\leq L_{\rm AGN}(M_{{\rm BH}})\right]~,
\end{equation}
where $L_{\rm AGN}(M_{\rm BH})$ is the maximum luminosity corresponding to a
final BH mass $M_{\rm BH}$, that can be written as
\begin{equation}\label{eq|Lmax}
L_{\rm AGN}(M_{\rm BH}) = {\lambda\, M_{\rm BH} c^2\over t_{\rm Edd}}\,
\left[1+f_\epsilon\,{\tau_{\rm D}\over \tau_{\rm ef}}\,(1-e^{-\zeta})\right]^{-1}~;
\end{equation}
the expression stresses the relevance of the mass accretion during the AGN
descending phase. This implies that the time spent in a luminosity bin is
longer by a factor $\tau_{\rm D}$ than on assuming a simple growing
exponential curve, and that Eq.~(\ref{eq|Lmax}) is implicit since $\tau_{\rm
D}/\tau_{\rm ef}$ is itself a function of the luminosity.

Using Eq.~(\ref{eq|dtaudL}) in the continuity equation (neglecting dry
merging, i.e., no source term) yields
\begin{equation}
L_{\rm AGN}\,N(L_{\rm AGN},t) = \int_{M_{{\rm BH}}(L_{\rm AGN})}^\infty{\rm d}M_{{\rm BH}}\,
\partial_t N(M_{{\rm BH}},t)\, \left[\tau_{\rm ef}+\tau_{\rm D}\right]~,
\end{equation}
where the minimum final mass that have shone at $L_{\rm AGN}$ is given by the
inverse of Eq.~(\ref{eq|Lmax}). We proceed by differentiating both sides with
respect to $L_{\rm AGN}$ and rearranging terms to find
\begin{equation}
{L_{\rm AGN}\over f_{{\rm BH},L_{\rm AGN}}}\,{\partial_{L_{\rm
AGN}}\left[L_{\rm AGN}\,N(L_{\rm AGN},t)\right]\over \tau_{\rm ef}+\tau_{\rm
D}} = - [\partial_t N(M_{{\rm BH}},t)\, M_{{\rm BH}}]_{|M_{{\rm BH}}(L_{\rm
AGN})}~;
\end{equation}
in deriving this equation we have defined $f_{{\rm BH},L_{\rm AGN}}\equiv
{\rm d}\log M_{{\rm BH}}/{\rm d}\log L_{\rm AGN}$, which is not equal to
unity since $\tau_{\rm D}/\tau_{\rm ef}$ in Eq.~(\ref{eq|Lmax}) depends on
$L_{\rm AGN}$.

Finally, we integrate over cosmic time and pass to logarithmic bins. The
outcome reads
\begin{equation}\label{eq|BH_solution}
N(\log M_{{\rm BH}},t) = -\int_{0}^t{{\rm d}t'\over f_{{\rm BH},L_{\rm
AGN}}}\, {\partial_{\ln L_{\rm AGN}} \left[N(\log L_{\rm AGN})\right]\over
\tau_{\rm ef}(L_{\rm AGN},t')+\tau_{\rm D}(L_{\rm AGN},t')}_{|L_{\rm
AGN}(M_{{\rm BH}},t')}~.
\end{equation}
Note that in practice we have started the integration at $z_{\rm in}=10$
assuming that the BH mass function at that time was negligibly small. This
solution constitutes a novel result. In the case when $\tau_{\rm D}=0$, and
when $\lambda$ and $\epsilon$ are constant with redshift and luminosity, the
above equation reduces to the form considered by Marconi et al. (2004).

\subsubsection{BHs: results}\label{sec|BH_results}

\begin{figure}
\epsscale{1.}\plotone{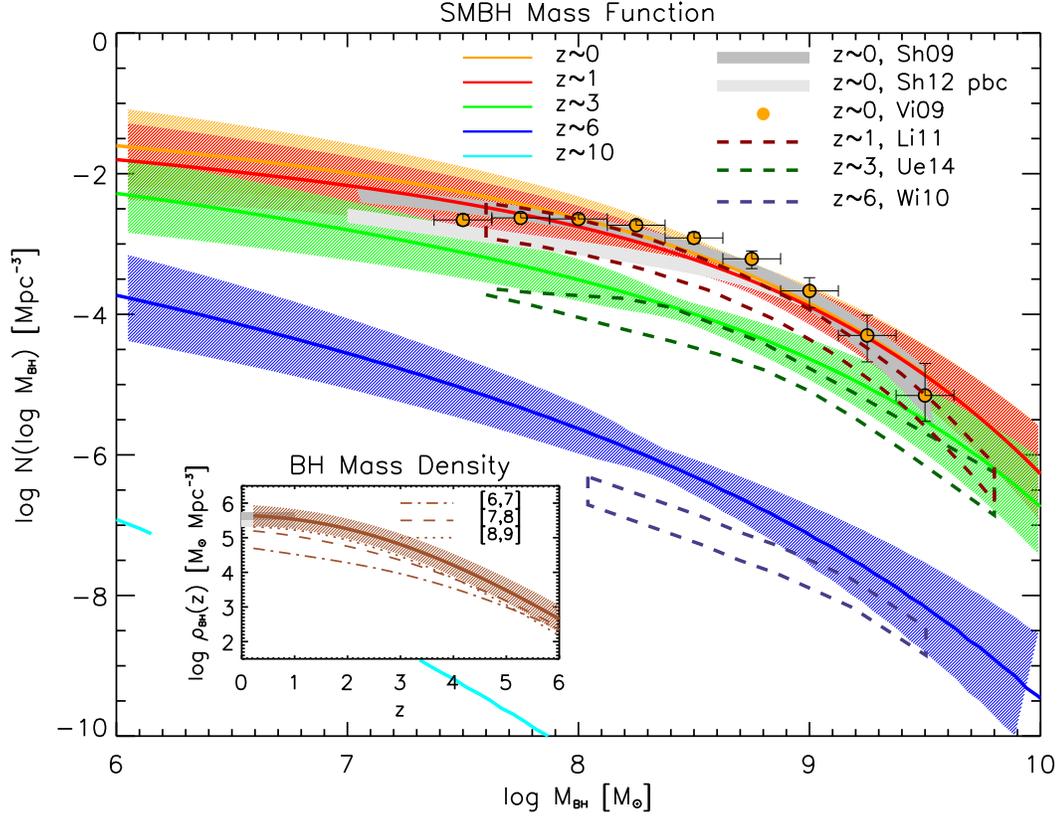}\caption{The supermassive BH mass function
$N(\log M_{{\rm BH}})$ as a function of final BH mass $M_{{\rm BH}}$. Results
from the continuity equation (see Sect.~\ref{sec|BH_solution}) at redshift
$z=0$ (orange), $z=1$ (red), $3$ (green), and $6$ (blue) are plotted as solid
lines, with the hatched areas illustrating the associated uncertainty; the
cyan line is the extrapolation to $z=10$ plotted for illustration. The dark
grey shaded area illustrates the collection of estimates by Shankar et al.
(2009) built by combining the stellar mass or velocity dispersion function
with the $M_{\rm BH}-M_\star$ or $M_{\rm BH}-\sigma$ relations of elliptical
galaxies; the light shaded area is the determination by Shankar et al. (2012)
corrected to take into account the different relations followed by
pseudobulges. The orange circles illustrate the determination at $z=0$ by
Vika et al. (2009). The red dashed area illustrate the determination at
$z\sim 1$ by Li et al. (2011), the green dashed area shows the range of
models by Ueda et al. (2014) at $z\sim 3$, and the blue dashed area the
estimate by Willott et al. (2010b) at $z\sim 6$. The inset shows the BH mass
density as a function of redshift computed from the continuity equation, for
the overall mass range (solid line with hatched area), and for BH masses
$\log M_{\rm BH}/M_\odot$ in the ranges $[6,7]$ (dot-dashed line), $[7,8]$
(dashed line), $[8,9]$ (dotted line). The grey shaded areas illustrate the
observational constraints from the above $z=0$ mass function by Shankar et
al. (2009, 2012).}\label{fig|BH_MF}
\end{figure}

In Fig.~\ref{fig|BH_MF} we illustrate our results on the supermassive BH mass
function at different representative redshifts. The outcomes of the
continuity equation can be fitted by the functional shape of
Eq.~(\ref{eq|AGN_LF}) with $L_{\rm AGN}$ replaced by $M_{\rm BH}$, and with
the parameter values reported in Table~1; the resulting fits are accurate to
within $5\%$ in the redshift range from $0$ to $6$ and over the BH mass range
$M_{\rm BH}$ from a few $10^7$ to a few $10^9\, M_\odot$.

We also illustrate two determinations of the local mass function. One is from
the collection of estimates by Shankar et al. (2009), that have been built by
combining the stellar mass or velocity dispersion functions with the
corresponding $M_{\rm BH}-M_\star$ (Haring \& Rix 2004) or $M_{\rm
BH}-\sigma$ (Tremaine et al. 2002) relations of elliptical galaxies and
classical bulges. The other is the determination by Shankar et al. (2012)
corrected to take into account the different relations followed by
pseudobulges. In addition, we present the determination at $z=0$ by Vika et
al. (2009) based on an object-by-object analysis and on the $M_{\rm BH}-L$
(McLure \& Dunlop 2003) relationship.

The BH mass function at $z\approx 0$ from the continuity equation provides
and almost perfect rendition of the local estimates by Shankar et al. (2009)
and Vika et al. (2009) when $\epsilon_{\rm thin}=0.1$ is adopted. At
$z\approx 1$ we find a BH mass function which is very similar to the local
determination. Our result is in good shape with, though on the high side of,
the determination by Li et al. (2011), based on luminosity (or stellar) mass
functions and mild evolution of the $M_{\rm BH}-L$ (or $M_{\rm BH}-M_\star$)
relationship. The same also holds at $z\approx 2$, which is not plotted for
clarity.

At $z\approx 3$ we find a BH mass function which at the knee is a factor
about $10$ below the local data. We are in good agreement with the
determination by Ueda et al. (2014) based on continuity equation models. This
is expected since we adopt similar bolometric luminosity functions, and
around $z\approx 3$ we have similar values of the Eddington ratio and
radiative efficiency. At $z\approx 6$ we find a BH mass function which is
about $3$ orders of magnitude smaller than the local data. We compare our
result with the estimate by Willott et al. (2010b) in the range $M_{\rm
BH}\sim 10^8-3\times 10^9\, M_\odot$. This has been derived by combining the
Eddington ratio distribution from single-epoch BH mass estimates to the
optical quasar luminosity functions corrected for obscured objects. At the
knee of the mass function we find a good agreement with our result based on
the continuity equation, while at lower masses we predict a slightly higher
number of objects.

The reasonable agreement with previous determinations in the redshift range
$z\sim 0-6$ validates our prescriptions for the lightcurves, the redshift
evolution of $\lambda_0(z)$, and the $\epsilon-\lambda$ relation of
Eq.~(\ref{eq|epslambda}). Besides, we recall that these were already
independently tested against the observed fractions of AGNs hosted in sub-mm
galaxies and related statistics (Lapi et al. 2014; see
Sect.~\ref{sec|intro}).

Note that during the slim-accretion regime, where most of the BH mass is
accumulated, the effective efficiency amounts to $\epsilon\la 0.05$ given our
assumed value $\epsilon_{\rm thin}\approx 0.1$ in Eq.~(\ref{eq|epslambda}),
see also Fig.~\ref{fig|parameters}. This requires a bit of discussion. In
principle, during a coherent disk accretion, the BH is expected to spin up
very rapidly, and correspondingly the efficiency is expected to attain values
$\epsilon\ga 0.15$ (Madau et al. 2014), corresponding to $\epsilon_{\rm
thin}\approx 0.3$ after Eq.~(\ref{eq|epslambda}). However, such a high value
of the efficiency would produce a local BH mass function in strong
disagreement with the data. This can be understood just basing on the
standard Soltan argument. In fact, the BH mass density inferred from the AGN
luminosity density would amount to $\rho_{\rm BH} \approx 2\times 10^4\,
(1-\epsilon)/\epsilon \, M_\odot$ Mpc$^{-3}\la 10^5\, M_\odot$ Mpc$^{-3}$.
Plainly, the $z=0$ result would fall short of the local observational
determinations, that yields a fiducial mass density of $\rho_{\rm BH}\approx
4.5\times 10^5\, M_\odot$ Mpc$^{-3}$ (using the Shankar et al. 2009 local
mass function). The discrepancy is even worse if one considers the local mass
function obtained by combining the velocity dispersion or stellar mass
function with the recently revised $M_{{\rm BH}}-\sigma$ or $M_{{\rm
BH}}-M_{\star}$ relations by McConnell \& Ma (2013) and Kormendy \& Ho
(2013), which feature a higher overall normalization.

In App.~B we have also tested the relevance of dry merging processes
(contributing via the source term in the continuity equation) in shaping the
BH mass function. At $z\ga 1$ BH merging effects are found to be
statistically negligible (see also Shankar et al. 2009), although smaller
mass BHs may undergo substantial merging activity with possible impact on the
seed distribution (for a review, see Volonteri 2010). At $z\la 1$ our tests
indicate that the mass function is mildly affected only for $M_{\rm BH} \ga
10^9\, M_\odot$.

Thus an average slim-disc efficiency $\epsilon\la 0.05$ is required. During
the slim-disc accretion, such a low efficiency can be maintained by, e.g.,
chaotic accretion, efficient extraction of angular momentum by jets, or
similar processes keeping the BH spin to low levels (King \& Pringle 2006;
Wang et al. 2009; Cao 2010; Li et al. 2012; Barausse 2012; Sesana et al.
2014). We also remark that an efficiency $\epsilon\la 0.05$ eases the
formation of supermassive BHs at very high redshift $z\ga 6$, so alleviating
any requirement on initial massive seeds (Volonteri 2010). On the other hand,
the supermassive BH mass function only poorly constrains the values of the BH
spin during the final thin-disc phase, which the current estimates suggest to
be rather high (Reynolds et al. 2013).

Bolometric corrections and obscured accretion can concur to alleviate the
requirement of a low slim-disc efficiency. Bolometric corrections are based
on studies of SEDs for large samples of AGNs (e.g., Elvis et al. 1994;
Richards et al. 2006; Hopkins et al. 2007; Lusso et al. 2010, 2012; Vasudevan
et al. 2010; Hao et al. 2014). In fact, the SEDs depend on the main selection
of the objects (e.g., X-ray, UV, optical, IR), possibly on the Eddington
ratio (Vasudevan et al. 2010; Lusso et al. 2012), and on bolometric
luminosity (Hopkins et al. 2007). The recent analysis of Hao et al. (2014)
finds no significant dependencies on redshift, bolometric luminosity, BH mass
and Eddington ratio of the mean SEDs for a sample of about $400$ X-ray
selected type-1 AGNs, although a large dispersion is signalled. A large
fraction of objects with accretion obscured at wavelengths ranging from X-ray
to optical bands has been often claimed, also in connection with their
contribution to the X-ray background (Comastri et al. 1995). The fraction
compatible with it at substantial X-ray energies has been recently discussed
by Ueda et al. (2014) and properly inserted in our AGN bolometric luminosity
functions.

Concerning the overall evolution of the BH mass function, we find that most
of the BH mass growth occurs at higher redshifts for the more massive objects
(see the inset of Fig.~\ref{fig|BH_MF}). The overall BH mass density at $z=0$
amounts to $\rho_{\rm BH}\approx 4.5\times 10^5\,M_\odot$ Mpc$^{-3}$, in
excellent agreement with observational determinations.

\begin{figure}
\epsscale{.85}\plotone{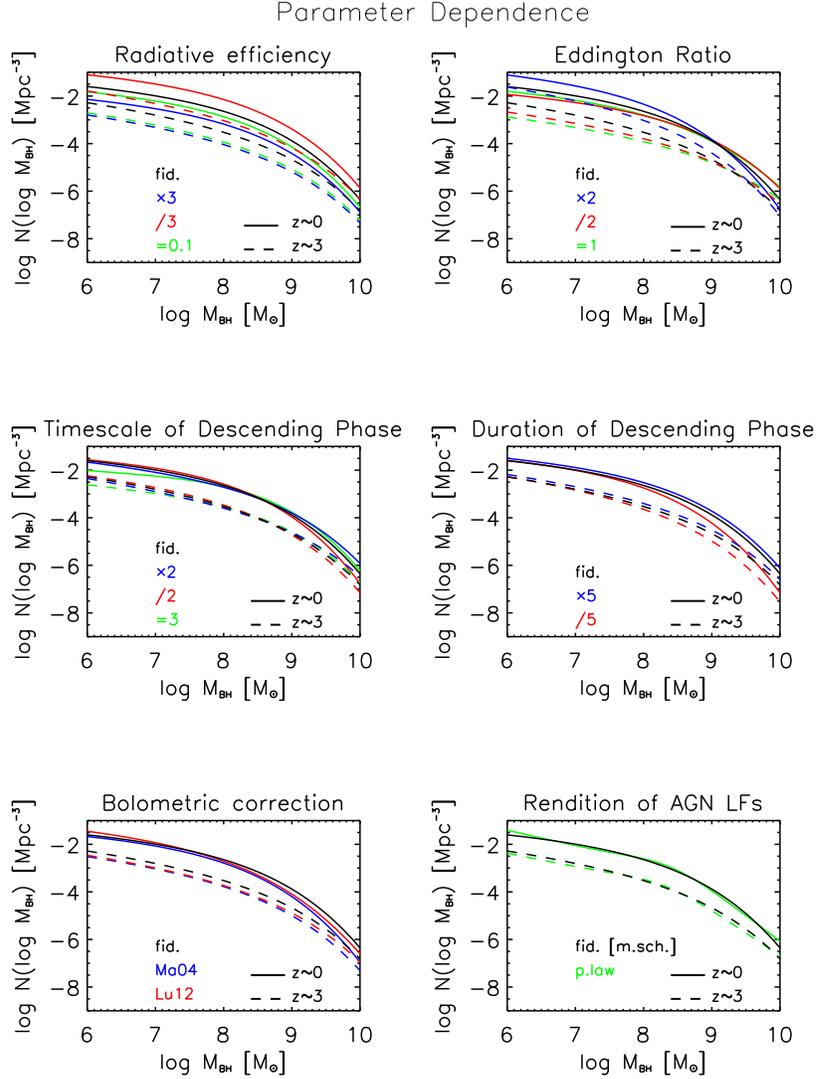}\caption{Comparison plot showing the
dependence of the supermassive BH mass function on various assumptions; for
clarity only results at $z=0$ (solid lines) and at $z=3$ (dashed lines) are
plotted. In the top and middle panels, we show the effects of changing the
values of the parameters describing the AGN lightcurve. The black lines are
for our fiducial values, the red and blue lines refer to values decreased or
increased of the amount specified in the legend, and the green lines to
constant values in redshift and luminosity. In the bottom left panel
we illustrate the effect of changing the optical/X-ray bolometric
corrections: the black lines refer to our reference one by Hopkins et al.
(2007), while the blue and red lines refer to the ones proposed by Marconi et
al. (2004) and by Lusso et al. (2012), respectively. In the bottom right
panel, we illustrate the effect of changing the functional form used to
analytically render the AGN luminosity functions: the black lines refer to
our fiducial rendition via a modified Schechter function (cf.
Eq.~\ref{eq|AGN_LF}), while the green lines refer to a double powerlaw
representation.}\label{fig|BH_MF_comp}
\end{figure}

In Fig.~\ref{fig|BH_MF_comp} we show how our results on the mass function
depend on various assumptions. The top and middle panels illustrate the
effect of changing the parameters of the lightcurve: radiative efficiency
$\epsilon$, Eddington ratio $\lambda$, timescale of the descending phase
$\tau_{\rm D}$, and duration of the descending phase $\zeta$. For clarity we
plot results only at $z=0$ and $z=3$. We illustrate our fiducial model, and
compare it with the outcome for values of the parameters decreased or
increased relative to the reference ones.

To understand the various dependencies, it is useful to assume a simple,
piecewise powerlaw shape of the luminosity function in the form $N(\log
L_{\rm AGN})\propto L_{\rm AGN}^{-\eta}$, with $\eta\la 1$ at the faint and
$\eta>1$ at the bright end. Then it is easily seen from
Eq.~(\ref{eq|BH_solution}) that the resulting mass function behaves as
\begin{equation}
N(\log M_{{\rm BH}})\propto {1-\epsilon\over \epsilon}\, \lambda^{1-\eta}\,
{[1+(\tau_{\rm D}/\tau_{\rm ef})\,(1-e^{-\zeta})]^\eta\over 1+\tau_{\rm
D}/\tau_{\rm ef}}\, \eta\,M_{{\rm BH}}^{-\eta}~.
\end{equation}
Thus the BH mass function features an almost inverse dependence on $\epsilon$
at given BH mass. The dependence on $\lambda$ is inverse at the high-mass
end, which is mostly contributed by high luminosities where $\eta>1$. On the
other hand, it is direct at the low-mass end, mainly associated to faint
sources with $\eta\la 1$. The opposite applies to the dependence on
$\tau_{\rm D}/\tau_{\rm ef}$, since roughly $N(\log M_{\rm BH})\propto
(\tau_{\rm D}/\tau_{\rm ef})^{\eta-1}$. Finally, the dependence on $\zeta$ is
mild, and practically irrelevant for $\zeta\ga 3$ since the exponential
$e^{-\zeta}$ in Eq.~(\ref{eq|BH_mass}) tends rapidly to zero. Differences in
the results are more evident in the $z=0$ than in the $z=3$ mass function,
since this is an integrated quantity, as expressed by
Eq.~(\ref{eq|BH_solution}).

In the bottom left panel we illustrate the effect of changing the
optical/X-ray bolometric corrections: the black lines refer to our reference
one by Hopkins et al. (2007), while the blue and red lines refer to the ones
proposed by Marconi et al. (2004) and by Lusso et al. (2012), respectively.
It is easily seen that the impact on the BH mass function is limited,
actually well within the uncertainties associated to the input luminosity
functions, and to the observational determinations of the local BH mass
function.

In the bottom right panel, we illustrate the effect of changing the
functional form used to analytically render the AGN luminosity functions: the
black lines refer to our fiducial rendition via a modified Schechter function
(cf. Eq.~\ref{eq|AGN_LF}), while the green lines refer to a standard double
powerlaw representation (e.g., Ueda et al. 2014; Aird et al. 2015). It is
seen that our results on the BH mass function are marginally affected; this
is because both shapes render comparably well the input AGN luminosity
functions.

\begin{figure}
\epsscale{0.5}\plotone{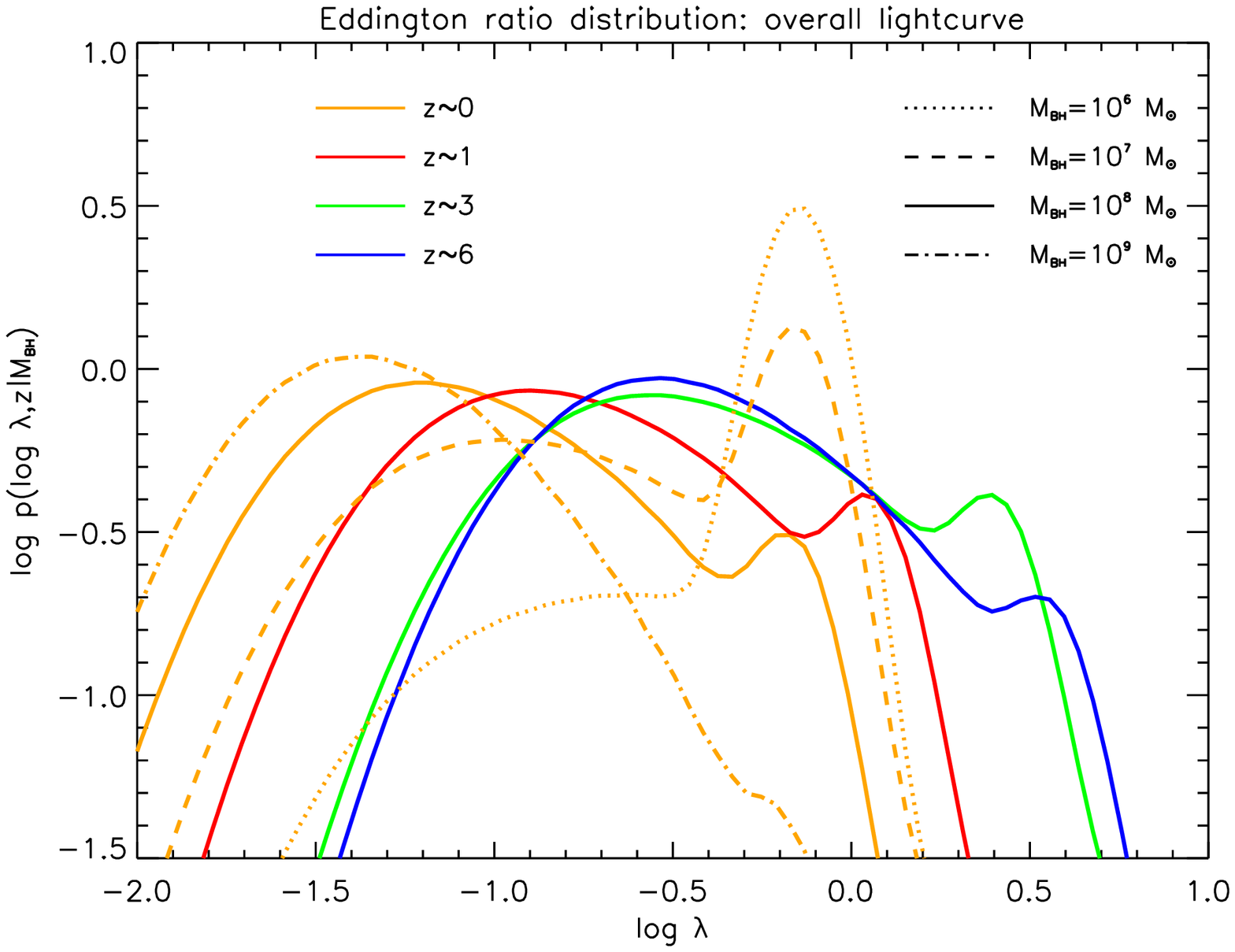}\\\plotone{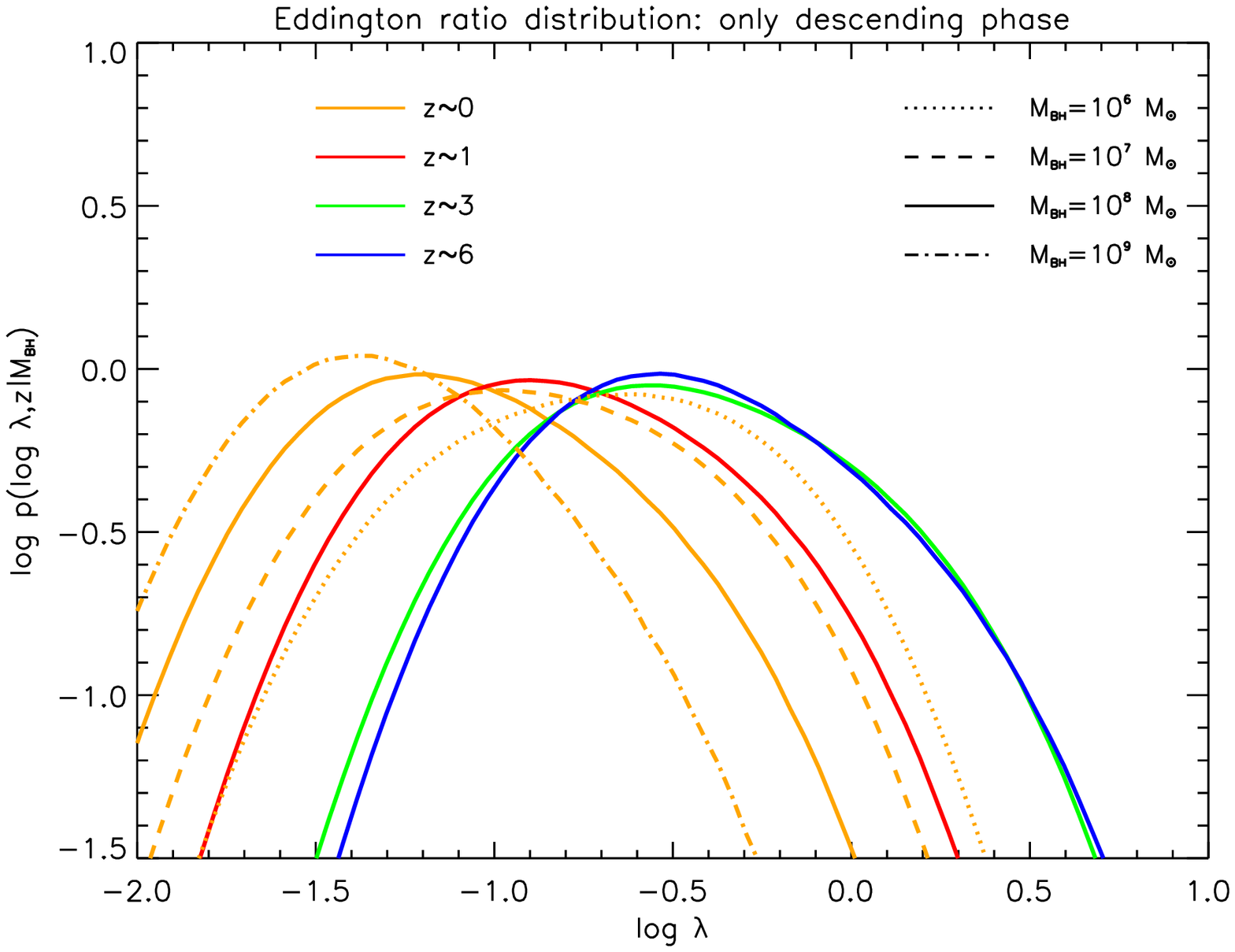}\\\plotone{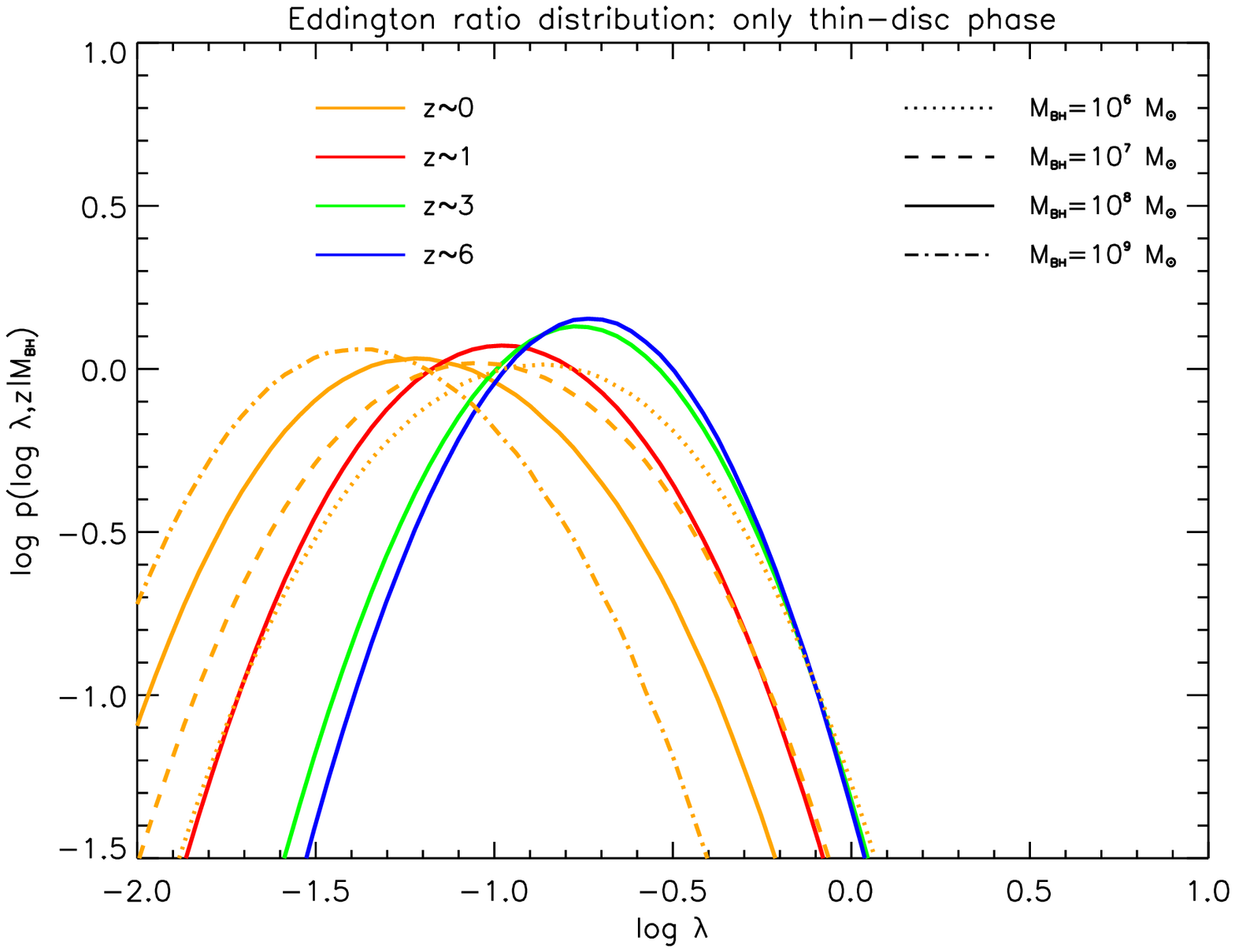}\caption{The
Eddington ratio distribution $P(\log\lambda|M_{\rm BH},z)$ associated to the overall
lightcurve (top panel), only to the descending phase (middle panel), and only
to the thin-disc phase (bottom panel), at different redshift $z=0$ (orange),
$1$ (red), $3$ (green), and $6$ (blue) and for different BH masses $M_{\rm
BH}=10^6$ (dotted), $10^7$ (dashed), $10^8$ (solid), and $10^9\, M_\odot$
(dot-dashed); for clarity the results relative to masses $10^6$, $10^7$ and
$10^9\, M_\odot$ are plotted only at $z=0$.}\label{fig|Plambda}
\end{figure}

In Fig.~\ref{fig|Plambda} we illustrate the Eddington ratio distribution
$P(\log\lambda|M_{\rm BH},z)$ associated to the overall adopted lightcurve at
different redshift and for different final BH masses. Typically at given
redshift and BH mass, the distribution features a Gaussian peak at high
values of $\lambda$, and then a powerlaw increase toward lower values of
$\lambda$ before an abrupt cutoff. The peak reflects the value of $\lambda$
in the ascending part of the lightcurve. Actually since $\lambda(\tau)$ is
constant there, the peak should be a Dirac $\delta-$function. However, small
variations around the central value and observational errors will broaden the
peak to a narrow Gaussian as plotted here; a dispersion of $0.3$ dex has been
safely adopted. The powerlaw behavior reflects the decrease of
$\lambda(\tau)$ during the declining part of the lightcurve at late times,
and the cutoff in the distribution mirrors that of the lightcurve. The
relative contribution of the Gaussian peak at high $\lambda$ and of the
powerlaw increase at low $\lambda$ depends on the relative duration of the
declining and ascending phases. Thus at given redshift, small mass BHs
feature a much more prominent peak and a less prominent powerlaw increase
relative to high mass ones. This is because in small mass objects the
descending phase is shorter. At given BH mass, the distributions shift to the
left, i.e., toward smaller values of $\lambda$, as the redshift decreases.
This is because the initial value $\lambda_0(z)$ decreases with redshift, as
prescribed by Eq.~(\ref{eq|lambda_z}).

Such a distribution has been computed under the assumption that the overall
lightcurve can be sampled. However, from an observational perspective, the
Eddington ratio distribution is usually determined via single-epoch BH mass
estimates of type-1 AGNs. This implies that only a portion of the descending
phase can be sampled. To ease the comparison with observations, we present in
the middle and bottom panels of Fig.~\ref{fig|Plambda} the expected
distribution considering only the descending phase (including both the final
portion of the slim-disc phase and the while thin-disc phase, with
$\lambda\ga 0.3$), and only the thin-disc phase (i.e., the portion with
$\lambda\la 0.3$). The resulting distributions feature a powerlaw shape,
whose slope depends on the portion of the declining phase that can be
effectively sampled: the shorter this portion, the steeper the powerlaw. The
result is roughly consistent with the observational determinations by, e.g.,
Kelly \& Shen (2013), although a direct comparison is difficult due to
observational selection effects. In fact, different observations are likely
to sample diversely the initial part of the declining phase, and this will
possibly make the expected and the observed distributions even more similar.
Note that especially at $z\la 1$, BH reactivations, which are not included in
our treatment (both in terms of lightcurve descriptions and of stochasticity
of the events), can contribute to broaden the Eddington ratio distribution
toward very low values of $\lambda\la 10^{-2}$ as estimated in the local
Universe (e.g., Kauffmann \& Heckman 2009; Brandt \& Alexander 2015).

\begin{figure}
\epsscale{1.0}\plotone{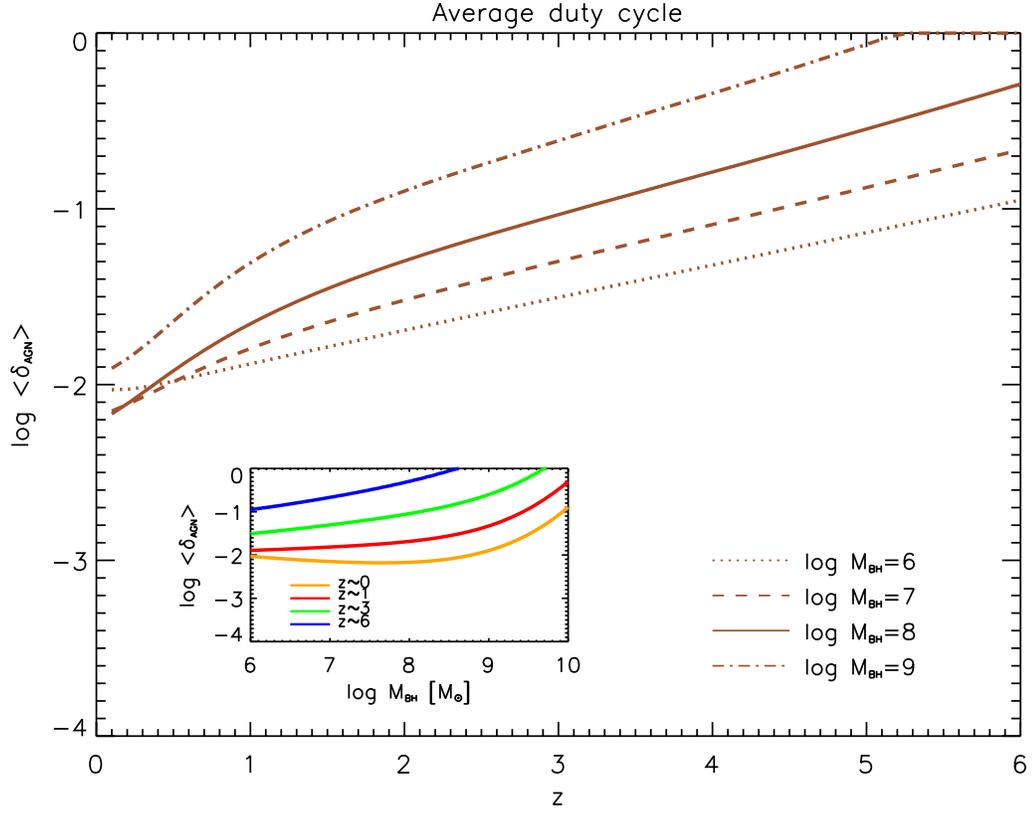}\caption{The average AGN duty cycle
$\langle\delta_{\rm AGN}\rangle$ as a function of redshift $z$, for different
BH masses $M_{\rm BH}=10^6$ (dotted), $10^7$ (dashed), $10^8$ (solid), and
$10^9\, M_\odot$ (dot-dashed). The inset illustrates the AGN duty cycle as a
function of the BH mass at different redshift $z=0$ (orange), $z=1$ (red),
$3$ (green), and $6$ (blue).}\label{fig|AGN_duty}
\end{figure}

In Fig.~\ref{fig|AGN_duty} we present the AGN duty cycle $\langle \delta_{\rm
AGN}\rangle$ averaged over the Eddington ratio distribution associated to the
adopted lightcurve. Specifically, this has been computed as
\begin{eqnarray}
\nonumber \langle \delta_{\rm AGN}\rangle(M_{\rm BH},t)\equiv {N_{\rm AGN}(\log M_{\rm BH},t)\over
N(\log M_{\rm BH},t)} = {1\over N(\log M_{\rm BH},t)}\, \int{\rm d}\log
\lambda~P(\log\lambda|M_{\rm BH},z)\times\\
\\
\nonumber \times N(\log L_{\rm AGN},t)_{|L_{\rm AGN}(M_{\rm BH},\lambda)}
\end{eqnarray}\label{eq|AGN_duty}
where $L_{\rm AGN}(M_{\rm BH},\lambda)$ is given by Eq.~(\ref{eq|Lmax}). In
our approach based on the continuity equation, the duty cycle is a quantity
derived from the luminosity and mass functions. It provides an estimate for
the fraction of active BHs relative to the total. At given redshift, the
average duty cycle increases with the BH mass, since more massive BHs are
typically produced by more luminous objects, that feature the descending
phase of the lightcurve. On the contrary, small mass BHs are originated
mainly by low-luminosity objects for which the descending phase is absent. At
given BH mass, the duty cycle increases with the redshift, essentially
because to attain the same final mass, BHs stay active for a larger fraction
of the shorter cosmic time. This is especially true for BHs with high masses,
up to the point that they are always active ($\delta_{\rm AGN}\approx 1$) for
$z\ga 3$. This agrees with the inferences from the strong clustering observed
for high-redshift quasars (Shen et al. 2009; Shankar et al. 2010; Willott et
al. 2010b; Allevato et al. 2014); we will further discuss the issue in
Sect.~\ref{sec|abmatch_BHvshalo}. The increase of the duty cycle with BH mass
is consistent with the active fraction measured by Bundy et al. (2008), Xue
et al. (2010), although the issue is still controversial and strongly
dependent on obscuration-corrections (see Schulze et al. 2015). On the other
hand, we again remark that our approach does not include AGN reactivations,
which may strongly enhance the duty cycle for low-luminosity objects
especially at $z\la 1$, accounting for the estimates by, e.g., Ho et al.
(1997), Green \& Ho (2007), Goulding \& Alexander (2009), Schulze \& Wisotzki
(2010).

\begin{figure}
\epsscale{0.5}\plotone{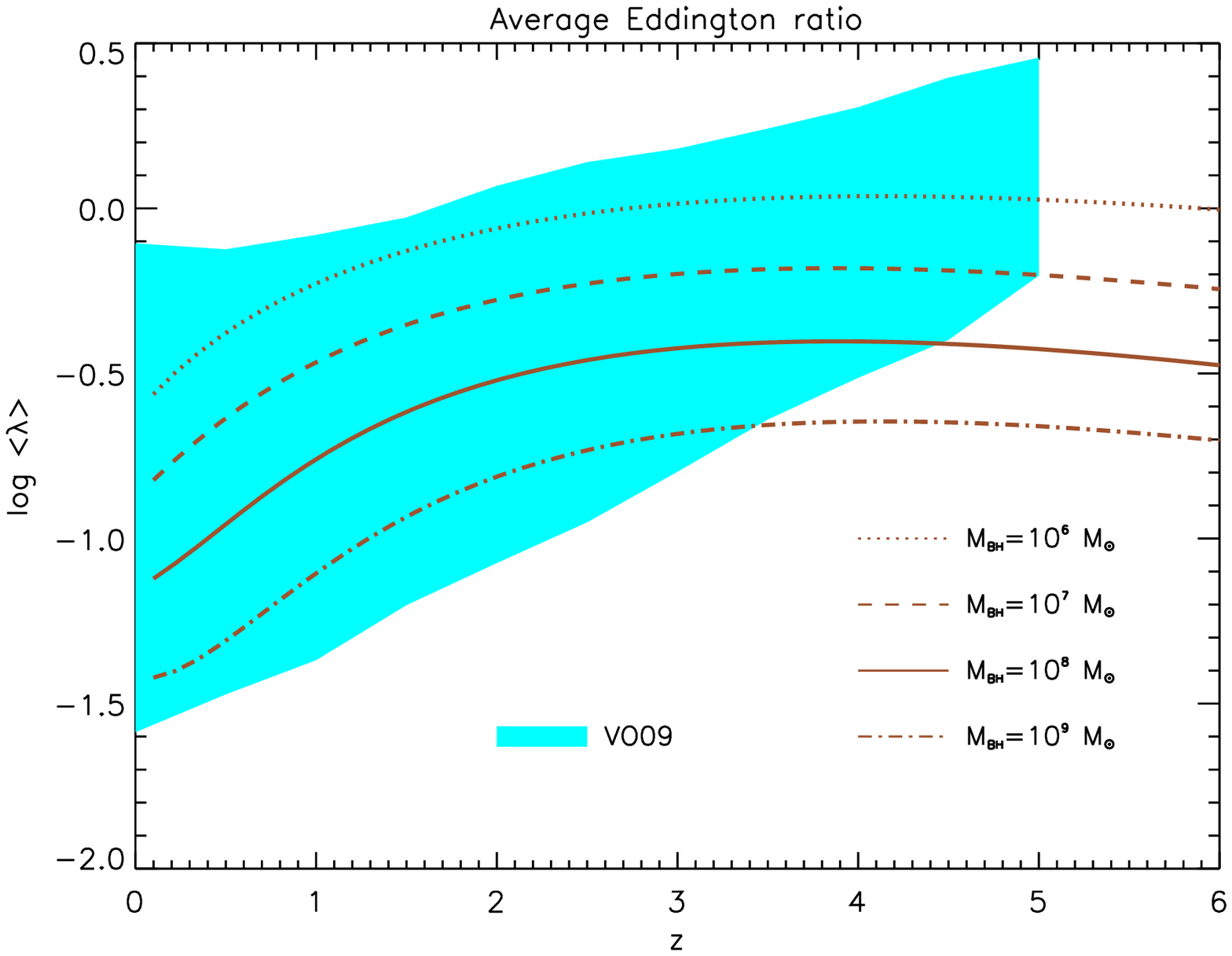}\\\epsscale{0.48}\plotone{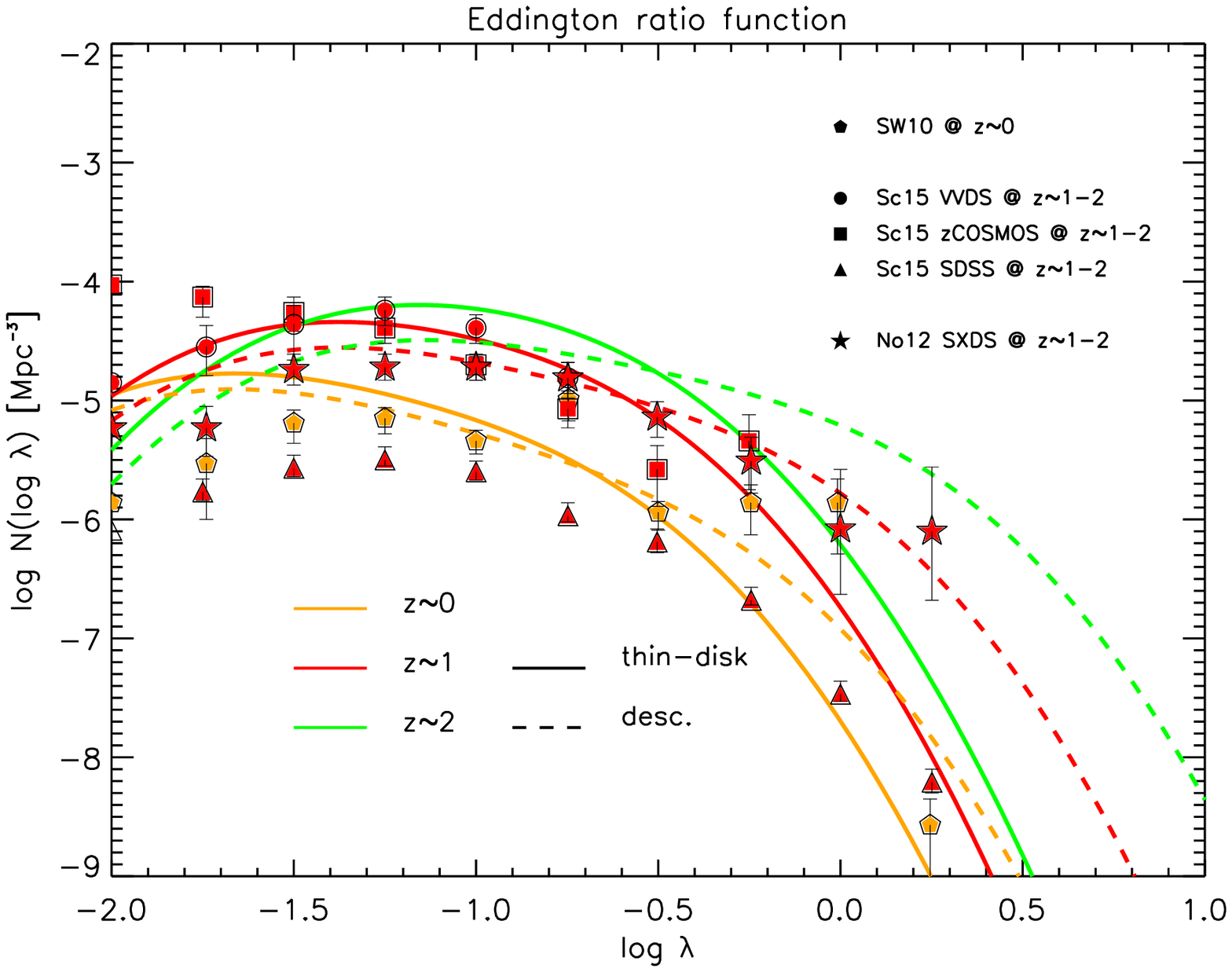}\plotone{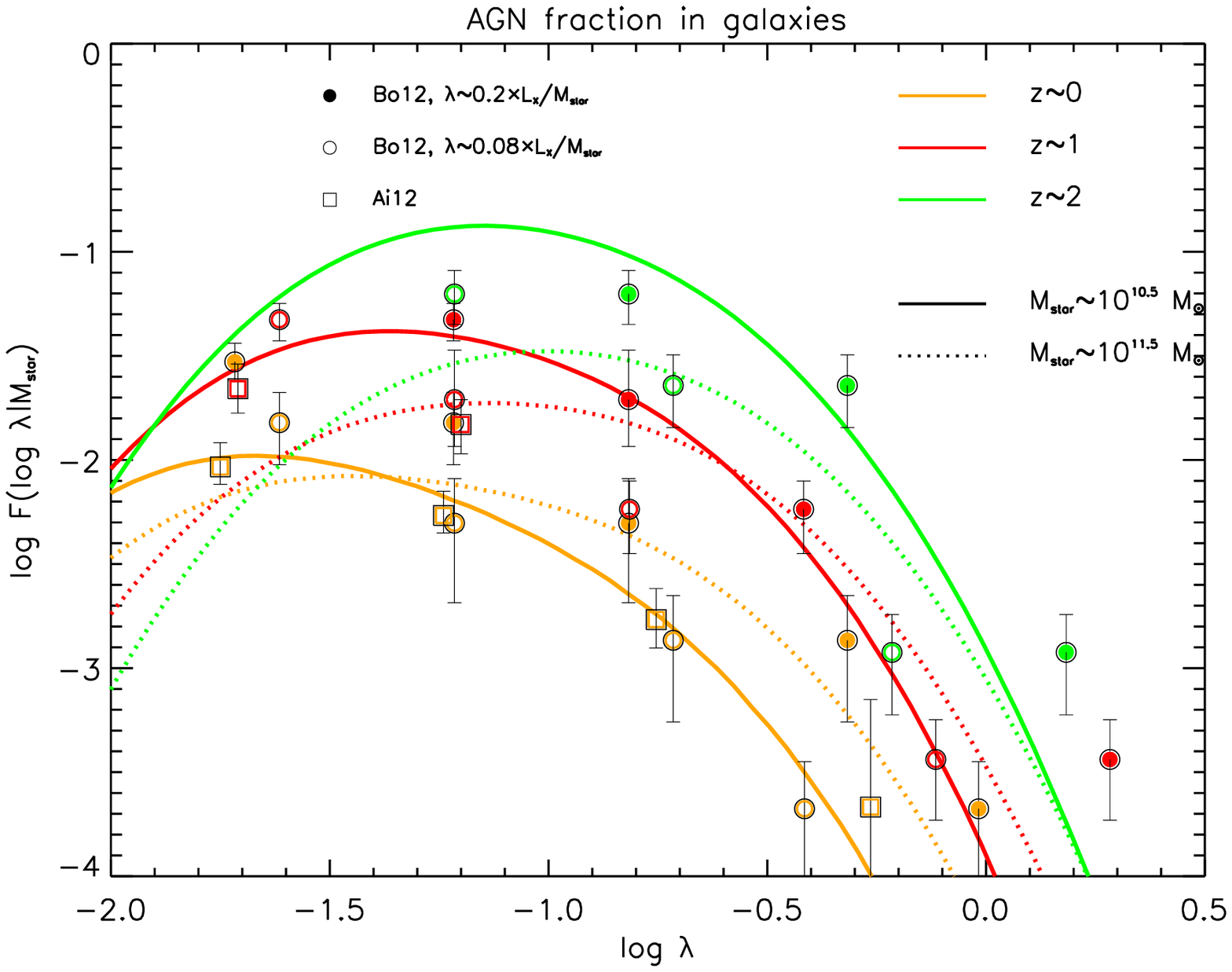}
\caption{Top panel: the average Eddington ratio $\langle\lambda_{\rm
AGN}\rangle$ as a function of redshift $z$, for different BH masses $M_{\rm
BH}=10^6$ (dotted), $10^7$ (dashed), $10^8$ (solid), and $10^9\, M_\odot$
(dot-dashed); the cyan shaded area covers the range of measured values by
Vestergaard \& Osmer (2009). Bottom left panel: the Eddington ratio function at
redshift $z=0$ (orange), $1$ (red), and $2$ (green) associated to the
thin-disk phase (solid lines) or to the descending phase (dashed lines; the
outcome when considering the overall lightcurve is very similar); observational
estimates at $z\sim 0$ are from Schulze \& Wisotzki (2010; orange diamonds),
at $z\sim 1-2$ from Schulze et al. (2015; red circles for VVDS, squares for
zCOSMOS, and triangles for SDSS) and from Nobuta et al. (2012; red stars for
SXDS). Bottom right panel: the fraction of galaxies of given stellar mass
(solid lines for $M_\star\sim 10^{10.5}\, M_\odot$ and dotted lines for
$M_\star\sim 10^{11.5}\, M_\odot$) hosting an AGN with given Eddington ratio
at redshift $z=0$ (orange), $1$ (red), and $2$ (green);
data are from Aird et al. (2012; orange squares) and from Bongiorno et al.
(2012; red and green circles), where the latter have been converted with the
relation $\lambda\approx 0.2\, L_X/M_\star$ (filled circles) or
$\lambda\approx 0.08\, L_X/M_\star$ (empty circle), see text for
details.}\label{fig|AGN_lambda}
\end{figure}

In Fig.~\ref{fig|AGN_lambda} (top panel) we present the AGN Eddington ratio
$\langle\lambda\rangle$ averaged over the lightcurve, computed as
\begin{equation}
\langle\lambda\rangle(M_{\rm BH},t)\equiv {1\over N(\log M_{\rm BH})}\,
\int{\rm d}\log \lambda~\lambda\,P(\log\lambda|M_{\rm BH},z)\, N(\log
L_{\rm AGN})_{|L_{\rm AGN}(M_{\rm BH},\lambda)}~.
\end{equation}
At given final BH mass, the Eddington ratio decreases with the redshift, as a
consequence of the dependence adopted in Eq.~(\ref{eq|lambda_z}). The average
values are consistent with those observed for a sample of quasars by
Vestergaard \& Osmer (2009). Note that to take into account the observational
selection criteria, we have used the Eddington ratio distribution associated
to the descending phase, presented in the middle panel of
Fig.~\ref{fig|Plambda}.

In Fig.~\ref{fig|AGN_lambda} (bottom left panel) we show the Eddington ratio
function, that has been computed as
\begin{equation}\label{eq|AGN_lambda}
N(\log\lambda,z)\equiv \int{\rm d}\log M_{\rm BH}~P(\log\lambda|M_{\rm
BH},z)\, N_{\rm AGN}(\log M_{\rm BH},z)~;
\end{equation}
the outcome is mildly sensitive to the lower integration limit, and a value
$M_{\rm BH}\approx 10^8\, M_\odot$ has been adopted to compare with data (see
Schulze et al. 2015). For the sake of completeness, we present the results
when using the Eddington ratio distribution associated to the thin disk phase
(cf. bottom panel of Fig.~\ref{fig|Plambda}) or to the whole descending phase
(cf. middle panel of Fig.~\ref{fig|Plambda}), being the outcome for the
overall lightcurve very similar to this latter case. Our results from the
continuity equation are confronted with the estimates from Schulze \&
Wisotzki (2010) at $z\sim 0$, and from Schulze et al. (2015) and Nobuta et
al. (2012) at $z\sim 1-2$, finding a nice agreement within the observational
uncertainties and the clear systematic differences between datasets.

In Fig.~\ref{fig|AGN_lambda} (bottom right panel) we present a related
statistics, i.e., the fraction $F(\log \lambda|M_\star)$ of galaxies with
given stellar mass hosting an AGN (active fraction) with a given Eddington
ratio. This has been computed simply on dividing the quantity
$P(\log\lambda|M_{\rm BH},z)$ $N_{\rm AGN}(\log M_{\rm BH},z)$ by the number
of galaxies $N(\log M_\star,z)$ with given stellar mass $M_\star$ (the
stellar mass function, cf. Sect.~\ref{sec|STAR_MF}). Plainly, $M_{\rm BH}\sim
2\times 10^{-3}\, M_\star$ must be set to the BH mass corresponding to
$M_\star$, the result being rather insensitive to the $M_{\rm BH}/M_\star$
ratio adopted; we further take into account a scatter of $0.3$ dex in this
relationship, whose effect is to make the active fraction
$F(\log\lambda|M_\star)$ to depend on $M_\star$ more weakly than the
Eddington ratio distribution $P(\log\lambda|M_{\rm BH},z)$ depends on $M_{\rm
BH}$. We illustrate the outcome for a range of stellar masses from
$M_\star\sim 10^{10.5}\, M_\odot$ to $M_\star\sim 10^{11.5}\, M_\odot$; it
turns out to be only mildly dependent on $M_\star$, and especially so at low
redshift $z\la 2$, as also indicated by current observations.

In fact, our results can be compared with the observational estimates at
$z\sim 0-2$ by Aird et al. (2012) and Bongiorno et al. (2012). The latter
authors actually provide the active fraction as a function of the observable
quantity $L_X/M_\star$; this can be converted into an Eddington ratio by
assuming an X-ray bolometric correction and a value for the $M_{\rm
BH}/M_\star$ ratio. Bongiorno et al. (2012) suggest an overall conversion
factor $\lambda\approx 0.2\, L_X/M_\star$ (here cgs units are used for the
quantities on the r.h.s.). We also plot their data points when using a
conversion $\lambda\approx 0.08\, L_X/M_\star$ (corresponding, e.g., to a
larger ratio $M_{\rm BH}/M_\star$ or a lower bolometric correction), giving
more consistency with the determination by Aird et al. (2012).

All in all, our results from the continuity equation are found to be in good
agreement with the observational estimates, reproducing their mild dependence
on stellar mass and their shape for $z\la 2$ down to an Eddington ratio
$\lambda\approx$ a few $10^{-2}$. On the other hand, AGNs at $z\la 1$ with
tiny accretion rates corresponding to $\lambda< 10^{-2}$ are likely triggered
by reactivations, which are not included in our lightcurve, and can
contribute to maintain a powerlaw shape of the Eddington ratio function and
of the active fraction down to $\lambda\sim 10^{-4}$ as observed by Aird et
al. (2012).

\subsection{The stellar mass function}\label{sec|STAR_MF}

Now we turn to the evolution of the stellar mass function from the
SFR-luminosity function.

\subsubsection{Stars: SFR function from UV and FIR luminosity}\label{sec|STAR_LF}

SFR can be inferred by lines (mainly Ly${\alpha}$ and H${\alpha}$) and by
continuum emission (mainly UV, FIR, radio and X-ray) of star forming galaxies
(for a review, see Kennicutt \& Evans 2012). The SFR is directly proportional
to the UV (chiefly far-UV, FUV) luminosity, which traces the integrated
emission by young, massive stars. On the other hand, even a small amount of
dust causes large extinction of the UV emission. The absorbed luminosity is
re-emitted at longer wavelengths, mostly in the $4-1000\, \mu$m interval.
Therefore ideal estimates would be based on both the UV ($L_{\rm UV}$) and
the FIR ($L_{\rm IR}$) observed luminosities for large galaxy samples at
relevant redshifts. This would allow to derive the total luminosity
proportional to the SFR
\begin{equation}
L_{\rm SFR}=L_{\rm UV}^{\rm corr}=L_{\rm UV}+f\, L_{\rm IR}~;
\end{equation}
here the fraction $f$ is meant to subtract from the budget the FIR luminosity
contributed by diffuse dust (cirrus) absorbing the light from less massive
older stars.

Actually, the SFR functions are inferred from UV or FIR-selected samples. In
both cases calibrations and \emph{corrections} come in (Calzetti et al. 2000;
Hao et al. 2011; Murphy et al. 2011; Kennicutt \& Evans 2012). The
calibration constants between SFR and luminosity in UV and FIR are
practically the same, as expected on energy conservation arguments (Kennicutt
1998; Kennicutt \& Evans 2012); for FIR luminosity we have
\begin{equation}
\log {{\rm SFR}\over M_\odot~{\rm yr}^{-1}} \approx -9.81+\log f\, {L_{\rm
IR}\over L_\odot}~,
\end{equation}
while for extinction-corrected UV luminosity we have
\begin{equation}\label{eq|LUV_sfr}
\log {{\rm SFR}\over M_\odot~{\rm yr}^{-1}}\approx -7.42-0.4\, M_{\nu_{\rm
UV}}^{\rm corr} \approx -9.76+\log {\nu_{\rm UV} L_{\nu_{\rm UV}}^{\rm
corr}\over L_\odot}~,
\end{equation}
$\nu_{\rm UV}$ being the frequency corresponding to $1550$ \AA.
\footnote{Some UV data are given at a restframe wavelength $\lambda$
different from $1550$ \AA; Eq.~(\ref{eq|LUV_sfr}) still holds provided that
on the right hand side the correction $-\log \lambda/1550$ \AA~is added.
E.g., for $\lambda=1350$ \AA~the correction amounts to $0.06$ and the zero
point calibration becomes $-7.36$.}.

The FIR luminosity ascribable to the diffuse dust emission (cirrus) depends
on several aspects such as stellar content (mass, age and chemical
composition), dust content and spatial distribution. The cirrus emission is
characterized by dust temperature lower than the emission associated with
star formation in molecular clouds (Silva et al. 1998; Rowlands et al. 2014).
There are local galaxies with quite low SFR, whose FIR luminosity is
dominated by cirrus emission. E.g., Hao et al. (2011) found $1-f\sim 0.5$ for
a sample of nearby starforming galaxies with SFR $\dot M_\star\la 30\,
M_{\odot}$ yr$^{-1}$. However, the fraction $1-f$ strongly reduces with
increasing star formation (e.g., Clemens et al. 2013). For strong local
starbursting galaxies with $\dot M_\star\ga 100\,M_{\odot}$ yr$^{-1}$ and
$L_{\rm IR}\ga 10^{12}\, L_{\odot}$, such as Arp 220, we get $1-f\la $ a few
percent (Silva et al. 1998; Rowlands et al. 2014). Hereafter we will assume
that $f=1$ for $L_{\rm IR}\ga 10^{12}\, L_{\odot}$ and that at such large
luminosities the SFR can be soundly inferred from the FIR luminosity.

At low luminosity, the SFR is better estimated from UV emission. For this
purpose it is essential to allow for dust absorption, that may drastically
reduce the UV luminosity to a few percent or even less of its intrinsic
value. When only UV data are available, the correlation between the UV slope
$\beta$ and the IRX ratio $L_{\rm IR}/L_{\rm UV}$ is largely used to infer
the dust attenuation (Meurer et al. 1999). While initially proposed only for
low redshift galaxies, the method has been tested and applied also to high
redshifts (Reddy et al. 2010; Overzier et al. 2011; Hao et al. 2011; Bouwens
et al. 2013, 2015). However, for large values of the slope $\beta$ and of the
attenuation, the spread around the correlation becomes huge (Overzier et al.
2011; Reddy et al. 2010) and the estimate of attenuation becomes quite
uncertain even in local samples (e.g., Howell et al. 2010). On the other
hand, the estimate of attenuation for UV-selected samples is less dispersed
for galaxies with SFRs $\dot M_\star\la 1\, M_{\odot}$ yr$^{-1}$. In such
instances the correction to UV luminosity is more secure and relatively small
on average (Bouwens et al. 2013). In fact, this is also suggested by the UV
attenuation inferred by combining the H$\alpha$ attenuation and the Calzetti
extinction curve (Hopkins et al. 2001; Mancuso et al. 2015).

\begin{figure}
\epsscale{1.}\plotone{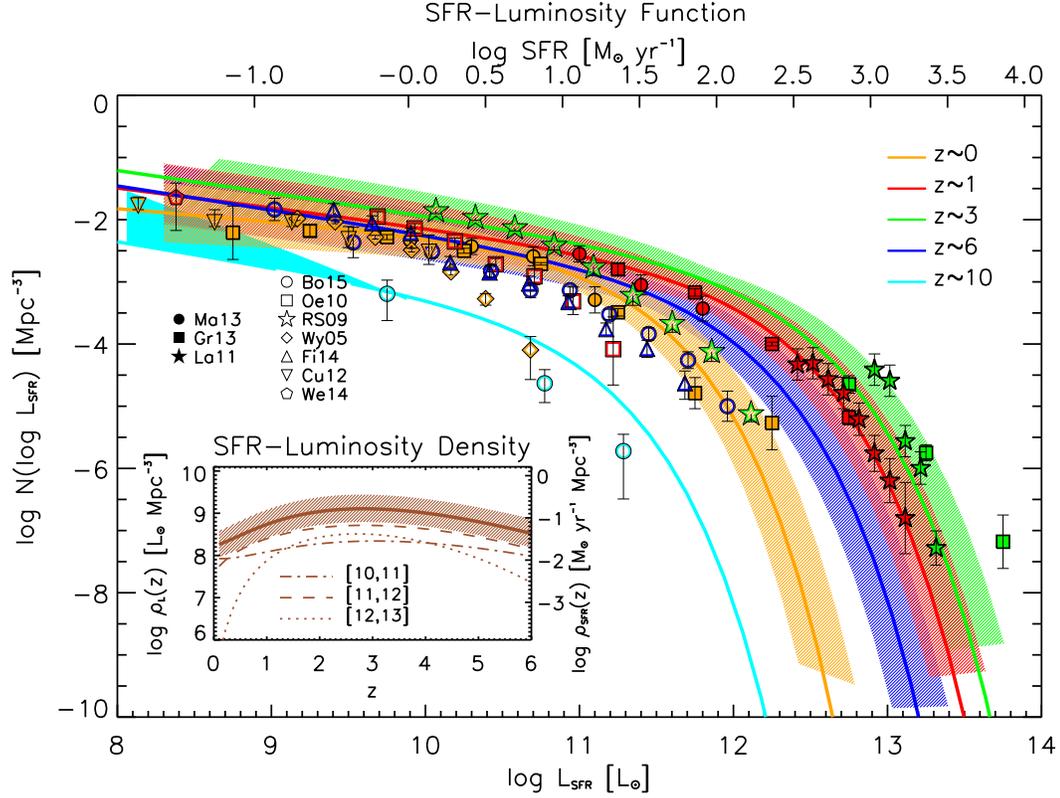}\caption{The SFR-luminosity function
$N(\log L_{\rm SFR})$ at redshift $z=0$ (orange), $1$ (red), $3$ (green), and
$6$ (blue), vs. the bolometric luminosity $L_{\rm SFR}$ associated to the SFR
(lower axis) and vs. the SFR (upper axis). Infrared data are from Magnelli et
al. (2013; filled circles), Gruppioni et al. (2013; filled squares), and Lapi
et al. (2011; filled stars); UV data (dust corrected, see text) are from
Bouwens et al. (2015; open circles), Oesch et al. (2010; open squares), Reddy
\& Steidel (2009; open stars), Wyder et al. (2005; open diamonds),
Finkelstein et al. (2014; open triangles), Cucciati et al. (2012; open
reversed triangles), Weisz et al. (2014; pentagons). The solid lines
illustrate the analytic rendition of the luminosity functions as described in
Sect.~\ref{sec|STAR_LF}, and the hatched areas represent the associated
uncertainty; the cyan line is the extrapolation to $z=10$ plotted for
illustration, with the shaded area representing the uncertainty on the
faint-end slope. The inset shows the SFR-luminosity density as a function of
redshift, for the overall luminosity range probed by the data (solid line
with hatched area), and for bolometric luminosity $\log L_{\rm SFR}/L_\odot$
in the ranges $[10,11]$ (dot-dashed line), $[11,12]$ (dashed line), $[12,13]$
(dotted line).}\label{fig|STAR_LF}
\end{figure}

Given all these considerations, we build up the overall SFR-luminosity
$L_{\rm SFR}$ function as follows. We start from the luminosity function at
different redshifts observed in the FIR band by Magnelli et al. (2013),
Gruppioni et al. (2013), Lapi et al. (2011), and in the UV band by Bouwens et
al. (2015), Oesch et al. (2010), Reddy \& Steidel (2009), and Wyder et al.
(2005). The data are reported in Fig.~\ref{fig|STAR_LF}. In passing note that
the SFR and the SFR-luminosity $L_{\rm SFR}$ scales on the upper and lower
axis have been related assuming the approximate relation $\log {\rm
SFR}/M_\odot~{\rm yr}^{-1} \approx -9.8+\log {L_{\rm SFR}/L_\odot}$, and so
the number density per unit SFR or per unit luminosity is the same.

For the FIR samples we assume $f=1$, while for the UV samples at redshift
$z\ga 2$ we have exploited the dust correction suggested by Meurer et al.
(1999) and Bouwens et al. (2013, 2015). At $z\la 2$ the attenuation has been
kept to the values found by Bouwens et al. (2013) for $z\approx 2.5$
galaxies. This assumption at $z\la 1$ produces an UV attenuation somewhat in
between those proposed by Wyder et al. (2005) and Cucciati et al. (2012), and
the one proposed by Hopkins et al. (2001). However, we stress that for
galaxies with $L_{\rm UV}\la 10^{10}\, L_\odot$ the correction is anyway
smaller than a factor $\sim 2$.

Fig.~\ref{fig|STAR_LF} shows that at any redshift we lack a robust
determination of the SFR-luminosity function at intermediate luminosities;
this occurs for two reasons: first, UV data almost disappear above $L_{\rm
UV}\approx 10^{11}\, L_\odot$ (see also Reddy et al. 2010) because of dust
extinction, while FIR data progressively disappear below $L_{\rm UV}\approx
10^{12}\, L_\odot$ because of current observational limits. Second, the UV
correction for $L_{\rm UV}\ga 10^{10}\, L_\odot$ or intrinsic SFR $\dot
M_\star\ga 1\, M_\odot$ yr$^{-1}$ becomes progressively uncertain, as
discussed above.

To fill in the gap, we render the overall SFR distribution with a continuous
function, whose shape is basically determined at the bright end by the FIR
data and at the faint end by the UV data. Specifically, we exploit the same
modified-Schechter functional shape of Eq.~(\ref{eq|AGN_LF}), with $L_{\rm
AGN}$ replaced by $L_{\rm SFR}$ and the parameter values reported in Table~1.
The UV data at the faint and FIR data at the bright are smoothly connected by
our analytic renditions at various redshifts. We also illustrate an estimate
of the associated $1\sigma$ uncertainty. In the inset we plot the evolution
with redshift of the SFR-luminosity density and the contribution to the total
by specific luminosity ranges.

It happens that our rendition of the data closely follows the model proposed
by Mao et al. (2007) and Cai et al. (2014), wherein the extinction is
strongly differential with increasing SFR (and gas metallicity). In such
models, the faint end of the UV luminosity function at high redshift is
dictated by the rate of halo formation, while the bright end is modeled by
the dust content in rapidly starforming galaxies. At $z\ga 6$ reliable
statistics concern only UV-selected galaxies endowed with low SFR. At high
luminosity we have extrapolated the behavior from lower redshift $z\la 4$,
finding a good agreement with the model proposed by Cai et al. (2014). This
extrapolation implies, at $z\ga 6$, a significant fraction of dusty galaxies
with SFR $\dot M_\star\ga 10^2\, M_{\odot}$ yr$^{-1}$, which are missed by UV
selection. Clues of such a population are scanty, but not totally missing.
Riechers et al. (2014) detected a dust obscured galaxy at $z\approx 6.34$
with SFR $\dot M_\star\approx 2900\,M_{\odot}$ yr$^{-1}$, and Finkelstein et
al. (2014) a second one at $z\approx 7.51$ with SFR $\dot M_\star\approx
300\, M_{\odot}$ yr$^{-1}$. The large SFR end at $z\ga 6$ will be probed in
the near future by \textsl{ALMA} and \textsl{JWST} observations.

In passing, we have also reported the extrapolation of the SFR-luminosity
function to $z=10$ (cyan line in Fig.~\ref{fig|STAR_LF}). It is interesting
to compare this with the recent estimates from UV observations by Bouwens et
al. (2015). At $L_{\rm UV}\approx 10^{9.7}\, L_{\odot}$ the extrapolation
matches the observed number density around $10^{-3}$ Mpc$^{-3}$. For smaller
$L_{\rm UV}\approx 10.^{9.7}\, L_{\odot}$ we remark that the slope of the
luminosity function is highly uncertain; data extrapolation suggests a slope
in the range from $-1.65$ to $-2$, as illustrated by the cyan shaded area. At
the other end, for $L_{\rm UV}\approx 10^{10.4}\, L_{\odot}$, the
extrapolated number density is around $10^{-4}$ Mpc$^{-3}$, a factor around
$3$ times larger than that observed in the UV. This possibly suggests that
dust already at $z\approx 10$ affects the UV data toward the bright end, as
it happens at lower redshift.

\subsubsection{Stars: lightcurve}\label{STAR_lightcurve}

There are three time-honored assumptions regarding the behavior of the SFR as
a function of galactic age: exponentially increasing (up to a ceiling value),
exponential decreasing, constant.

Here we specialize to a very simple, constant lightcurve, motivated by the
recent FIR data from the \textsl{Herschel} satellite concerning
high-redshift, luminous starbursting galaxies, and their physical
interpretation on the basis of the BH-galaxy coevolution model by Lapi et al.
(2014). Hence we adopt
\begin{eqnarray}\label{eq|SFRlightcurve}
L_{\rm SFR}(\tau|M_\star,t)\nonumber &=& \kappa_\star\,\dot{M}_{\star} ~~~~~~~~~~~~ \tau\le \tau_{\rm burst}\\
\\
\nonumber &=& 0 ~~~~~~~~~~~~~~~~~~~\tau > \tau_{\rm burst}
\end{eqnarray}
where $\kappa_\star$ is a dimensional constant converting SFR into bolometric
luminosity. For a Chabrier IMF we have $\kappa_\star\approx 2.5\times
10^{43}$ {\rm yr} erg s$^{-1}/M_\odot\approx 6.5\times 10^9$ yr
$L_\odot/M_\odot$ (see Sect.~\ref{sec|STAR_LF}). The constant SFR $\dot
M_\star=M_{\star,{\rm burst}}/\tau_{\rm burst}$ represents an average over
the fiducial period of the burst $\tau_{\rm burst}$, with the total mass of
formed stars amounting to $M_{\star,{\rm burst}}$.

Since the more massive stars restitute most of their mass to the ISM, the
total amount of surviving mass is $M_\star=(1-\mathcal{R})\, M_{\star,{\rm
burst}}$, where $\mathcal{R}$ is the restituted fraction that depends on the
IMF and on the time elapsed from the burst. For the Chabrier IMF the mass in
old, less massive stars approaches to $1-\mathcal{R}\approx 0.5$ when the
time elapsed is larger than a few Gyrs. Since we shall exploit the continuity
equation also at relatively short cosmic times at $z\ga 1$ we adopt the value
$1-\mathcal{R}\approx 0.6$ corresponding to $\tau_{\rm burst}\sim 1$ Gyr (see
below).

The most recent observations by \textsl{ALMA} have undoubtedly confirmed that
the SFR in massive high-redshift galaxies must have proceeded over a
timescale around $\la 0.5$ Gyr at very high rates $\ga$ a few $\times 10^2\,
M_\odot$ yr$^{-1}$ under heavily dust-enshrouded conditions (e.g., Scoville
et al. 2014, 2015, their Table~1). The observed fraction of FIR detected host
galaxies in X-ray (e.g., Page et al. 2012; Mullaney et al. 2012b; Rosario et
al. 2012) and optically selected (e.g., Mor et al. 2012; Wang et al. 2013b;
Willott et al. 2015) AGNs points toward a SFR abruptly shutting off after
this period of time. In the analysis by Lapi et al. (2014) this rapid
quenching is interpreted as due to the energy feedback from the supermassive
BH growing at the center of the starbursting galaxy. In the first stages of
galaxy evolution the BH is still rather small and the nuclear luminosity is
much less than that associated to the star formation in the host. The SFR is
then regulated by feedback from SN explosions, and stays roughly constant
with time, while the AGN luminosity increases exponentially. After a period
$\la 1$ Gyr in massive galaxies the nuclear luminosity becomes dominant,
blowing away most of the gas and dust from the ambient medium and hence
quenching abruptly the star formation in the host. On the other hand,
longstanding data on stellar population and chemical abundances of galaxies
with final stellar masses $M_\star \la 10^{10}\, M_\odot$ indicate that star
formation have proceeded for longer times regulated by supernova feedback
(see reviews by Renzini et al. 2006; Conroy 2013; Courteau et al. 2014; and
references therein).

On this basis, we adopt a timescale for the duration of the starburst given
by
\begin{equation}\label{eq|tau_burst}
\tau_{\rm burst}(t) = 1\, {\rm Gyr}\,\left({1+z\over 3.5}\right)^{-3/2}\,
\left[1+2\,{\rm erfc}\left({4\over 3}\,\log{L_{\rm SFR}\over 10^{10.5}\,
L_\odot}\right)\right]~;
\end{equation}
the dependence on the cosmic time mirrors that of the dynamical/condensation
time, in turn reflecting the increase of the average density in the ambient
medium. In addition, the ${\rm erfc}$-smoothing connects continuously the
behavior for bright and faint objects expected from the discussion above. We
tested that our results are insensitive to the detailed shape of the
smoothing function. At high redshift, as noted by Lapi et al. (2014), such a
timescale is around $15-20$ $e-$folding time of the hosted BH (i.e., $\la
0.5-1$ Gyr). The luminosity scale $3\times 10^{10}\, L_{\odot}$ corresponds
to SFR $\dot M_\star\approx 5\, M_{\odot}$ yr$^{-1}$.

\subsubsection{Stars: solution}\label{sec|STAR_solution}

Given the lightcurve in Eq.~(\ref{eq|SFRlightcurve}), the fraction of the
time spent per luminosity bin reads
\begin{equation}
\sum_i{{\rm d}\tau_i\over {\rm d}L_{\rm SFR}} = \tau_{\rm
burst}\,\delta_D\left[L_{\rm SFR}-L_{\rm SFR}(M_{\star})\right]
\end{equation}
with $L_{\rm SFR}(M_{\star})=\kappa_\star\,M_{\star,{\rm burst}}/\tau_{\rm
burst}$ the SFR-luminosity associated to the final stellar mass $M_\star$;
the Dirac delta-function $\delta_D(\cdot)$ specifies that, since the
lightcurve is just a constant, the luminosity associated to a stellar mass
$M_\star$ must be in the luminosity bin ${\rm d}L_{\rm SFR}$.

Using this expression in the continuity equation Eq.~(\ref{eq|continuity})
without source term yields
\begin{equation}
{L_{\rm SFR}\,N(L_{\rm SFR},t)\over \tau_{\rm burst}\, f_{\star,L_{\rm SFR}}} =
\left[\partial_t N(M_\star,t)\, M_\star\right]_{|M_\star(L_{\rm SFR},t)} ~,
\end{equation}
where the final stellar mass that have shone at $L_{\rm SFR}$ is given by
\begin{equation}\label{eq|Mstarmax}
M_\star(L_{\rm SFR},t)={(1-\mathcal{R})\, L_{\rm SFR}\, \tau_{\rm burst}\over
\kappa_\star}~.
\end{equation}
In deriving these equation, we have used ${\rm d}M_\star/{\rm d}L_{\rm
SFR}\equiv f_{\star,L_{\rm SFR}}\, M_\star/L_{\rm SFR}$. On the same line of
Sect.~\ref{sec|BH_solution}, we integrate over cosmic time, and turn to
logarithmic bins. The outcome reads
\begin{equation}\label{eq|STAR_solution}
N(\log M_\star,t) = \int_0^t{\rm d}t'\, \left[{N(\log L_{\rm SFR})\over
f_{\star,L_{\rm SFR}}\,\tau_{\rm burst}}\right]_{|L_{\rm SFR}(M_\star,t')}~.
\end{equation}
This solution constitutes a novel result. Note that our approach exploits in
the continuity equation the full SFR-luminosity function, and is almost
insensitive to initial conditions; in these respects it differs from the
technique developed by Leja et al. (2015; see also Peng et al. 2010) to
evolve the stellar mass function backwards from $z\la 2$ basing on the
observed SFR$-M_\star$ relationship and the starforming fraction.

Interestingly, if $\tau_{\rm burst}$ is independent of $L_{\rm SFR}$, a
Soltan-type argument can be extended to the stellar content. It can be easily
found on multiplying Eq.~(\ref{eq|STAR_solution}) by $M_\star$ and
integrating over it, to obtain
\begin{equation}
\int{\rm d}M_\star\, M_\star\, N(M_\star,t) = {1-\mathcal{R}\over
\kappa_\star}\,\int_0^t{\rm d}t'\, \int{\rm d}L_{\rm SFR}\,L_{\rm SFR}\, N(L_{\rm SFR},t')~;
\end{equation}
comparing with the classic expression for the BH population, it is seen that
the role of the efficiency combination $(1-\epsilon)/\epsilon\, c^2\approx
7\times 10^{-14}\, (1-\epsilon)/\epsilon\,$ yr$^{-1}$ $M_\odot/L_\odot$ is
played by the quantity $(1-\mathcal{R})/\kappa_\star\approx 9\times
10^{-11}\,$ yr$^{-1}$ $M_\odot/L_\odot$, which mainly depends on the IMF
(here the constant refers to the Chabrier's IMF).

In passing, we notice that for conventional IMFs most of the stellar mass in
galaxies resides in stars with mass $\la 1\, M_{\odot}$. Since these stars
emit most of their luminosity in the near IR, the galaxy stellar mass
$M_{\star}$ can be inferred by the near-IR luminosity functions. At variance
with the BH case, the so called 'remnants' are not dark but luminous red
stars. This provides a significant vantage point to estimate the mass
function of these 'remnants'. In fact, the stellar mass function $N(M_\star)$
is worked out via the statistics of the stellar luminosity function
$N(L_\star)$, not to be confused with the SFR-luminosity function $N(L_{\rm
SFR})$ used above.

\subsubsection{Stars: results}\label{STAR_results}

\begin{figure}
\epsscale{1.}\plotone{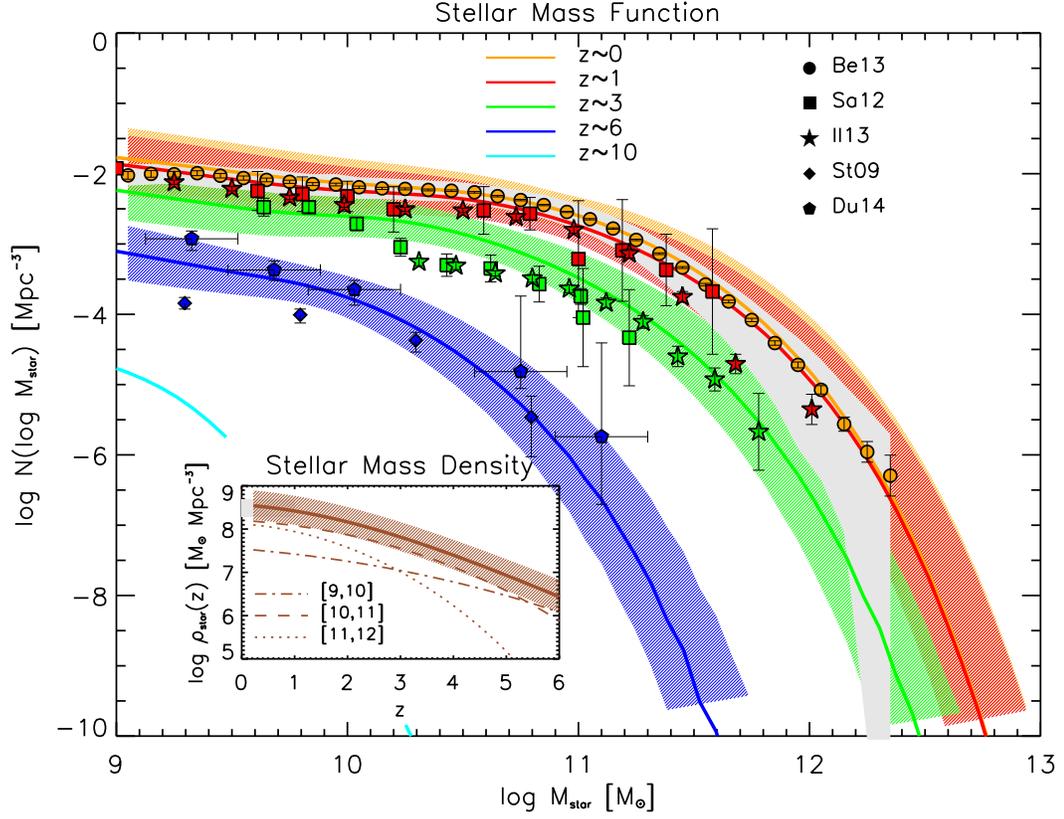}\caption{The stellar mass function $N(\log
M_{\star})$ as a function of the (survived) final stellar mass $M_{\star}$ in
solar units. Results from the continuity equation (see
Sect.~\ref{sec|STAR_solution}) at redshift $z=0$ (orange), $z=1$ (red), $3$
(green), and $6$ (blue) are plotted as solid lines, with the hatched areas
illustrating the associated uncertainty; the cyan line is the extrapolation
to $z=10$ plotted for illustration. High redshift data are from Ilbert et
al. (2013; filled stars), Santini et al. (2012; filled squares),
Stark et al. (2009; filled diamonds) and Duncan et al. (2014, filled
pentagons). Local data at $z=0$ are from Bernardi et al. (2013):
filled circles with errorbars illustrate their fiducial measurements with the
associated statistical uncertainty, while the shaded area shows the
systematic uncertainty related to light profile fitting. The inset shows the
stellar mass density as a function of redshift computed from the continuity
equation, for the overall mass range (solid line with hatched area), and for
stellar masses $\log M_{\star}/M_\odot$ in the ranges $[9,10]$ (dot-dashed
line), $[10,11]$ (dashed line), $[11,12]$ (dotted line). The grey shaded area
illustrates the observational constraints from the $z=0$ mass
function.}\label{fig|STAR_MF}
\end{figure}

In Fig.~\ref{fig|STAR_MF} we illustrate our results on the stellar mass
function at different representative redshifts. The outcomes of the
continuity equation can be fitted with the functional shape of
Eq.~(\ref{eq|AGN_LF}) with $L_{\rm AGN}$ replaced by $M_\star$, and the
parameter values given in Table~1. The resulting fits are accurate within
$5\%$ in the redshift range from $0$ to $6$, and in the stellar mass range
$M_\star$ from a few $10^9$ to a few $10^{12}\, M_\odot$.

The outcome at $z\approx 0$ is compared with the determination of the local
mass function by Bernardi et al. (2013). The outcomes at $z\approx 1$ and
$z\approx 3$ are compared with the determinations by Santini et al. (2012)
and Ilbert et al. (2013), while the result at $z\approx 6$ is compared with
the measurements by Stark et al. (2009) and Duncan et al. (2014). The results
of the continuity equation and the observational estimates at different
redshifts are in very good agreement, considering the associated
uncertainties and systematic differences among different datasets.

Concerning the overall evolution, the high-mass end of the mass function is
mainly built up at $z\ga 1.5$, while the low-mass end is still forming at low
$z$. The inset shows the progressive buildup of the stellar mass density as a
function of redshift. The global stellar mass density at $z=0$ reads
$\rho_{\star}\approx 3\times 10^8\, M_\odot$ Mpc$^{-3}$, in good agreement
with observational determinations, and a factor about $10^{3}$ above the
total BH mass density. The stellar mass densities at $z\approx 1$ is already
very close to the local value.

\begin{figure}
\epsscale{0.5}\plotone{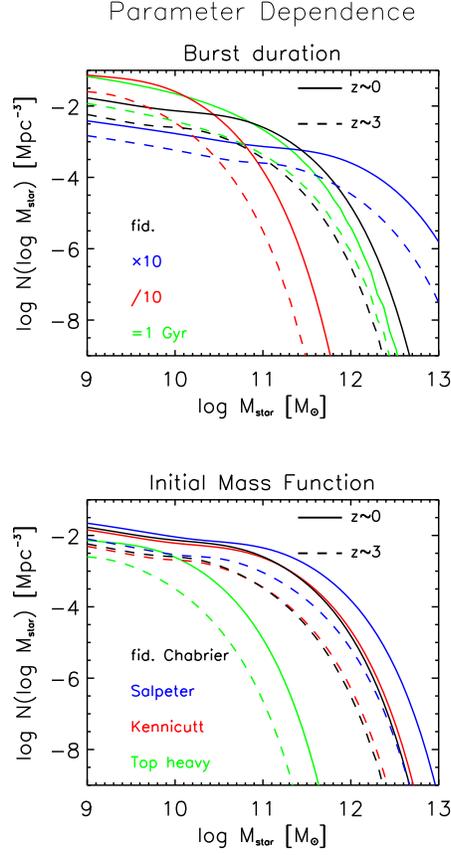}\caption{Comparison plot showing the
dependence of the stellar mass function on the parameters of the assumed
stellar lightcurve; for clarity only results at $z=0$ (solid) and $z=3$
(dashed lines) are plotted. In the top panel, the black line is our fiducial model,
while the red and blue lines refer to values decreased or
increased relative to the reference one; the green lines refers to a constant
(in redshift and luminosity) value. In the bottom panel, the black line refers to our
fiducial Chabrier IMF, while the colored lines are for Kennicutt (1983; red),
Salpeter (1955; blue) and a top-heavy (Lacey et al. 2010; green) IMF.}\label{fig|STAR_MF_comp}
\end{figure}

In Fig.~\ref{fig|STAR_MF_comp} we show how our results on the stellar mass
function depends on the parameters of the lightcurve: the timescale of burst
duration $\tau_{\rm burst}$ and the adopted IMF. To understand the various
dependencies, it is useful to assume a simple, piecewise powerlaw shape of
the luminosity function in the form $N(\log L_{\rm SFR})\propto L_{\rm
SFR}^{-\eta}$, with $\eta\la 1$ at the faint and $\eta>1$ at the bright end.
Then it is easily seen from Eq.~(\ref{eq|STAR_solution}) that the resulting
stellar mass function behaves as
\begin{equation}
N(\log M_\star) \propto\left({1-\mathcal{R}\over
\kappa_\star}\right)^\eta\,\tau_{\rm burst}^{\eta-1}\,\eta\,M_\star^{-\eta}~.
\end{equation}

Thus the stellar mass function features an almost direct dependence on
$\tau_{\rm burst}$ at the high-mass end, which is mostly contributed by high
luminosities where $\eta>1$. On the other hand, the dependence is inverse at
the low-mass end, mainly associated to faint sources with $\eta\la 1$. The
dependence on the IMF is related to the ratio $(1-\mathcal{R})/\kappa_\star$;
e.g., passing from the Chabrier to the Salpeter (1955) IMF, the ratio
increases by a factor of $2$. More significant variations are originated when
passing from Chabrier to a top-heavy IMF (e.g., Lacey et al. 2010) which
implies the ratio to be reduced by a factor $\sim 8$.

We have also tried to parameterize the stellar lightcurve with a decreasing
or increasing exponential like $L_{\rm SFR}\propto e^{-\tau/\tau_\star}$; the
solution of the continuity equation in these instances can be derived on the
same route used for BHs. The net result is that to reproduce the observed
stellar mass function at different redshifts the typical timescale of the
exponential $\tau_\star$ must be of the order of $\tau_{\rm burst}$, i.e.,
the lightcurve is required to be approximately constant over such a timescale
as we have indeed assumed.

\begin{figure}
\epsscale{1.0}\plotone{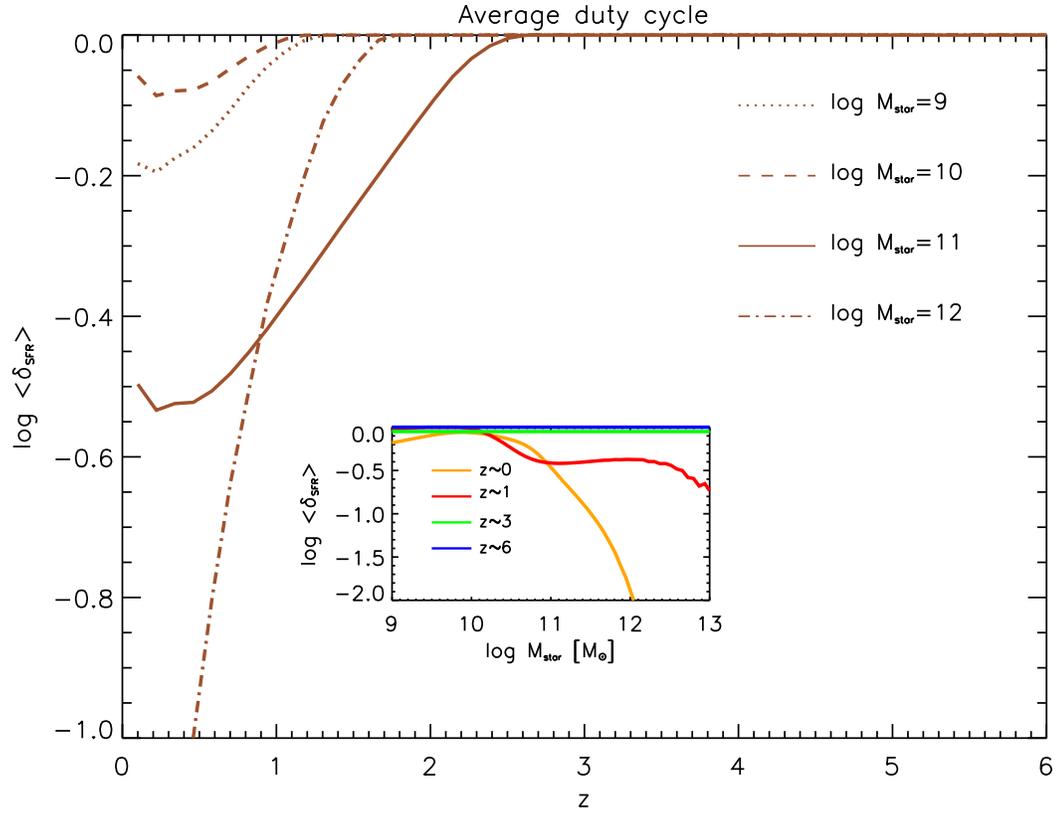}\caption{The average stellar duty cycle
$\langle \delta_{\rm SFR}\rangle$ as a function of redshift $z$, for
different stellar masses $M_\star=10^9$ (dotted), $10^{10}$ (dashed),
$10^{11}$ (solid), and $10^{12}\, M_\odot$ (dot-dashed). The inset
illustrates the duty cycle as a function of the stellar mass at different
redshift $z=0$ (orange), $z=1$ (red), $3$ (green), and $6$
(blue).}\label{fig|STAR_duty}
\end{figure}

Fig.~\ref{fig|STAR_duty} shows the average duty cycle $\langle\delta_{\rm
SFR}\rangle$ of star formation in galaxies. In analogy to
Eq.~(\ref{fig|AGN_duty}), this has been computed as $\langle\delta_{\rm
SFR}\rangle(M_\star,z) = N[\log M_{\star}(L_{\rm SFR}),z]/N(\log M_\star,z)$,
where the relation between $L_{\rm SFR}$ and $M_\star$ is given by
Eq.~(\ref{eq|SFRlightcurve}). At $z\ga 1$ the duty cycle is almost unity,
reflecting the build up of the stellar mass function in real time. On the
other hand, at $z\la 1$ the duty cycle progressively drops down, dramatically
for stellar masses $M_\star\ga$ a few $10^{11}\, M_\odot$. This is because
the mass added by in situ star formation becomes negligible.

Two related outcomes presented in the Appendices are extremely relevant in
this context, the first concerning dust formation, and the second concerning
the role of dry merging. In App.~C we highlight the fundamental role of the
dust, by confronting our fiducial result with that derived basing on the
UV-selected luminosity functions. Fig.~\ref{fig|STAR_UVMF} directly shows
that the UV-selected luminosity function, even corrected for dust extinction,
produces a stellar mass function much lower than the observed one for
$M_{\star}\ga 2\times 10^{10}\, M_\odot$ at any redshift $z\la 6$. In
particular, we stress that our extrapolated FIR portion of the SFR-luminosity
function at $z\sim 6$ is validated by the good comparison with the stellar
mass function observed around that redshift. This implies that massive
galaxies formed most of their stars in a dusty environment. We expect a large
fraction of massive galaxies to be already passively-evolving (i.e., with
quite low SFR and `red' colors) at $z\ga 1$, as indeed increasingly observed
even at substantial redshift $z\sim 3$ (Man et al. 2014; Marchesini et al.
2014).

\begin{figure}
\epsscale{1.}\plotone{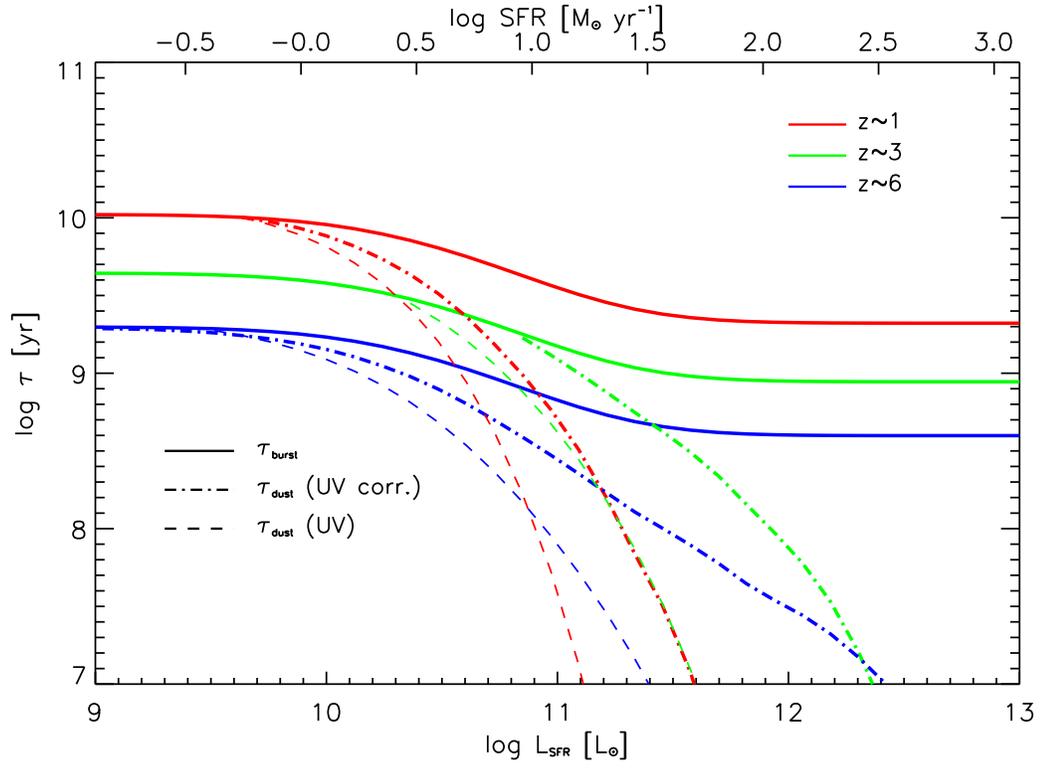}\caption{An estimate of (actually an
upper bound to) the dust formation timescale as a function of the
SFR-luminosity (lower scale) and of the SFR (upper scale) at redshifts
$z=1$ (red), $3$ (green) and $6$ (blue), computed from dust-corrected
UV data (dot-dashed lines), and dust-uncorrected UV data (dotted lines); for
comparison the timescale of the burst duration is also shown (solid
lines).}\label{fig|dustime}
\end{figure}

The point is strengthened by Fig.~\ref{fig|dustime}, which shows an estimate
of (actually an upper bound to) the dust `formation' time $\tau_{\rm dust}$,
computed multiplying the starformation timescale $\tau_{\rm burst}$ by the
ratio of the UV-selected to the total SFR-luminosity functions. At $z\approx
6$, galaxies with SFRs $\dot M_\star\approx 100\, M_\odot$ yr$^{-1}$ and
final stellar masses $M_\star\ga 3\times 10^{10}\, M_\odot$ have a dust
formation time of $\tau_{\rm dust}\approx 3\times 10^7$ yr, implying a quite
rapid metal/dust enrichment. Interestingly, this is much shorter than the
fiducial time $\approx 15\, \tau_{\rm ef}\approx $a few $10^8$ yr to grow the
hosted final BH mass (cf. Eq.~[\ref{eq|AGNlightcurve}]). Therefore, most of
the BH growth must occur in dusty galaxies (e.g., Mortlock et al. 2011). At
redshift $z\ga 2-3$ the constraints on $\tau_{\rm dust}$ for strongly
starforming objects stays almost constant. In moving toward lower redshift
$z\la 2$ the dust formation time becomes shorter, even for moderately
starforming objects with SFR $\dot M_\star\la 30\, M_\odot$ yr$^{-1}$. This
can be interpreted as star formation episodes mainly occurring within
dust-rich molecular clouds, within galaxies already evolved as to the
chemical composition of their ISM.

In App.~B we investigate the impact of dry merging on the evolution of the
stellar mass function. In this context it is worth stressing that the effect
of dry merging is negligible at redshift $z\ga 1$ and it can play some role
only at lower redshift (see Fig.~\ref{fig|STAR_merging}). These outcomes
statistically ascertain that most of the stellar content in massive galaxies
with $M_\star\ga 3\times 10^{10}\, M_\odot$ is formed \emph{in situ}.
However, we caution that the observed stellar mass function cannot currently
be assessed for $M_\star\ga 3\times 10^{11}\, M_\odot$ given the substantial
systematic uncertainties in the data (see discussion by Bernardi et al.
2013), and that the role of dry mergers can be of some relevance in the
growth of such extremely massive galaxies (see Liu et al. 2015; Shankar et
al. 2015).

All in all, we stress the capability of the continuity equation in
reconstructing the star formation history in the Universe from the past SFR
activity.

\section{Abundance matching}\label{sec|abundance}

Having obtained a comprehensive view of the bolometric luminosity and mass
functions for stars and supermassive BHs at different redshift, we now aim at
establishing a link among them and the gravitationally dominant dark matter
component. To this purpose, we exploit the abundance matching technique, a
standard way of deriving a monotonic relationship between galaxy and halo
properties by matching the corresponding number densities (Vale \& Ostriker
2004, Shankar et al. 2006, Moster et al. 2010, 2013; Behroozi et al. 2013).

When dealing with stellar or BH mass $M$, we derive the relation $M(M_{\rm
H},z)$ with the halo mass $M_{\rm H}$ by solving the equation (e.g., White et
al. 2008; Shankar et al. 2010)
\begin{equation}\label{eq|abmatch}
\int_{\log M}^\infty{\rm d}\log M'\, N(\log M',z) =
\int_{-\infty}^{+\infty}{\rm d}\log M_{\rm H}'\, N(\log M_{\rm H}',z)\,
{1\over 2}\, {\rm erfc}\left\{{\log[M_{\rm H}(M)/M_{\rm H}']\over
\sqrt{2}\,\tilde\sigma_{\log M}}\right\}~,
\end{equation}
which holds when a lognormal distribution of $M$ at given $M_{\rm H}$ with
dispersion $\sigma_{\log M}$ is adopted. In the above expression we have
defined $\tilde\sigma_{\log M}=\sigma_{\log M}/\mu$ with $\mu\equiv {\rm
d}\log M/{\rm d}\log M_{\rm H}$. On the basis of the investigation by Lapi et
al. (2006, 2011, 2014) on the high-redshift galaxy and AGN luminosity
function, we expect the $M_{\rm BH}-M_{\rm H}$ correlation to feature a quite
large scatter $\sigma_{\log M_{{\rm BH}}}\approx 0.4$ dex, while a smaller
value $\sigma_{\log M_\star}\approx 0.15$ dex is expected for the correlation
with the stellar component. We shall compare the correlations $M-M_{\rm H}$
obtained when such values for the scatter are considered with those obtained
by assuming one-to-one relationships, i.e., taking $\sigma_{\log M}=0$.

In Eq.~(\ref{eq|abmatch}) the quantity $N(\log M_{\rm H})$ is the galaxy halo
mass function (GHMF), i.e., the mass function of halos hosting one individual
galaxy. We do not simply rely on the overall halo mass function (HMF),
because we aim at obtaining relationships valid for one single galaxy, not
for a galaxy system like a group or a cluster. In a nutshell, we build up the
GHMF on correcting the overall HMF from cosmological $N-$body simulations, by
adding to it the contribution of subhalos, but by probabilistically removing
from it the contribution of halos corresponding to galaxy systems. We defer
the reader to App.~A for the detailed description of this procedure. The
resulting GHMF is plotted at different redshifts in Fig.~\ref{fig|GHMF}; we
stress that the determination of the GHMF as a function of redshift
constitutes in its stand a novel result. The outcomes can be fitted with the
functional shape of Eq.~(\ref{eq|AGN_LF}) with $L_{\rm AGN}$ replaced by
$M_{\rm H}$, and with the parameter values reported in Table~1. The resulting
fits are accurate within $5\%$ in the redshift range from $0$ to $10$.

In the same Figure we also compare the GHMF to the overall HMF. The
difference between the two, i.e. the galaxy system halo mass function at
$z=0$, is compared with local data to cross-check our approach. At the bright
end the GHMF drastically falls off, so that even at $z\la 1$ galactic halo
masses of $\approx 10^{14}\, M_\odot$ are very rare, since these masses
pertain to galaxy systems. These findings agree with galaxy-galaxy weak
lensing measurements (Kochanek et al. 2003; Mandelbaum et al. 2006, van
Uitert et al. 2011, Leauathaud et al. 2012, and Velander et al. 2014) and
dynamical observations in nearby galaxies (Gerhard et al. 2001, Andreon et
al. 2014; see also the review by Corteau et al. 2014).

The same abundance matching technique may also be applied to the stellar or
AGN bolometric luminosity $L$, looking for a relation $L(M_{\rm H},z)$
specifying the typical luminosity to be expected in a halo of mass $M_{\rm
H}$ at given redshift $z$. However, when dealing with luminosities, one has
to take into account that galaxies and AGNs shine only for a fraction of the
cosmic time. In practice, we use a modified abundance matching of the form
\begin{equation}
\int_{\log L}^\infty{\rm d}\log L'\, {N(\log L',z)\over
\langle\delta\rangle\times t} = \int_{-\infty}^{+\infty}{\rm d}\log M_{\rm
H}'\, \partial_t^+ N(\log M_{\rm H}',z)\, {1\over
2}\, {\rm erfc}\left\{{\log[M_{\rm H}(L)/M_{\rm H}']\over
\sqrt{2}\,\tilde\sigma_{\log L}}\right\}~,
\end{equation}
where $\langle\delta\rangle\times t$ is the duty cycle $\delta$ averaged over
the lightcurve multiplied by the cosmic time $t$, and $\partial_t^+ N(\log
M_{\rm H},z)$ is the formation rate of galactic halos computed according to
Lapi et al. (2013).

\subsection{Abundance matching results}\label{sec|abmatch_results}

We turn to present the results of the abundance matching technique among
various statistical properties of BH, galaxies, and host halos. Analytic fits
to such outcomes can be found in App.~D and Table~2.

\subsubsection{BH vs. halo properties}\label{sec|abmatch_BHvshalo}

In Fig.~\ref{fig|Mbh_MH} we show the relationship between the final BH mass
$M_{{\rm BH}}$ and halo mass $M_{\rm H}$ from the abundance matching
technique, at different redshifts.

Since we are comparing BH and halo statistics at fixed $z$, the resulting
relationship constitutes a snapshot, that can be operationally exploited in
numerical works to properly populate halos at the reference redshift. On the
other hand, the evolution of BHs and halos due to accretion is expected to
modify, though on different timescales, the relation as the cosmological time
passes. For example, if the cosmological growth of halos is dominant, then
the relation would shift along the $M_{\rm H}$ axis. The relationship at a
subsequent redshift takes into account such an evolution, although the number
of evolved BHs and halos is generally subdominant with respect to the newly
formed objects.

The top panel of Fig.~\ref{fig|Mbh_MH} shows the results when a one-to-one
(i.e., no scatter) $M_{{\rm BH}}-M_{\rm H}$ relationship is assumed, while
bottom panel shows the resulting average relationship when a Gaussian
distribution in $M_{{\rm BH}}$ at given $M_{\rm H}$ with a scatter of $0.4$
dex is adopted. The presence of scatter is particularly relevant at high
redshift. Assuming a one-to-one relationship would yield at $z\approx 6$
average BH masses $M_{{\rm BH}}\ga 10^{10}\, M_\odot$ within halos of $M_{\rm
H}\ga 5\times 10^{12}\, M_\odot$, much larger than at $z\approx 3$. This
would imply a significant change in the physical mechanisms establishing the
$M_{\rm BH}-M_{\rm H}$ relation over a relatively short timescale of $\sim 1$
Gyr. In the presence of scatter instead such large BH masses constitute only
extreme instances, relative to much smaller average values $M_{{\rm
BH}}\approx 10^9\,M_\odot$, not very different from the lower redshift ones.
In this scenario such peculiar instances are precisely those picked by
current observations at high-redshift, which are biased toward the more
luminous AGNs powered by the more massive BHs. Thus the one-to-one
relationship offers a view of the observed properties at the given redshift,
while the average relationship (with scatter) is helpful to provide a
physical interpretation. With scatter included, taking into account the
considerable uncertainties, one can estimate that the logarithmic slope of
the average relationship at $z\ga 1$ is around $M_{\rm BH}\propto M_{\rm
H}^{1-2}$, so encompassing the range suggested for AGN feedback processes
(Silk \& Rees 1998; Fabian 1999; King 2005; for a recent review see King
2014). The average relationship is practically unchanged within the
uncertainties over the range $z\sim 1-6$. Plainly, at $z=0$ we find very good
agreement with the relation inferred from the BH mass function by Shankar et
al. (2009).

In Fig.~\ref{fig|Lagn_MH} we show the relationship between the AGN luminosity
$L_{\rm AGN}$ and halo mass $M_{\rm H}$, both with and without scatter.
Concerning the scatter, the same comments of the previous Figure apply. The
flattening in the relation toward lower redshift is mainly driven by the
evolution of the AGN luminosity function, especially at the bright end. The
one-to-one relationship, together with the duty cycle, is required to
properly populate halos and derive the clustering properties of AGNs.

In Fig.~\ref{fig|AGN_bias} we show the luminosity- and BH mass-averaged AGN
bias as a function of redshift $z$. This has been computed as follows: we
start from the linear halo bias model $b(M_{\rm H},z)$ including excursion
set peak prescriptions (Lapi \& Danese 2014) for halos of mass $M_{\rm H}$ at
redshift $z$ (see also Sheth et al. 2001). Then we associate to each halo
mass $M_{\rm H}$ an AGN of luminosity $L_{\rm AGN}$ as prescribed according
to the $L_{\rm AGN}-M_{\rm H}$ one-to-one relationship discussed above.
Finally, we compute the luminosity-weighted bias as a function of redshift
\begin{equation}
\bar b(z) = {\int_{\log L_{\rm min}}^\infty{\rm d}\log L_{\rm AGN}\, N(\log
L_{\rm AGN},z)\, b(L_{\rm AGN},z)\over \int_{\log L_{\rm min}}^\infty{\rm
d}\log L_{\rm AGN}\, N(\log L_{\rm AGN},z)}~,
\end{equation}
where $L_{\rm min}$ is a minimum bolometric luminosity. The same procedure
can be followed to obtain the $M_{\rm BH}$-averaged bias through the
one-to-one $M_{\rm BH}-M_{\rm H}$ relation and the average over the BH mass
function.

The resulting bias as a function of redshift compares well with the
observational data points from large optical and X-ray survey samples.
Typically, the optical data refer to quasars with luminosities $L_{\rm
AGN}\ga$ a few $10^{12}\, L_\odot$, while X-ray data refer to AGNs with
$L_{\rm AGN}\ga$ a few $10^{11}\, L_\odot$; however, the selection of the
datasets reported in the plot are diverse, and the reader is deferred to the
original papers for details. Note that while at $z\ga 2$ X-ray selected AGNs
appear to be less clustered than optical quasars, the opposite holds true at
low $z\la 2$. This fact is somewhat puzzling since X-ray AGNs feature
generally lower bolometric luminosities, and is often interpreted in terms of
a different accretion mode becoming dominant at low $z$ (e.g., sporadic
reactivation episodes in place of continuous accretion, see discussion by
Allevato et al. 2011, 2014). As a reference, the halo bias $b(M_{\rm H},z)$
for various $M_{\rm H}$ is also shown. It is evident that typical host halos
feature $M_{\rm H}\ga 10^{12}\, M_\odot$ with a clear tendency for more
massive halos to host more luminous AGNs and more massive BHs. In the inset
it is seen that even the mild trend of the bias with luminosity at given
redshift $z\approx 2$ from optical surveys is reproduced. On the other hand,
the dependence on luminosity is expected to significantly increase at higher
$z\ga 4$.

We stress that the clustering properties constitute a byproduct of our
approach, and the comparison with observations validate our results on BH
mass function and duty cycle (see also Shankar et al. 2010). Note that past
studies (Martini \& Weinberg 2001) have instead exploited the clustering
properties to constrain the AGN duty cycle. In comparing with previous works
related to the AGN bias (e.g., White et al. 2008; Hopkins et al. 2007; Wyithe
\& Loeb 2009; Bonoli et al. 2010; Shankar et al. 2010), a few remarks are in
order: (i) we stress that our adoption of the GHMF in place of the
routinely-used HMF appreciably improves the agreement with observations of
the bias for luminous AGNs/massive BHs at $z\ga 3$; (ii) we confirm that
values $\lambda\ga $ a few at $z\ga 3$, implying a quite rapid growth of the
BH during the ascending portion of the AGN lightcurve, are also required to
meet the observational constraints; (iii) we find that the weak dependence of
the bias on luminosity at $z\sim 2$ is rather insensitive to the presence of
the descending portion of the AGN lightcurve, that we recall is instead
indicated in luminous objects by the observed fraction of starforming hosts
in optically-selected quasars (see Sect.~\ref{sec|AGNlightcurve}).

\subsubsection{Stellar vs. halo properties}\label{sec|abmatch_STARvshalo}

In Fig.~\ref{fig|Mstar_MH} we show the relationship between the final stellar
mass $M_{\star}$ vs. the halo mass $M_{\rm H}$, for different redshift. The
result at $z=0$ is compared with the relationship inferred from the local
stellar mass function by Bernardi et al. (2013). We find a good agreement
within the associated uncertainties. The $M_{\star}$ vs. $M_{\rm H}$ at given
redshift can be described by a powerlaw with slope around $1$ at the
high-mass end, then steepening for halo masses $M_{\rm H}\la$ a few
$10^{12}\, M_\odot$. The presence of the scatter around $0.15$ dex does not
affect appreciably the correlation.

At $z\ga 1$ the statistics of both stellar and halo masses are dominated by
newly-created objects, so that the evolution in both masses of the older
individuals is irrelevant. From this perspective, the little if no evolution
of the $M_\star-M_{\rm H}$ relationship can be interpreted in the light of
similar, in-situ processes regulating the star formation at different
redshifts (Moster et al. 2013). This may be seen more clearly in the inset,
showing the efficiency $M_\star/f_{\rm b}\, M_{\rm H}$ for the conversion
into stellar component of the original baryon content within the halo $f_{\rm
b}\, M_{\rm H}$, having adopted a cosmic initial baryon to DM ratio $f_{\rm
b}=0.2$. The efficiency rises from values $\la 10\%$ for halo masses $M_{\rm
H}\la 10^{11}\, M_\odot$ to a roughly constant values $\la 25\%$ around halo
masses $M_{\rm H}\approx $ a few $10^{12}\, M_\odot$. All in all, the star
formation process in halos is highly inefficient. From a physical point of
view, this is usually interpreted in terms of competition between cooling and
heating processes. In low-mass halos, the latter is provided by energy
feedback from SN explosions. In massive halos, cooling rates are not
significantly depressed by SN feedback, and the star formation can proceed at
much higher levels until the AGN attains enough power to quench it abruptly.

At $z\la 1$ the interpretation is more complex, since the statistics of stars
and halos are no longer dominated by newly-formed objects and significant
evolution in one of the two components may occur. Specifically, for high
masses the halo evolution dominates, and the $M_{\star}-M_{\rm H}$ evolves
shifting toward higher halo masses at almost constant stellar mass;
contrariwise, for small masses, stellar mass evolution dominates over the
halo's, and the relationship shifts upward at almost constant halo mass.

In Fig.~\ref{fig|Mstar_MH_z0comp} we present a comparison of our
$M_\star-M_{\rm H}$ relationship at $z=0$ with literature determinations. Our
result when the GHMF is exploited for the abundance matching (same as in
previous Figure) can be directly compared with the determination by Shankar
et al. (2006) based on the same abundance matching technique. The difference
is mainly due to the dynamical $M_\star/L_{\star, K}$ adopted by Shankar et
al. in building the stellar mass function from the $K-$band luminosity
function.

On the other hand, our result when the overall HMF is adopted can be directly
compared to the determinations based on the abundance matching by Behroozi et
al. (2013) and by Moster et al. (2013). These are quite similar to ours at
the low-mass end, while appreciably steeper at the high-mass end (see also
Kravtsov et al. 2014; Shankar et al. 2014), where the Bernardi et al. (2013)
stellar mass function we adopt contains relatively more objects.

These results based on the overall HMF can also be directly compared with the
data from gravitational lensing measurements in groups and clusters of
galaxies by Han et al. (2014), Velander et al. (2014), and Mandelbaum et al.
(2006). The agreement is very nice. Note that since gravitational lensing
probes the mass projected along the line of sight, it is sensitive to the
presence of groups and/or clusters surrounding the individual galactic halo.

In Fig.~\ref{fig|Lstar_MH} we show the relationship between the luminosity
$L_{\rm SFR}$ associated to the SFR vs. the halo mass $M_{\rm H}$, for
different redshifts. The presence of the scatter around $0.15$ dex only
marginally affects the average relationship. We show both the outcome based
on the overall SFR-luminosity function, and the one based on the
dust-corrected UV luminosity function only. This has been determined by
matching the GHMF with the SFR-luminosity function of UV-selected galaxies
corrected for dust extinction (see App.~C and Fig~\ref{fig|STAR_UVLF}). It is
evident that the typical UV data substantially underestimate the luminosities
associated to the SFR. We stress once again that FIR data are crucial for an
unbiased view of the star formation process.

In Fig.~\ref{fig|Lstar_MH} we also plot the relationship expected at
$z\approx 10$, although we caution that for halo masses $M_{\rm H}\la
10^{11}\, M_\odot$ the relationship strongly depends on the faint-end slope
of the luminosity function. To illustrate the variance, we plot as a lower
bound the relation corresponding to a faint-end slope $-1.65$, and an upper
bound corresponding to $-2$ (Bouwens et al. 2015; see also
Sect.~\ref{sec|STAR_LF}). The latter instance is required to keep the
Universe ionized out to $z\la 8.8$, corresponding to an electron-scattering
optical depth $\tau_{\rm es}\approx 0.066$ as recently measured by the
\textsl{Planck} Collaboration (2015). Our SFR vs. $M_{\rm H}$ relationship
suggests that this can be afforded by galaxies starforming at rates $\ga
10^{-2}\, M_\odot$ yr$^{-1}$, with UV magnitudes $M_{\rm UV}\ga -13$ hosted
within halos of masses $M_{\rm H}\ga 10^9\, M_\odot$ (see also Cai et al.
2014).

In Fig.~\ref{fig|STAR_bias} we show the galaxy bias, both luminosity (or
SFR)-averaged and stellar mass-averaged, for different values of minimum SFR
or $M_\star$. These quantities have been computed following the same
procedure for the AGN bias as described in Sect.~\ref{sec|abmatch_BHvshalo}.
For reference we also report the halo bias for different halo masses. It is
seen that the bias computed from the abundance matching reproduces very well
the determination at different redshifts for various populations of objects.
In particular, UV-selected objects like Lyman Break Galaxies and
Lyman-$\alpha$ emitters feature low stellar masses $M_\star\la 10^9\,
M_\odot$ and SFRs less than a few $M_\odot$ yr$^{-1}$, while FIR-selected
objects are associated to much more violent SFRs $\ga 10^2\, M_\odot$
yr$^{-1}$ and constitute the progenitors of massive galaxies with final
stellar content $M_\star\ga 10^{11}\, M_\odot$.

\subsubsection{SFR and sSFR vs. stellar mass and
redshift}\label{sec|abmatch_sSFR}

In Fig.~\ref{fig|sSFR_z} we plot the cosmic specific (sSFR) defined as the
ratio between the SFR density $\rho_{\rm SFR}\equiv \int{\rm d}\log \dot
M_\star\, \dot M_\star\, N(\log \dot M_\star)$ and the stellar mass density
$\rho_{\star}\equiv \int{\rm d}\log M_\star\, M_\star\, N(\log M_\star)$. The
resulting cosmic sSFR reflects the behavior for typical SFR and stellar
masses at the knee of the corresponding distributions, and it includes all
galaxies, even the passively evolving ones (see also Madau \& Dickinson
2014).

We report both the outcome based on the overall SFR-luminosity function, and
the one based on the dust-corrected UV luminosity function only. It is
apparent that the latter case underestimates the cosmic sSFR at any redshift
(cf. Wilkins et al. 2008). We also illustrate the result by Madau \&
Dickinson (2014), which is similar to ours up to $z\sim 2$, and then
approaches the UV-inferred result. As a matter of fact, their cosmic
star-formation history at $z\ga 3$ is based on UV data (see their Fig.~9).

The reported observational estimates refer to galaxy samples selected with
different criteria. Specifically, at $z\ga 3$ they mainly refer to
UV-selected samples. This explains why they are better reproduced by our
results for the UV dust-corrected case. On the other hand, at $z\la 1.5$ they
are mainly based on UV$+$near-IR data with ongoing star formation inferred
from $24\,\mu$m or radio fluxes. In this redshift range the sSFR estimated
from the ratio $\rho_{\rm SFR}/\rho_{\star}$ lies below most of the data
points, because it includes an increasing fraction of objects in passive
evolution, while observations refer to starforming galaxies only (see
discussion by Madau \& Dickinson 2014). On the other hand, Ilbert et al.
(2015) report values of the sSFR closer to the ratio $\rho_{\rm
SFR}/\rho_{\star}$, but a factor of $\sim 2-3$ lower than previous estimates
in literature. The authors attribute this difference to their more accurate
treatment of the selection effects, leading to inclusion of galaxies with
lower sSFR, and to their more accurate statistics.

In Fig.~\ref{fig|sSFR} we show the relationships between the SFR and the
stellar mass $M_\star$, at different redshifts; this is often referred to as
the `main sequence' of starforming galaxies (e.g., Elbaz et al. 2011;
Rodighiero et al. 2011). Note that the outcome is obtained by matching the
abundances of two observable quantities like the SFR-luminosity and stellar
mass functions (the halo mass is bypassed), including the star formation duty
cycle. From this point of view, the outcome is only mildly dependent on
assumptions on the star formation lightcurve and timescales. As in the
previous Figure, we report the outcome from the abundance matching based on
the overall SFR-luminosity functions, and the one based on dust-corrected UV
luminosity functions only.

We compare the abundance matching result with the recent observational
estimates by Rodighiero et al. (2014), Speagle et al. (2014), Salmon et al.
(2015), Renzini \& Peng (2015) based on large samples of individual
measurements of SFRs and stellar masses. We stress that, especially at $z\la
1$, determinations of the main sequence by various authors differ, mainly
because of the way galaxies are selected as being starforming (see discussion
by Renzini \& Peng 2015). Further observations and analysis are needed to
fully assess the main sequence, both regarding the the overall normalization
and the slopes and the high and low mass end (that can even be different, see
Whitaker et al. 2014).

At $z\ga 1$ our results from the abundance matching based on FIR+UV
luminosity functions well agree with the estimates by Rodighiero et al.
(2014) and Speagle et al. (2014) based on multiwavelength observations of
galaxy samples. At $z\approx 6$ our result from the abundance matching based
on UV luminosity function is in excellent agreement with the data for
UV-selected galaxies by Salmon et al. (2015).

On the other hand, for $z\la 1$ our results appear to be at variance with the
observational determinations for stellar masses $M_\star\la 5\times 10^{10}\,
M_\odot$. However, in this range the results from the abundance matching
becomes loosely constraining, because of the large uncertainties introduced
by the flatness of the stellar mass function (cf. Fig.~\ref{fig|STAR_MF}).
This suggests that the stellar mass and SFR luminosity function may not
sample the same galaxy population at their respective faint end.
Nevertheless, at high masses the abundance matching technique is consistent
with current data, and actually extends the main sequence in a range where
determinations from individual measurements are still scanty.

\subsubsection{BH mass vs. stellar mass}\label{sec|abmatch_BHvsSTAR}

In Fig.~\ref{fig|Mbh_Mstar} we illustrate the relationship between the BH
mass and stellar mass at different redshifts. The computation is performed by
the abundance matching of the BH and stellar mass function from the
continuity equation, thus bypassing the halo mass. We show results both for
the one-to-one case (top panel), and when a Gaussian scatter of $0.3$ dex
between $M_{\rm BH}$ and $M_\star$ is assumed (bottom panel). The presence of
the scatter is increasingly relevant at higher redshift in biasing
observations toward extreme values of the $M_{\rm BH}/M_\star$ ratio (shown
in the inset). It is worth noticing that the evolution of the relationship
and hence of the $M_{\rm BH}/M_\star$ ratio is quite small for $z\la 3$, at
variance with the claims by some authors (Peng 2007; Jahnke \& Macci\'o
2011). This signals once again that the BH and stellar mass growth occurs
\emph{in parallel} by \emph{in-situ} accretion and star formation processes.

Our results at $z=0$ agree with the relations inferred from the abundance
matching of the local determinations for the stellar and BH mass functions by
Bernardi et al. (2013) and Shankar et al. (2009). Our findings are in very
good agreement with the individual determinations of BH and stellar masses
based on dynamical measurements by Haring \& Rix (2004). On the other hand,
Kormendy \& Ho (2013) propose a relation which is systematically higher by a
factor $\approx 2.5$. The Soltan argument would then imply an extremely high
final BH mass density, and in turn a value $\epsilon\la 0.02$ of the average
efficiency during the slim-disc regime (see Sect.~\ref{sec|continuity}.).

In Fig.~\ref{fig|cosmacc} we illustrate the evolution with redshift of the
mass density in DM halos, stars, and BHs. The stellar mass density closely
mirrors that of galactic DM halos, because the star to DM mass ratio (i.e.,
the star formation efficiency) stays roughly constant with redshift for
typical galaxies at the knee of the mass function (see
Fig.~\ref{fig|Mstar_MH}). On the other hand, for $z\la 2$ the stellar mass
density progressively differs from that of the overall halo population
(including galaxy groups/clusters). Once again this strengthens the point
that star-formation at high redshift occurs via in-situ processes within the
central regions of galactic halos. The stellar mass density is a factor about
$\sim 30-50$ lower than the galactic halo mass density, reflecting the
inefficiency of galaxy formation due to feedback processes, as discussed in
Sect.~\ref{sec|abmatch_STARvshalo}.

The BH mass density has a considerably different shape, that toward higher
$z$ progressively steepens relative to the galactic halo and stellar mass
density. This is due to two effects: (i) the number density of halos able to
host massive BHs declines rapidly; (ii) the time needed to grow massive BHs
becomes comparable with the age of the Universe, so making apparent the delay
of about a few $10^8$ yrs between the BH and stellar formation. In the inset
we show that the observed density ratio between the SFR and the AGN
luminosity attains a minimum around $z\sim 1.5$ and it stays almost constant
toward lower redshift. This is because both luminosity densities decline in
parallel (cf. insets in Figs.~\ref{fig|AGN_LF} and \ref{fig|STAR_LF}). The
same trend also applies for the corresponding mass density ratio.

\section{Summary}\label{sec|conclusions}

We have investigated the coevolution of galaxies and hosted supermassive
black holes throughout the history of the Universe by a statistical approach
based on the continuity equation and the abundance matching technique. Our
main results are the following:

\begin{itemize}

\item We have demonstrated that the local supermassive BH mass function and
    the stellar mass functions at different redshift can be reconstructed
    from the SFR and AGN luminosity functions via a continuity equation
    approach without source term. This implies that the buildup of stars
    and BHs in galaxies occurs mainly via local, \emph{in-situ} processes,
    with dry mergers playing a marginal role at least for stellar masses
    $M_\star\la 3\times 10^{11}\, M_\odot$ and BH masses $M_{\rm BH}\la
    10^9\, M_\odot$, where the statistical data are more secure and less
    biased by systematic errors.

\item As to the AGN/BH component, our analysis nicely reproduces the
    observed Eddington ratio function and the observed fraction of galaxies
    with given stellar mass hosting an AGN with given Eddington ratio (see
    Sect.~\ref{sec|BH_results}). Such an agreement strongly suggests that
    the fraction of AGNs observed in slim-disc regime increases with
    redshift, and that anyway most of the BH mass is accreted in such
    conditions.

\item We have inferred relationships between the stellar, BH, and DM
    components of galaxies at various redshifts. These imply that stellar
    and AGN feedback cooperate with gas cooling in the star formation
    process within halos, whose binding energy at formation is the most
    relevant feature. Specifically, in low-mass halos SN explosions keep
    star formation low on long timescales, while in massive halos star
    formation can proceed at much higher levels until the AGN quenches it
    abruptly. These relationships between galaxy/BH and halo properties
    constitute \emph{testbeds} for galaxy formation and evolution models,
    and can be operationally implemented in numerical simulations to
    \emph{populate} dark matter halos or to gauge \emph{subgrid} physical
    prescriptions. Duty cycles for both the AGN and the stellar components
    are derived, and found to be close to unity at high-redshift.

\item We have derived the \emph{bias} as a function of redshift and
    luminosity, both for the AGN and for various galaxy populations. The
    clustering properties constitute a byproduct of our approach, and the
    nice agreement with observations \emph{validate} our results on BH and
    stellar mass functions, and related duty cycles from the continuity
    equation.

\item The specific SFR increases with redshift at least up to $z\sim 6$. In
    the range $z\ga 1$ the results from the abundance matching technique
    agree with the so called `main sequence' of starforming galaxies,
    although we underline that the comparison with observations critically
    depend on sample selection. For $z\la 1$ the results from abundance
    matching are reliable for stellar masses $M_\star\ga 5\times 10^{10}\,
    M_\odot$, where they are consistent and actually extend the
    observational determinations in a range where individual measurements
    are still scanty.

\item We show how strongly the presence of the \emph{dust} affects the view
    of the star formation process in galaxies with SFRs $\dot M_\star\ga
    10\, M_\odot$ yr$^{-1}$ at any redshift, even the quite large ones. In
    fact, we have shown that dust is formed on a timescale which is only a
    small fraction of the burst duration. Such a behavior is also mirrored
    in the estimated cosmic SFR and sSFR density.

\item The low efficiency $\la 20\%$ in star formation elucidates that a
    fraction $\ga 50\%$, up to $\sim 70\%$ depending on mass, of the gas
    associated to a galaxy halo is always in warm/hot form.

\item The BH to stellar mass ratio evolves mildly at least up to $z\la 3$,
    signaling that the BH and stellar mass growth occurs \emph{in parallel}
    by in-situ accretion and star formation processes.

\end{itemize}

The marginal role of dry merging and the inefficiency of star formation imply
that galaxy formation is basically a process inherent to the inner regions of
halos, where most of the gas mass resides.

These evidences strongly add motivation to the the development of
hydrodynamical simulations at very high spatial resolution, which allow
detailed studies of small-scale gravitational instabilities connected to gas
cooling and condensation, star formation, BH accretion, and associated
feedback processes (e.g., Ceverino et al. 2015; for a comprehensive review
Bournaud 2015). Our main results are listed in Table~3, where we also recall
their location in the paper, and cross-reference to the corresponding
sections and figures.

From the technical point of view, the novel achievements of the present work
can be summarized as follows:
\begin{itemize}

\item We have presented an analytical solution of the continuity equation
    for BHs that holds under quite general assumptions, including a
    redshift/luminosity dependence of the Eddington ratio, radiative
    efficiency, and lightcurve timescales.

\item We have developed the continuity equation for the stellar component,
    solving it under quite general assumptions about the lightcurve shape
    and timescales.

\item We have provided a continuous rendition of the overall SFR function,
    interpolating between the UV data at the faint and the FIR data at the
    bright end. A posteriori, our approach is validated by the agreement of
    the stellar mass function via continuity equation with the
    observational determinations over the redshift range $z\sim 0-6$.

\item We have developed a procedure to derive the galaxy halo mass function
    at different redshifts. This can be implemented in halo occupation
    distribution models.

\item We have generalized the abundance matching technique to deal with
    relationships between luminosity and mass, by considering the duty
    cycle of BHs and star formation in galaxies.

\end{itemize}

We stress that the added value of continuity equation and abundance matching
is to provide largely model-independent outcomes, which must be complied by
detailed physical models.

Finally, two remarks are in order. As to the AGN/BH component, large samples
of AGN with multiwavelength SEDs are crucial in testing the statistics of the
slim-disc fraction and in measuring the associated radiative efficiency (cf.
Raimundo et al. 2012). As to the stellar component, our analysis allows to
extrapolate the SFR, stellar mass, and sSFR functions to higher redshift, yet
unexplored but within the reach of future instrumentations like
\textsl{ALMA}, \textsl{JWST} and \textsl{SKA}. In particular, a crucial point
will be to estimate the bright end of the SFR-luminosity function at $z\ga
4$, to obtain direct constraints on the \emph{timescale} of dust formation in
high-redshift galaxies.

\begin{acknowledgements}
We thank F. Bianchini, A. Bressan, A. Cavaliere, A. Celotti, P.S. Corasaniti,
C. Mancuso, P. Salucci for helpful discussions. We acknowledge the anonymous
referee for valuable comments and suggestions. This work has been supported
in part by the MIUR PRIN 2010/2011 `The dark Universe and the cosmic
evolution of baryons: from current surveys to Euclid', by the INAF PRIN
2012/2013 `Looking into the dust-obscured phase of galaxy formation through
cosmic zoom lenses in the Herschel Astrophysical Terahertz Large Area
Survey', and by the ASI/INAF Agreement 2014-024-R.0 for the \textsl{Planck}
LFI activity of Phase E2. A.L. is grateful to SISSA for warm hospitality.
\end{acknowledgements}

\appendix

\section{Appendix A: Galactic Halo Mass Function}

In this Appendix we detail our procedure to derive the galactic halo mass
function, i.e., the mass function associated to halos hosting one individual
galaxy. The computation actually includes two steps: (i) we account for the
possibility that a halo contains various subhalos; (ii) we probabilistically
exclude halos corresponding to galaxy systems rather than to individual
galaxies.

Our starting point is the subhalo mass function, as recently determined by
Jiang \& van den Bosch (2014). The distribution of subhalos with mass
between $m$ and $m+{\rm d}m$ in a halo of mass $M_{\rm H}$ at redshift $z$
can be well fitted by the function
\begin{equation}
N(\log \psi) = \gamma\, \psi^{\alpha}\, e^{-\beta\,\psi^\omega}\, \ln{10}~,
\end{equation}
where $\psi=m/M_{\rm H}$. Actually if $m$ is taken as the subhalo mass at the
infall time, the resulting \emph{unevolved} subhalo mass function is
universal for any mass $M_{\rm H}$ and as such described by the parameter set
$[\gamma,\alpha,\beta,\omega]=[0.22,-0.91,6.00,3.00]$. This is plotted in
Fig.~A1.

However, we are more interested in taking $m$ as the mass of the surviving,
self-bound entity at redshift $z$, which is reduced with respect to that at
accretion due to mass stripping and dynamical friction. The resulting
\emph{evolved} subhalo mass function is then described by the same functional
shape in Eq.~(A1) but with modified parameter set
$[\gamma,\alpha,\beta,\omega]=[0.31\, f_s,-0.82,50.00,4.00]$ which depends on
the host halo mass and redshift through the quantity $f_s$. The latter may be
determined as follows: first one obtains the half-mass redshift $z_{0.5}$
solving $\delta_c(z_{0.5})=\delta_c(z)+1.19\,\sqrt{\sigma^2(M_{\rm
H}/2)-\sigma^2(M_{\rm H})}$, $\delta_c(z)$ being the linear threshold for
collapse at redshift $z$, and $\sigma^2(M)$ the mass variance at the scale
$M$. Then one computes $N_\tau=\int_{t(z)}^{t(z_{0.5})}{\rm d}t/\tau_{\rm
dyn}(t)$, $\tau_{\rm dyn}$ being the halo dynamical time. Finally, $f_s =
0.3563\,N_\tau^{-0.6}-0.075$ holds. The outcome is illustrated for different
redshift and host halo masses in Fig.~A1.

Now we can compute the overall subhalo contribution to the halo mass function
in the mass bin between $M_{\rm H}$ and $M_{\rm H}+{\rm d}M_{\rm H}$ as
\begin{equation}
N_{\rm subH}(\log M_{\rm H}) = \int_0^\infty{\rm d}\log M_{\rm H}'~
N_{\rm H}(\log M_{\rm H}')\,N(\log \psi)_{|\psi=M_{\rm H}/M_{\rm
H}'}~,
\end{equation}
where $N_{\rm H}(\log M_{\rm H})$ is the standard halo mass function (see
Sheth \& Tormen 1999; Tinker et al. 2008). Thus the total halo $+$ subhalo
mass function just reads
\begin{equation}
N_{\rm H+subH}(\log M_{\rm H})=N_{\rm H}(\log M_{\rm H})+N_{\rm subH}(\log
M_{\rm H})~.
\end{equation}
In Fig.~A2 we plot at different redshift the halo mass function, the overall
subhalo mass function, and the total halo plus subhalo mass function. It is
easily seen that the subhalo contribution is almost negligible for any
redshift in the mass range of interest for this work.

Now we turn to compute the probability distribution for a given halo to
contain one individual galaxy. The first step is to obtain the halo
occupation number (HON), i.e., the average number of subhalos inside a host
halo of mass $M_{\rm H}$; it writes
\begin{equation}
\langle N\rangle(M_{\rm H},z) = \int_{\log m_{\rm min}/M_{\rm H}}^0{\rm d}\log
\psi~N(\log \psi)~.
\end{equation}
Here $m_{\rm min}$ represents a minimum mass for subhalos, required to avoid
the divergence in the above integral. This will be set by comparison with
numerical simulations and observational datasets. The resulting HON as a
function of $M_{\rm H}$ and redshift, for different minimum subhalo masses
$m_{\rm min}$ is illustrated in Fig.~A3. For high $M_{\rm H}\gg m_{\rm min}$
the HON is well fitted by a power-law with logarithmic slope $\la 1$, going
into an abrupt cutoff for masses $M_{\rm H}\la 3-5\, m_{\rm min}$.

The HON represents the \emph{average} number of subhalos inside a host halo,
but we need instead the probability distribution $P(N|\langle N\rangle)$ of
having $N$ subhalos given the average number $\langle N\rangle(M_{\rm H},z)$.
Numerical simulations and HOD models aimed at reproducing various galaxy
observables (Zehavi et al. 2005, 2011; Zheng et al. 2007, 2009; Tinker et al.
2013) indicate that such a distribution is well approximated by a Poissonian.
Then one can easily compute the cumulative probability $P(<N|\langle
N\rangle)$ of having less than $N$ subhalos. This reads
\begin{equation}
P(<N|\langle N\rangle) = {\Gamma(\underline{N+1},\langle
N\rangle)\over \underline{N}!}~,
\end{equation}
where $\Gamma(a,x)=\int_x^\infty{\rm d}t~t^{a-1}\, e^{-t}$ is the incomplete
complementary $\Gamma$-function, $\underline{x}$ is the floor function (the
closest integer lower than $x$), and $n!=1\times 2\times...\times n$ the
factorial function. We stress that in such a probability the dependence on
host halo mass and redshift are encased into the HON $\langle N\rangle(M_{\rm
H},z)$.

\begin{figure}
\epsscale{1.0}\plotone{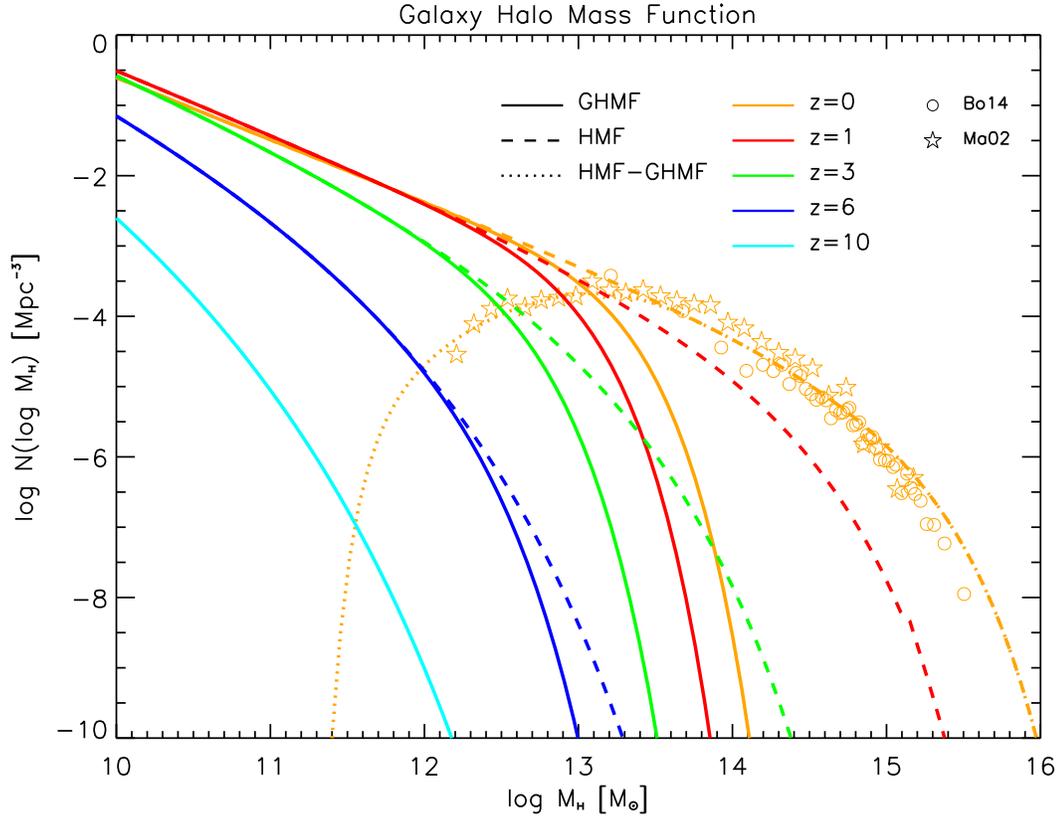}\caption{The galaxy halo mass
function $N(\log M_{\rm H})$ (solid lines) at redshift $z=0$ (orange), $1$
(red), $3$ (green), $6$ (blue) and $10$ (cyan), vs. the halo mass $M_{\rm H}$
is solar units. This is obtained from the halo mass function (dashed lines)
by adding the global subhalo mass function and subtracting the mass function
of multiply-occupied halos (or equivalently multiplying by the probability of
single occupation). More details are given in App.~A. At $z=0$ we also report
as a dotted line the resulting cluster and group halo mass function (obtained
by subtraction of the solid from the dashed line); this is compared with the
determinations by Boehringer et al. (2014; circles) from X-ray observations
of groups and clusters and by Martinez et al. (2002; stars) from optical
observations of loose groups.}\label{fig|GHMF}
\end{figure}

\begin{figure}
\epsscale{0.7}\plotone{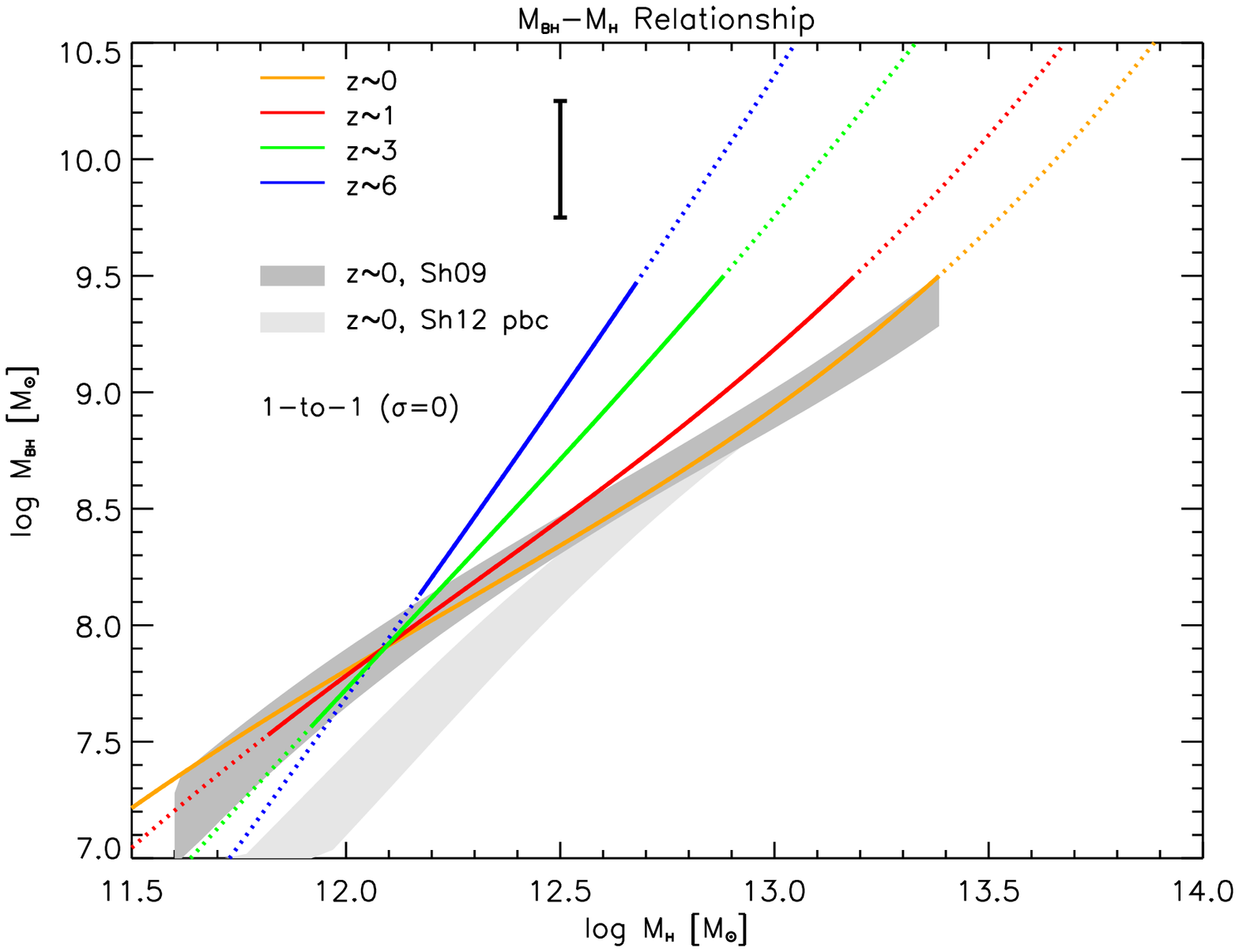}\\\plotone{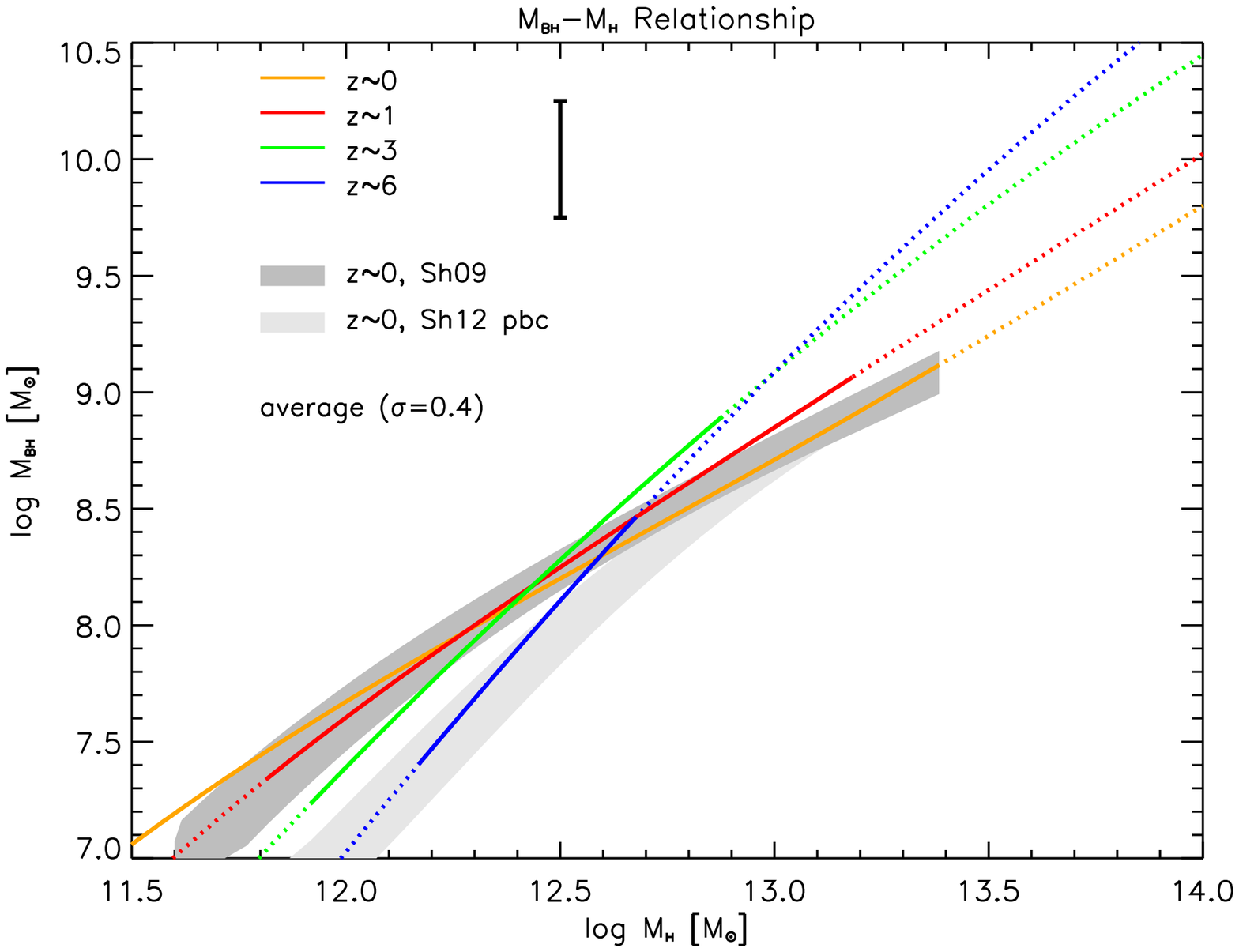}\caption{Relationship
between BH mass $M_{{\rm BH}}$ and halo mass $M_{\rm H}$ from the abundance
matching technique, at redshift $z=0$ (orange), $1$ (red), $3$ (green), and $6$
(blue). Top panel shows the results when a one-to-one
(i.e., no scatter) relationship $M_{{\rm BH}}$ vs. $M_{\rm H}$ is assumed,
while bottom panel shows the resulting average relationship when a Gaussian
distribution in $M_{{\rm BH}}$ at given $M_{\rm H}$ with a scatter of $0.4$
dex (see text) is assumed. In both panels the black errorbar illustrates the
typical associated uncertainty, and the dotted lines highlight the ranges
not covered by the current data on the BH mass function. The grey shaded
areas show the relations at $z = 0$ from the BH mass functions uncorrected
(dark grey) and corrected (light grey) for pseudobulges by Shankar et al.
(2009) and (2012), respectively.}\label{fig|Mbh_MH}
\end{figure}

\begin{figure}
\epsscale{0.7}\plotone{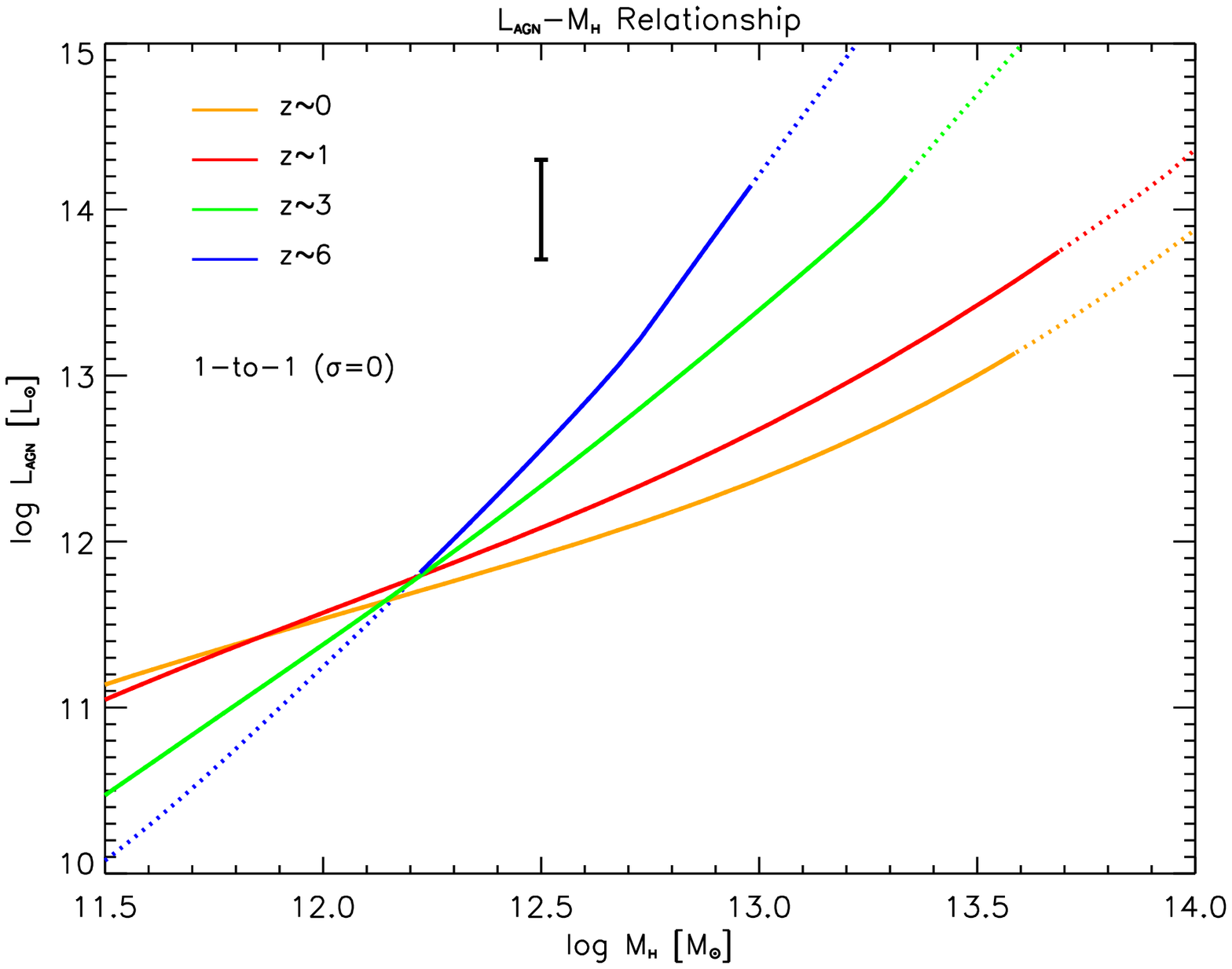}\\\plotone{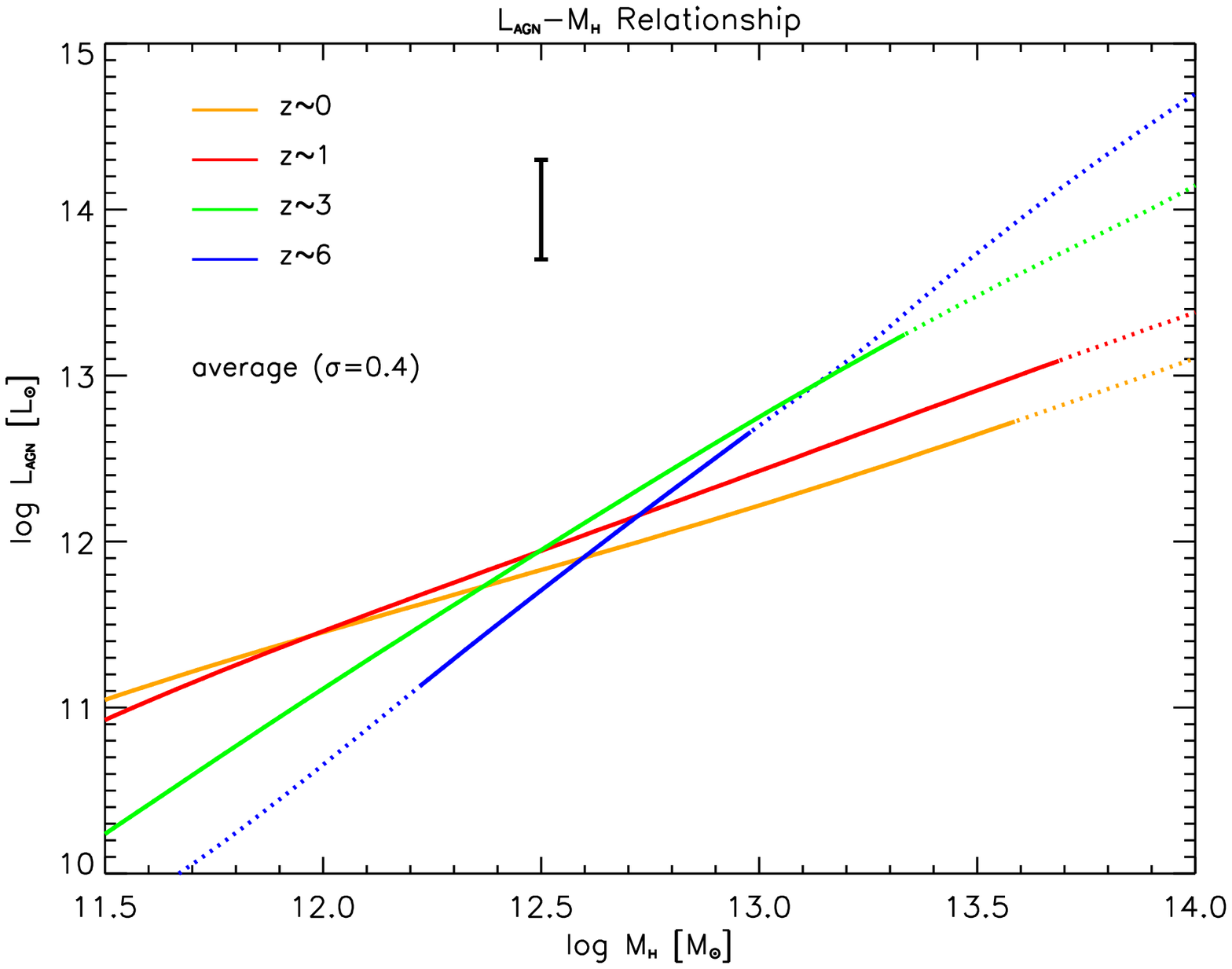}
\caption{Relationship between bolometric AGN luminosity $L_{\rm AGN}$ and
halo mass $M_{\rm H}$ from the abundance matching technique, at redshift
$z=0$ (orange), $1$ (red), $3$ (green), and $6$ (blue). Top panel shows the
results when a one-to-one (i.e., no scatter) relationship $L_{\rm AGN}$ vs.
$M_{\rm H}$ is assumed, while bottom panel shows the resulting average
relationship when a Gaussian distribution in $L_{\rm AGN}$ at given $M_{\rm
H}$ with a scatter of $0.4$ dex (see text) is assumed; in both panels the
black errorbar illustrates the typical associated uncertainty, and the
dotted lines highlight the ranges not covered by the current
data on the AGN luminosity function.}\label{fig|Lagn_MH}
\end{figure}

\begin{figure}
\epsscale{1.0}\plotone{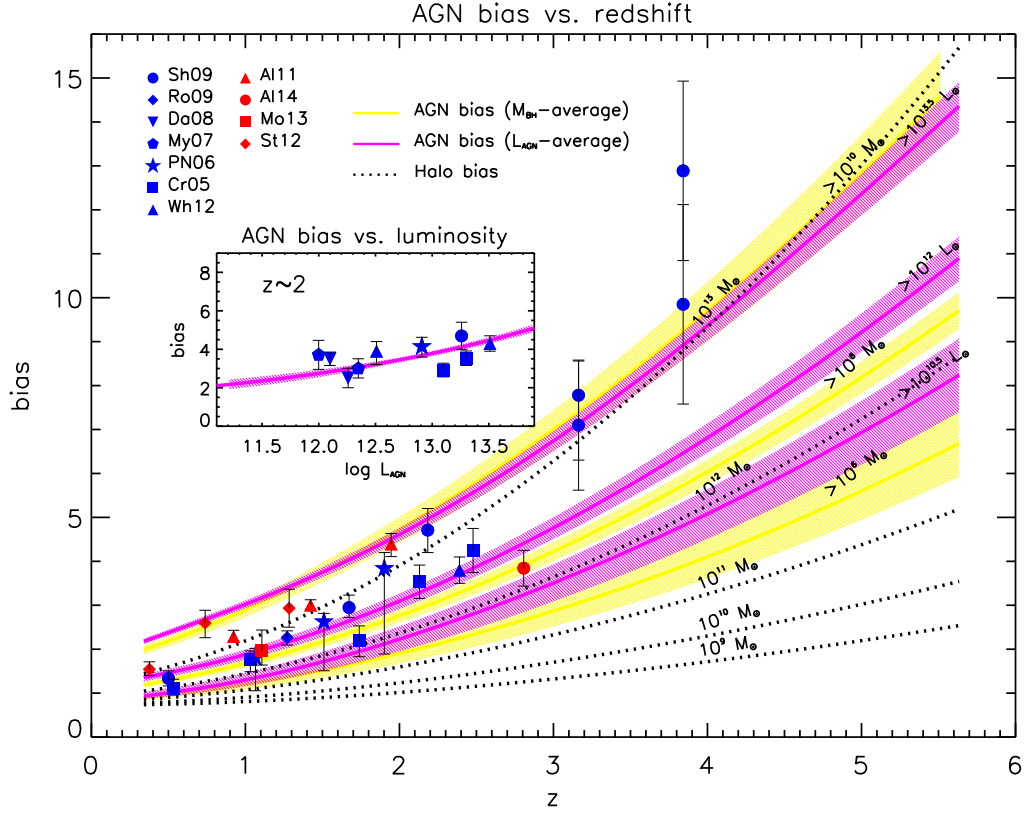}\caption{The AGN bias as a function of
redshift $z$. Results from the abundance matching technique are illustrated
by magenta ($L_{\rm AGN}$-average bias) and yellow ($M_{\rm BH}$-average bias)
continuous lines, with the hatched areas showing the associated
uncertainty; specifically, the magenta curves refer to different AGN
luminosities $L_{\rm AGN}>10^{10.5}$, $10^{12}$, and $10^{13.5}\, L_\odot$
and the yellow curves to different BH masses $M_{\rm BH}>10^6$, $10^8$, and
$10^{10}\, M_\odot$ as labeled. Black dotted lines illustrate for comparison
the halo bias referring to different halo masses from $10^9$ to $10^{13}\,
M_\odot$ as labeled. The inset shows the AGN bias from the abundance matching
technique at redshift $z=2$ as a function of the bolometric AGN luminosity
$L_{\rm AGN}$. Optical data are from Shen et al. (2009; blue circles), Ross
et al. (2009; blue diamonds), Da \^{Angela} et al. (2008; blue reversed
triangles), Myers et al. (2007; blue pentagons), Porciani \& Norberg (2006;
blue stars), Croom et al. (2005; blue squares), White et al. (2012; blue
triangles); X-ray data from Allevato et al. (2011; red triangles), Allevato
et al. (2014; red circles), Mountrichas et al. (2013; red squares), and
Starikova et al. (2012; red diamonds).}\label{fig|AGN_bias}
\end{figure}

\begin{figure}
\epsscale{1.}\plotone{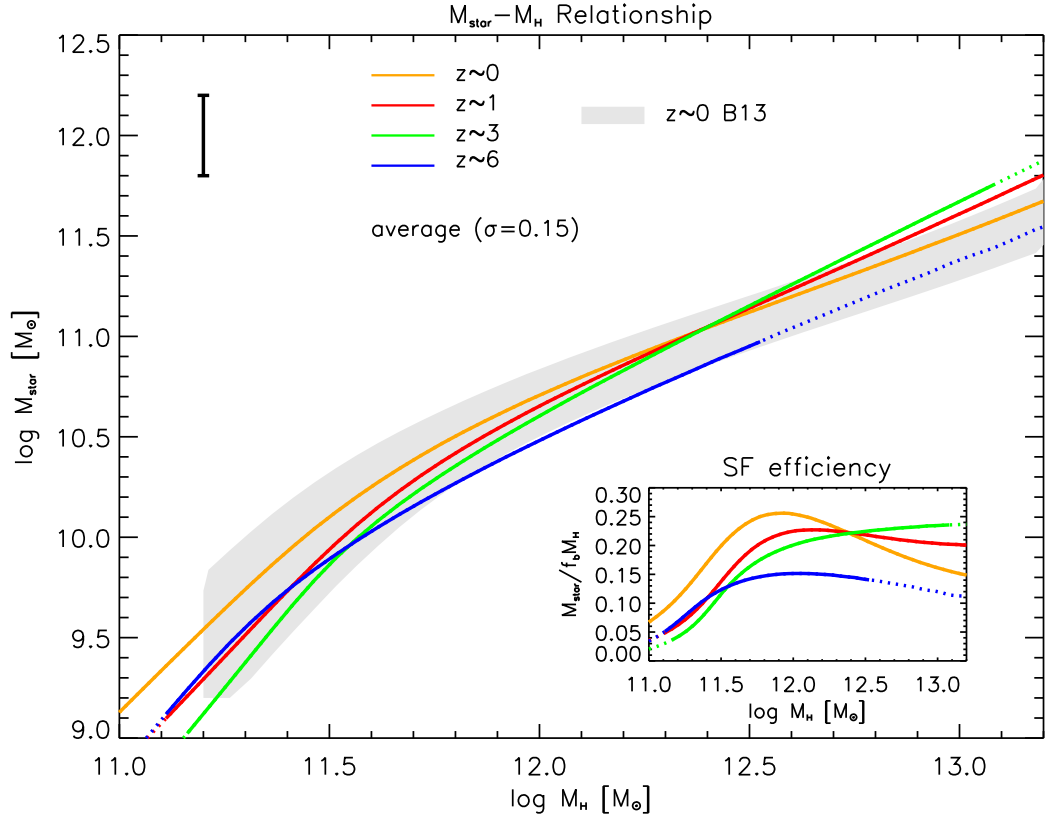}\caption{Relationship between stellar
mass $M_{\rm star}$ and halo mass $M_{\rm H}$ from the abundance matching
technique, at redshift $z=0$ (orange), $1$ (red), $3$ (green), and $6$
(blue). The results refer to the average relationship when a Gaussian
distribution in $M_{\rm \star}$ at given $M_{\rm H}$ with a scatter of $0.15$
dex is assumed (the one-to-one relationship is practically identical); the
black errorbar illustrates the typical associated uncertainty, and the dotted
lines highlight the ranges not covered by the current data on the stellar
mass function. The grey shaded area shows the relation at $z=0$ obtained from
the observed stellar mass function by Bernardi et al. (2013). The inset
illustrates the efficiency $M_{\star}/f_{\rm b}\, M_{\rm H}$ for the
conversion of the initial baryonic mass $f_{\rm b}\, M_{\rm H}=0.2\, M_{\rm
H}$ associated to the halo into the final stellar mass
$M_\star$.}\label{fig|Mstar_MH}
\end{figure}

\begin{figure}
\epsscale{1.}\plotone{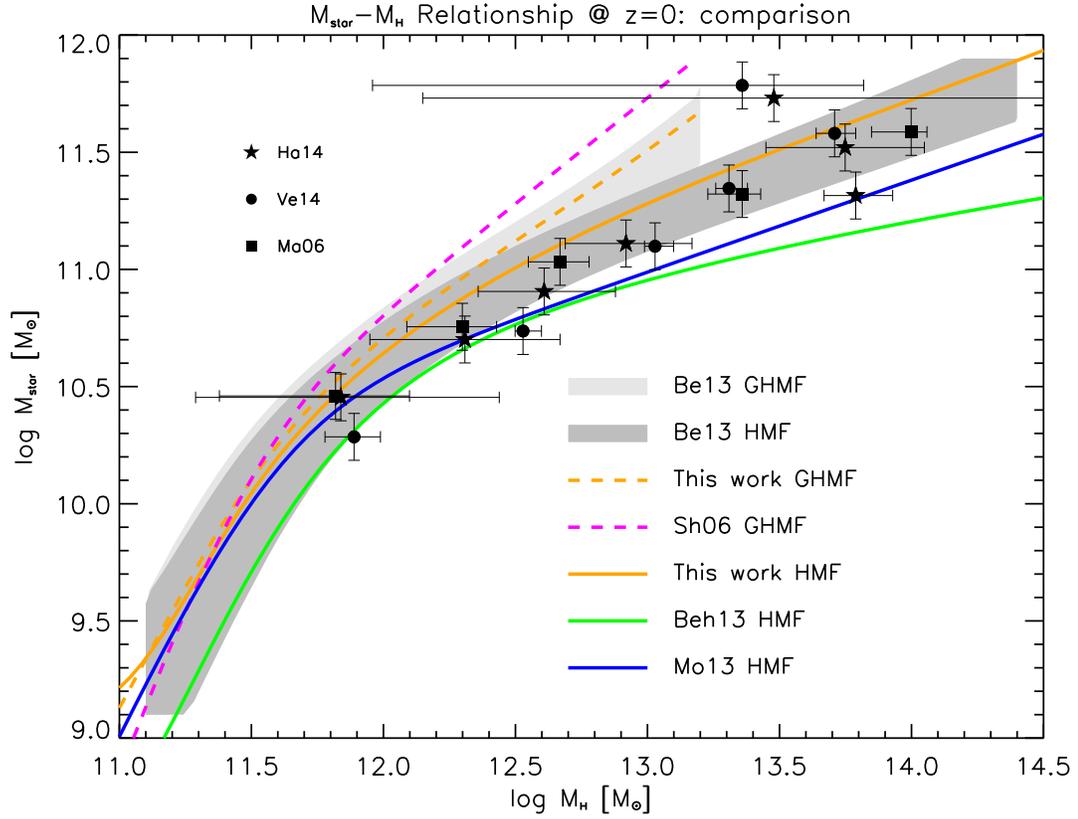}\caption{Relationship between
stellar mass $M_{\star}$ and halo mass $M_{\rm H}$ from the
abundance matching technique at redshift $z=0$ (orange line), when the galaxy
halo mass function (dashed) or the full halo mass function (solid) are
adopted.  The grey shaded areas show the relations at $z=0$ obtained from the
observed stellar mass function by Bernardi et al. (2013), matched with the
galaxy (light grey) or the overall (dark grey) halo mass function. The green
solid line refers to the result by Behroozi et al. (2013), the blue solid
line to that by Moster et al. (2013), and the magenta dashed line to that by
Shankar et al. (2006). Data from gravitational lensing measurements in groups
and clusters of galaxies are from Han et al. (2014; filled stars), Velander
et al. (2014; filled circles), and Mandelbaum et al. (2006; filled
squares).}\label{fig|Mstar_MH_z0comp}
\end{figure}

\begin{figure}
\epsscale{1.}\plotone{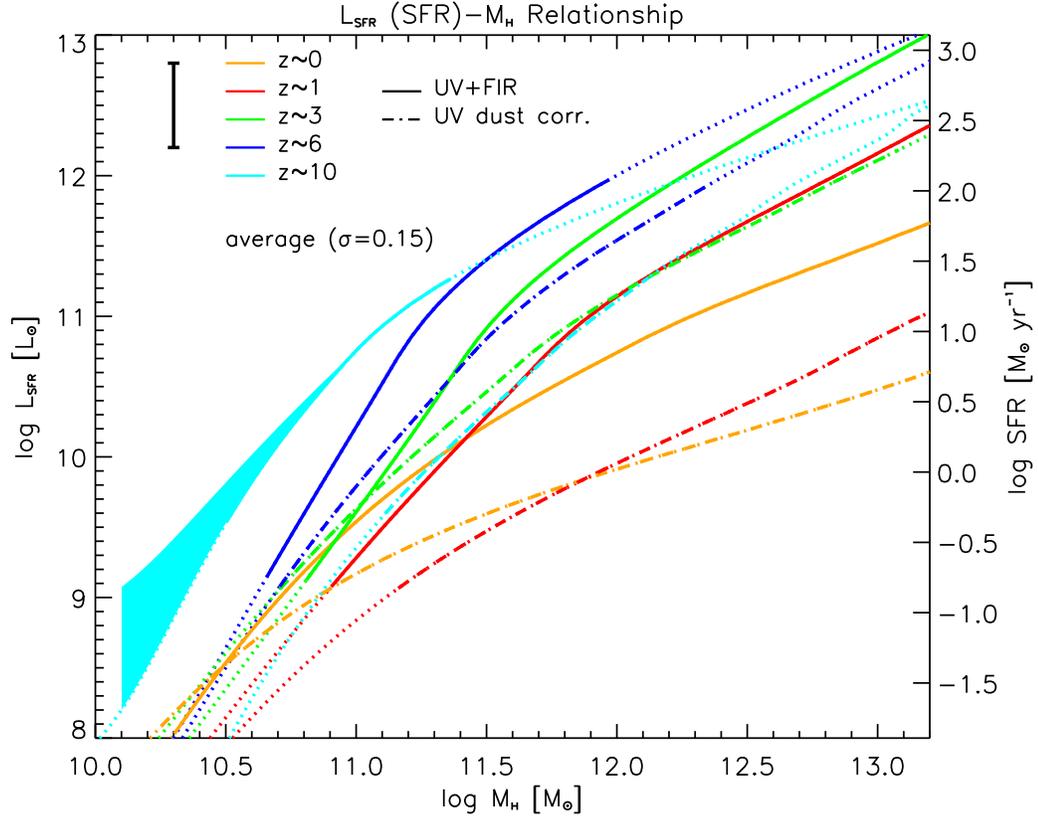}\caption{Average relationship between
bolometric SFR-luminosity $L_{\rm SFR}$ and halo mass $M_{\rm H}$ from the
abundance matching technique, at redshift $z=0$ (orange), $1$ (red), $3$
(green), $6$ (blue), and $10$ (cyan). A Gaussian distribution in $L_{\rm
SFR}$ at given $M_{\rm H}$ with a scatter of $0.15$ dex (see text) is assumed
(the one-to-one relationship is identical); the black errorbar illustrates
the typical associated uncertainty, and the dotted lines highlight the ranges
not covered by the current data on the SFR-luminosity function. Solid lines
refer to the overall SFR-luminosity function, while dot-dashed lines to
dust-corrected UV luminosity function only. At $z=10$ the cyan shaded area
for small halo masses illustrates the systematic uncertainty related to the
faint-end slope of the SFR-luminosity function.}\label{fig|Lstar_MH}
\end{figure}

\begin{figure}
\epsscale{1.0}\plotone{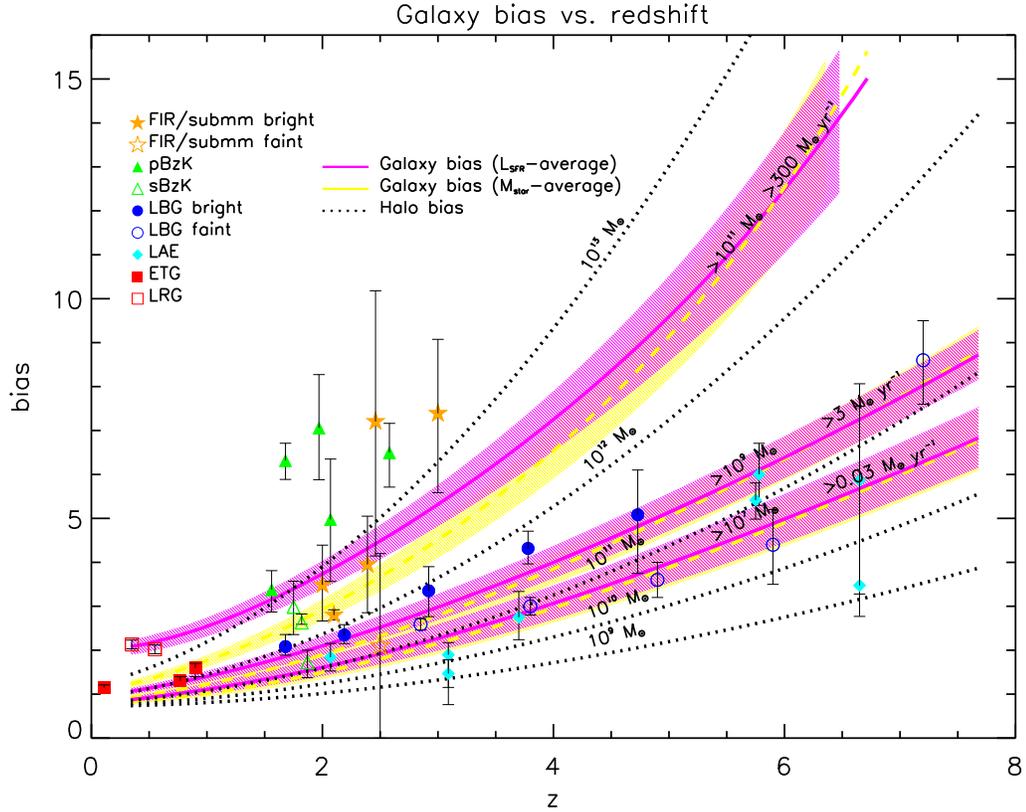}\caption{The galaxy bias as a function
of redshift $z$. Results from the abundance matching technique are
illustrated by magenta ($L_{\rm SFR}$-average bias) and yellow
($M_\star$-average bias) continuous lines, with the hatched areas showing the
associated uncertainty; specifically, the magenta curves refer to different
SFRs$>0.03$, $3$, and $300\, M_\odot$ yr$^{-1}$ and the yellow curves to
different stellar masses $M_\star>10^7$, $10^9$, and $10^{11}\, M_\odot$ as
labeled. Black dotted lines illustrate for comparison the halo bias referring
to different halo masses from $10^9$ to $10^{13}\, M_\odot$ as labeled. Data
for FIR/sub-mm bright galaxies (filled orange stars) are from Webb et al.
(2003), Blain et al. (2004), Weiss et al. (2009), Hickox et al. (2012),
Bianchini et al. (2014), for FIR/sub-mm faint galaxies (orange empty stars)
are from Ono et al. (2014), for passive BzK galaxies (green filled triangles)
are from Grazian et al. (2006), Quadri et al. (2007), Blanc et al. (2008),
Furusawa et al. (2011), Lin et al. (2012),  for starforming BzK galaxies
(green empty triangles) are from Hayashi et al. (2007), Blanc et al. (2008),
Furusawa et al. (2011), for bright Lyman Break Galaxies (blue filled circles)
are from Ouchi et al. (2004), Adelberger et al. (2005), Lee et al. (2006),
Overzier et al. (2006), for faint Lyman Break Galaxies (blue empty circles)
are from Bielby et al. (2013), Barone-Nugent et al. (2014), for
Lyman-$\alpha$ Emitters (cyan diamonds) are from Gawiser et al. (2007), Ouchi
et al. (2010), Guaita et al. (2010), for passively-evolving Early-Type
Galaxies (red filled squares) are from Hawkins et al. (2003), Guzzo et al.
(2008), Georgakakis et al. (2014), for Luminous Red Galaxies (red empty
squares) are from Tegmark et al. (2006), Ross et al.
(2007).}\label{fig|STAR_bias}
\end{figure}

\begin{figure}
\epsscale{1.0}\plotone{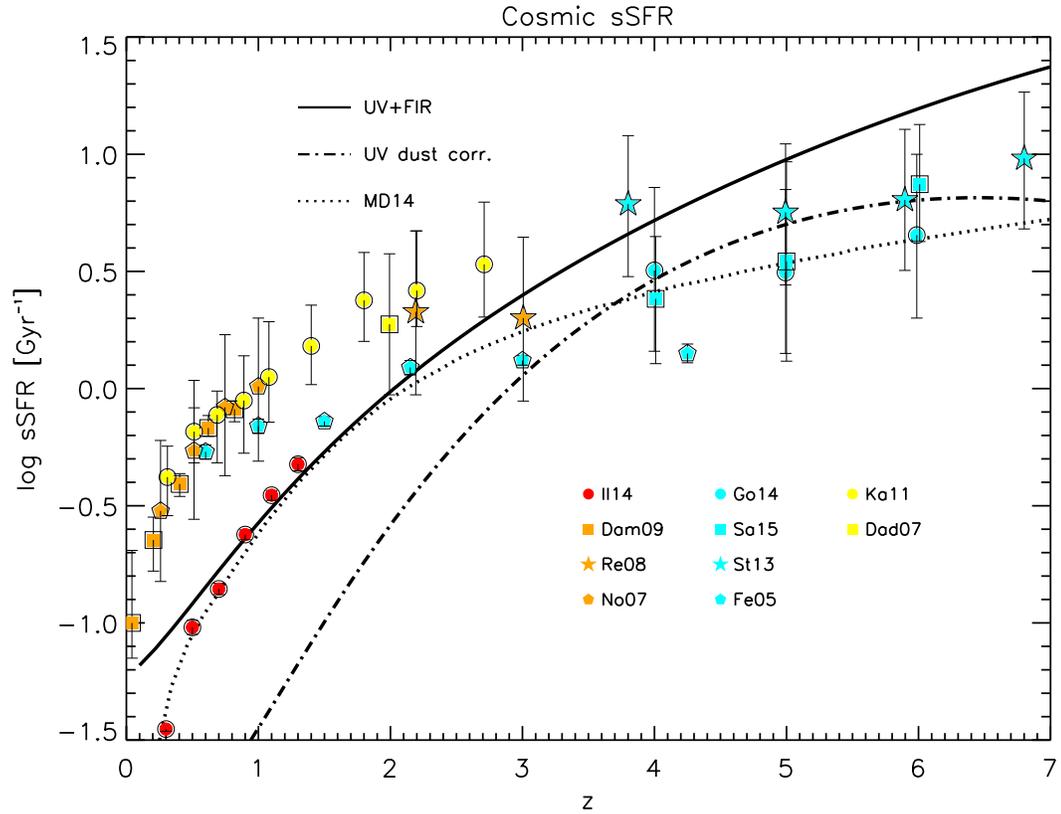}\caption{Cosmic sSFR as a function of
redshift. Solid line refers to the overall SFR-luminosity function, while
dot-dashed line to dust-corrected UV luminosity function only, and the dotted
line illustrates the model by Madau \& Dickinson (2014). IR data are from
Ilbert et al. (2014; red circles, referring to $M_\star\ga 10^{10.5}\,
M_\odot$), Damen et al. (2009; orange squares), Reddy et al. (2008; orange
stars), Noeske et al. (2007; orange pentagons); UV data are from Gonzalez et
al. (2014; cyan circles), Salmon et al. (2015; cyan squares), Stark et al.
(2013; cyan stars), Feulner et al. (2005; cyan pentagons); radio data are
from Karim et al. (2011; yellow circles), Daddi et al. (2007; yellow
squares).}\label{fig|sSFR_z}
\end{figure}

\begin{figure}
\epsscale{1.0}\plotone{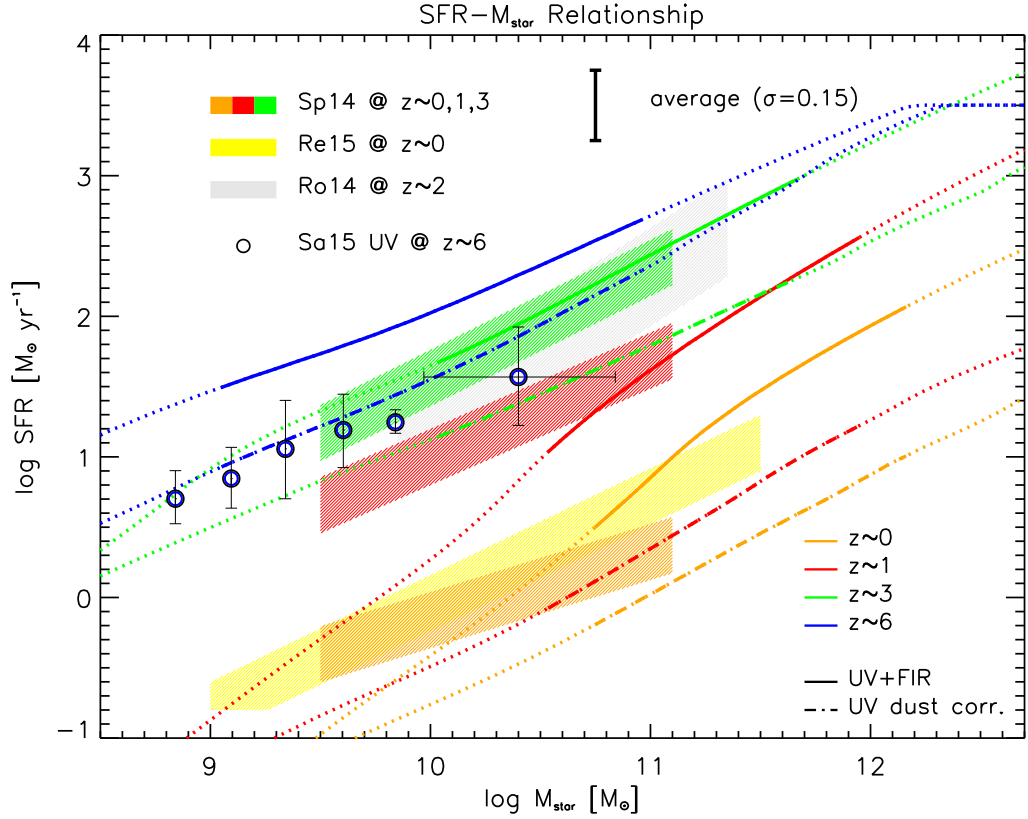}\caption{Relationship between the
SFR and the final stellar mass from the abundance matching technique (the so
called `main sequence' of starforming galaxies), at redshift $z=0$ (orange),
$1$ (red), $3$ (green), and $6$ (blue). Results with and without scatter are
almost undistinguishable. Solid lines refer to the overall SFR-luminosity
function, while dot-dashed lines to dust-corrected UV luminosity function
only. The black errorbar illustrates the typical associated uncertainty. The
dotted lines highlight the ranges not covered by the current data on the
SFR-luminosity and stellar mass functions, or where the determination
from the abundance matching technique is largely uncertain because of the
flatness at the faint end of the stellar mass function. Observational
estimates are in the range $z\sim 0-3$ by Speagle et al. (2014; orange, red
and green areas), at $z\sim 0$ by Renzini \& Peng (2015; yellow area), at
$z\sim 2$ by Rodighiero et al. (2014; light grey area), and at $z\sim 6$ from
UV determinations by Salmon et al. (2015; blue open circles).} \label{fig|sSFR}
\end{figure}

\begin{figure}
\epsscale{0.7}\plotone{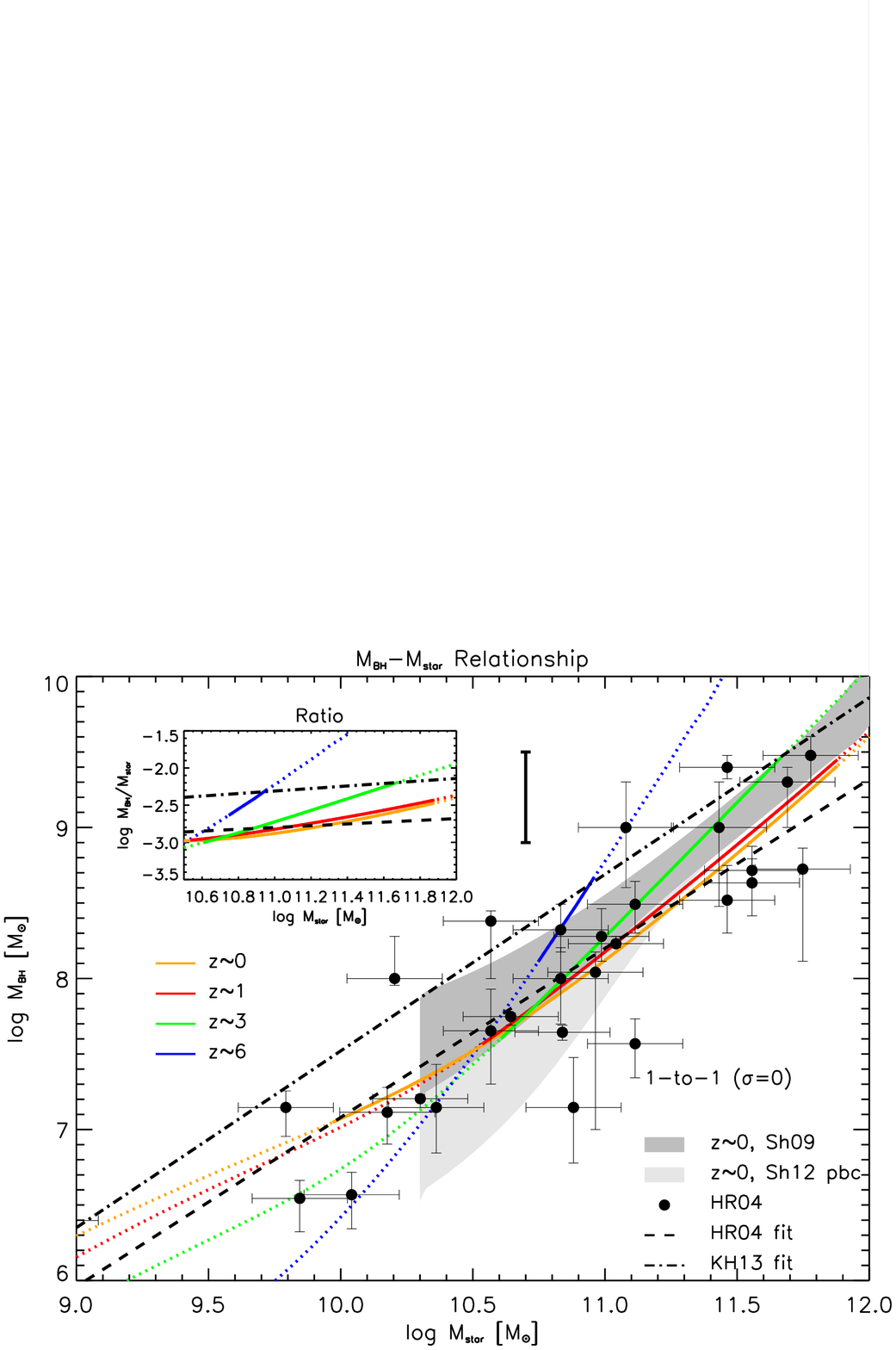}\\\plotone{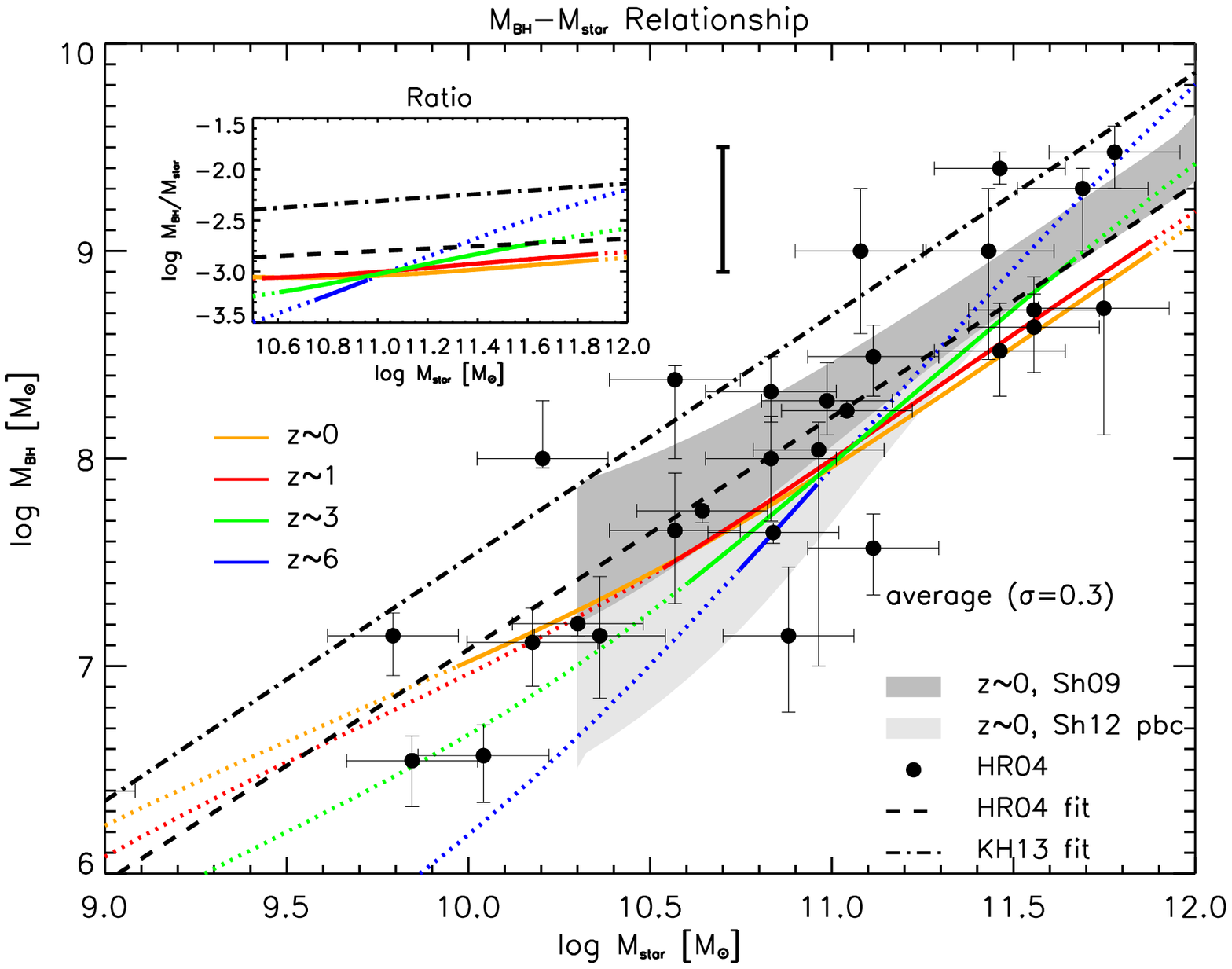}\caption{Relationship
between the BH mass $M_{{\rm BH}}$ and the stellar mass $M_{\star}$ from the
abundance matching technique, at redshift $z=0$ (orange), $1$ (red), $3$
(green), and $6$ (blue). Top panel shows the results when a one-to-one (i.e.,
no scatter) relationship $M_{{\rm BH}}$ vs. $M_{\rm H}$ is assumed, while
bottom panel shows the resulting average relationship when a Gaussian
distribution in $M_{{\rm BH}}$ at given $M_{\rm H}$ with a scatter of $0.3$
dex (see text) is assumed; in both panels the black errorbar illustrates the
typical associated uncertainty, and the dotted lines highlight the ranges not
covered by the current data on the BH and stellar mass functions. The shaded
areas show the relations at $z=0$ obtained from the matching between the
stellar and the BH mass functions, uncorrected (dark grey) and corrected
(light grey) for pseudobulges. Data points are from the compilation by Haring
\& Rix (2004), with the dashed line representing their best-fit relation; the
relation proposed by Kormendy \& Ho (2013) is also shown as a dot-dashed
line. The inset illustrates the corresponding BH-to-stellar mass ratio
$M_{{\rm BH}}/M_\star$ as a function of $M_\star$.}\label{fig|Mbh_Mstar}
\end{figure}

\begin{figure}
\epsscale{1.}\plotone{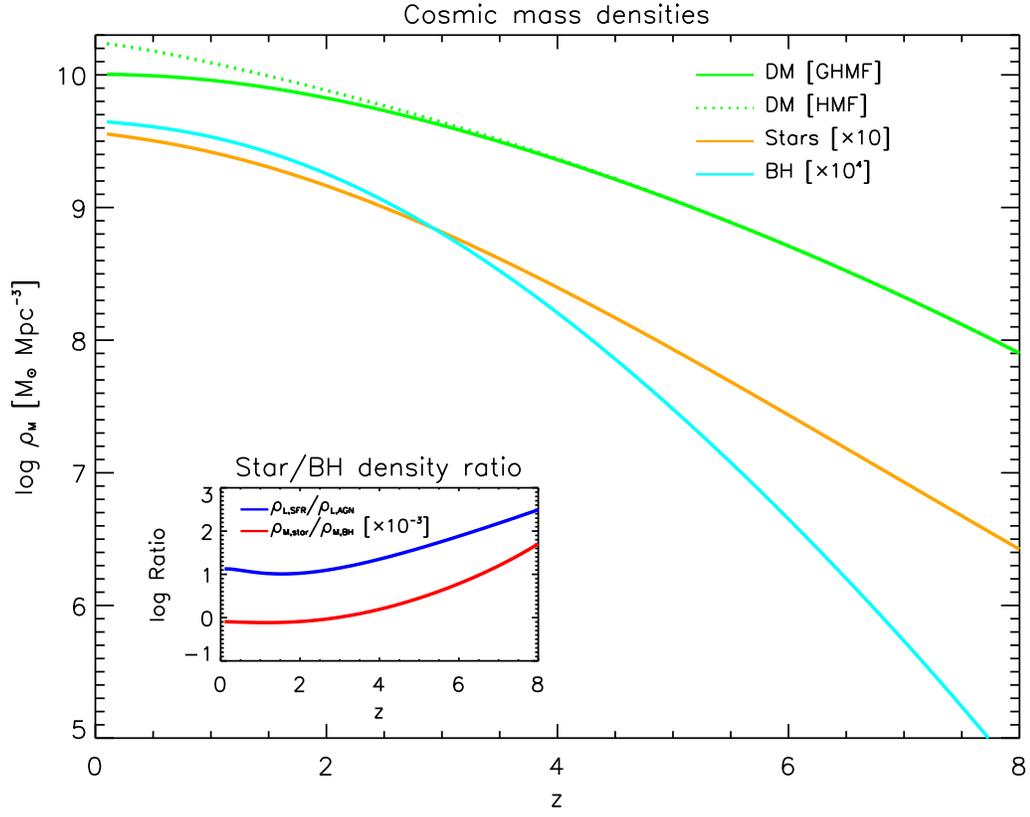}\caption{The evolution with redshift of
the mass density in the overall DM halo population (dotted green line), in
galactic DM halos (solid green line), in stars (solid orange line), and in
black holes (solid cyan line). In the inset the luminosity density ratio
$\rho_{L_{\rm SFR}}/\rho_{L_{\rm AGN}}$ (blue line) and the mass
density ratio $\rho_{M_{\star}}/\rho_{M_{\rm BH}}$ (red line) are
illustrated.}\label{fig|cosmacc}
\end{figure}

Finally, the galaxy halo mass function can be computed as
\begin{equation}
N_{\rm GHMF}(\log M_{\rm H})=N_{\rm H+subH}(\log M_{\rm H})\times
P(<N=1|\langle N\rangle)~.
\end{equation}
The outcomes at different redshifts, having adopted a minimum satellite mass
of $m_{\rm min}=10^{11}\, M_\odot$, are illustrated in Fig.~\ref{fig|GHMF} of
the main text. With respect to the full halo $+$ subhalo mass function, the
GHMF features a cutoff for host halo masses $M_{\rm H}\ga 1-3\times 10^{13}\,
M_\odot$, more pronounced at lower redshift. This is because the probability
of hosting subhalos (hence more than one galaxy) increases strongly for large
masses. In other words, such halos are more likely to host a galaxy group or
cluster than an individual galaxy.

We stress that both a minimum mass of $\sim$ a few $10^{11}\, M_\odot$ for
satellite halos (corresponding to the adopted $m_{\rm min}$), and a maximal
value $\sim$ a few $10^{13}\, M_\odot$ for a halo to host an individual
galaxy (corresponding to the resulting cutoff in the GHMF) are strongly
indicated by galaxy weak lensing observations (Mandelbaum et al. 2006, van
Uitert et al. 2011, Leauathaud et al. 2012, and Velander et al. 2014).
Furthermore, such maximum galactic halo masses are also strongly suggested by
dynamical observations of gas and stars in nearby galaxies (see Gerhard et
al. 2001, Andreon et al. 2014; see also the review by Corteau et al. 2014).

Finally, to provide a further observational test, we have computed the group
and cluster mass function by subtracting the GHMF from the overall halo $+$
subhalo mass function. This represents the halo mass function associated only
to galaxy systems, and as such is expected to show a cutoff for halo masses
$\la 10^{13}\, M_\odot$. The result at $z=0$ is plotted as a dotted line in
Fig.~\ref{fig|GHMF} of the main text, and compared with the determinations by
Boehringer et al. (2014) from X-ray observations of groups and clusters and
by Martinez et al. (2002) from optical observations of loose groups; the
agreement is quite impressive. We notice that the behavior in the range of
poor clusters and groups is particularly sensitive to the parameter $m_{\rm
min}$. Thus the agreement with the data is another indication that the
adopted value around $10^{11}\, M_\odot$ is appropriate.

The overall halo and the galaxy halo mass functions can be fitted with the
functional shape of Eq.~(\ref{eq|AGN_LF}) with $L_{\rm AGN}$ replaced by
$M_{\rm H}$, and the parameter values given in Table~1. The resulting fits
are accurate within $5\%$ in the redshift range from $0$ to $10$ and halo
mass range from $10$ to $14$ for the GHMF and from $10$ to $16$ for the HMF.

\section{Appendix B: Dry mergers}

Many recent works (e.g., Shankar et al. 2009, 2013; Moster et al. 2010, 2013)
have shown that the role of dry mergers (i.e., addition of the whole mass
content in stars or BHs of the merging halos without contributing
significantly to star formation or BH accretion) in building up the
BH/stellar mass functions is less important than accretion/in-situ star
formation at $z\ga 1$. This is simply because the evolutionary times
associated to mergers are much longer than those associated to the in-situ
BH/stellar mass growth. On the other hand, at $z\la 1$ the situation is
expected to reverse. This is because the cold accreting or starforming gas
within the DM halo gets progressively exhausted or is ejected/heated by the
energy feedback from supernovae or from the AGN itself. In fact, the
continuity equation with accretion only yields little evolution of the mass
functions from $z\sim 1$ to $z\sim 0$ (cf. Figs.~\ref{fig|BH_MF} and
\ref{fig|STAR_MF} and related insets). Thus the low redshift $z\la 1$
evolution could be in principle more affected by dry merging. This is a hotly
debated issue in the literature, especially in relation to the size evolution
of massive, passively-evolving galaxies (e.g., Naab et al. 2009; Fan et al.
2010; Nipoti et al. 2012; Shankar et al. 2015; Kulier et al. 2015).

In this Appendix we highlight the impact of dry mergers on the supermassive
BH/stellar mass functions at $z\la 1$. We start from the observed mass
functions at $z\sim 1$, then evolve them down to redshift zero by taking into
account both dry mergers and accretion in the continuity equation. The effect
of dry mergers is evaluated numerically with a mid-point scheme computation
that divides the overall timegrid in sufficiently small steps $\delta t$, and
then evolves the mass function at each timestep $t_{i}$ according to
\begin{equation}
N(\log M, t_{i}+\delta t) = N(\log M, t_i)+{\mathcal{P}\over 2}\,
N(\log M/2, t_{i})\,\delta t-\mathcal{P}\,N(\log M,t_{i})\, \delta t~,
\end{equation}
where $\mathcal{P}$ is the probability of dry mergers. We adopt the common
simplifying assumption that dry merging of the associated stellar and BH
components follows halo mergers of given mass ratio. We base on the DM
merging rates provided by Stewart et al. (2009) via high-resolution $N-$body
simulations, and write
\begin{equation}
\mathcal{P}(>\zeta)\approx 0.02\,{\delta t\over {\rm
Gyr}}\,(1+z)^{2.1}\,{(1-\zeta)^{0.72}\over \zeta^{0.54}}~,
\end{equation}
where $\zeta$ specifies the mass ratio above which mergers are considered;
thus $\mathcal{P}(>0.5)$ is the probability of major mergers, while
$\mathcal{P}(>0.1)-\mathcal{P}(>0.5)$ is that of minor mergers.

The results on the BH and stellar mass functions are illustrated in Fig.~B1
and B2. The impact of dry mergers on the mass functions is apparent only at
the high mass end. Dry mergers increase moderately the space densities of BHs
with mass $M_{{\rm BH}}\ga 10^9\, M_\odot$ and boost that of stellar masses
$M_{\star}\ga 10^{12}\, M_\odot$, ranges where data are still statistically
uncertain and/or affected by large systematics.

Specifically, assuming that a dry merger follows any DM halo merger (either
major or minor) yields a local BH mass function still consistent with data,
even considering the uncertainties on the bolometric corrections in
converting from luminosity to mass (cf. Sect.~\ref{sec|BH_results}). On the
other hand, while major dry merger produce a stellar mass function consistent
with the data (see also Liu et al. 2015), this is not the case for minor dry
mergers. All that implies that the addition of stellar mass by (minor) dry
mergers following the DM halos' must be only partial, possibly depending on
mass ratio, orbital parameters, tidal stripping, and structural properties
(see Naab et al. 2009; Krogager et al. 2014).

\section{Appendix C: The complementarity of UV and FIR data}

In this Appendix we stress the importance of the FIR, in addition to the UV,
data in probing the star formation process in high-redshift galaxies.

To this purpose, we present in Fig.~\ref{fig|STAR_UVLF} the SFR luminosity
function estimated on the basis of dust-uncorrected UV data, dust-corrected
UV data, and UV+FIR data. It is evident that, even when dust-corrected
according to the prescriptions described in Sect.~\ref{sec|STAR_LF}, UV data
strongly undersample the bright end of the luminosity function. For example,
at $z\sim 3$ the number of sources with $\dot M_\star \sim 300\, M_\odot$
yr$^{-1}$, which is not an extreme but rather a typical value, is estimated
to be $10^{-4}$ Mpc$^{-3}$ from UV+FIR data, while it is inferred to be $\la
10^{-6}$ Mpc$^{-3}$ from dust-corrected UV data, and would be $\la 10^{-10}$
Mpc$^{-3}$ from dust-uncorrected UV data. We stress that especially at $z\ga
1.5$, the dust corrections routinely applied to the UV data are unable to
fully account for the population of strongly starforming galaxies seen in the
FIR band.

In Fig.~\ref{fig|STAR_UVMF} we illustrate the stellar mass function obtained
via the continuity equation from the above input luminosity functions. We
keep the same lightcurve of our fiducial model, that for UV-bright,
low-luminosity objects yields a duration of the burst already close to the
Hubble time. As it can be seen, when basing only on UV data (even if
dust-corrected) the high mass end of the resulting stellar mass function is
substantially underpredicted relative to the FIR+UV results (which well
reproduces observational estimates, see Fig.~\ref{fig|STAR_MF}) at any
redshift. Note that this mismatch can hardly be recovered by mass additions
from dry merging events, since a factor $10$ in mass is needed from $z\approx
3$ to $z\approx 0$.

\section{Appendix D: Analytic fits to abundance matching
relationships}

Here we provide analytic fits to the relationships derived from the abundance
matching technique. To fit a relation of the form $Y-M$ we adopt a double
powerlaw shape:
\begin{equation}
Y(M,z) = N(z)\times \left\{\left[{M\over M_b(z)}\right]^{\alpha(z)}+\left[{M\over
M_b(z)}\right]^{\omega(z)}\right\}^{\theta}~,
\end{equation}
with $\theta=-1$ for a convex or $\theta=+1$ for a concave relationship.

The normalization $\log N(z)$, the mass of the break $\log M_b(z)$, and the
characteristic slopes $\alpha(z)$ and $\omega(z)$ evolve with the redshift
according to the same parametrization
\begin{equation}
p(z) = p_0 + k_{p1}\, \chi + k_{p2}\, \chi^2 + k_{p3}\, \chi^3
\end{equation}
with
\begin{equation}
\chi= \log \left(1+z\over 1+z_0\right)
\end{equation}
and $z_0=0.1$. The parameter values are reported in Table~2.

These expressions can be exploited to interpolate and/or extrapolate the
relationships all the way from $z\approx 0$ to $z\approx 6$. Interpolation is
helpful to produce mock galaxy and AGN/BH catalogs that can be used to
compute gravitational lensing effects, to investigate clustering properties,
to gauge sub-grid physics in numerical simulations, and to design
observational setups. On the other hand, extrapolation is particularly
helpful to obtain specific predictions in redshift and mass ranges not
currently probed by the data but within the reach of upcoming experiments.

\clearpage

\clearpage

\clearpage

\clearpage

\clearpage

\clearpage

\clearpage

\clearpage

\clearpage

\clearpage

\clearpage

\clearpage

\clearpage

\clearpage

\clearpage

\clearpage

\clearpage

\clearpage

\clearpage

\clearpage

\clearpage

\clearpage

\clearpage

\clearpage

\clearpage

\clearpage
\begin{figure}
\figurenum{A1}
\plotone{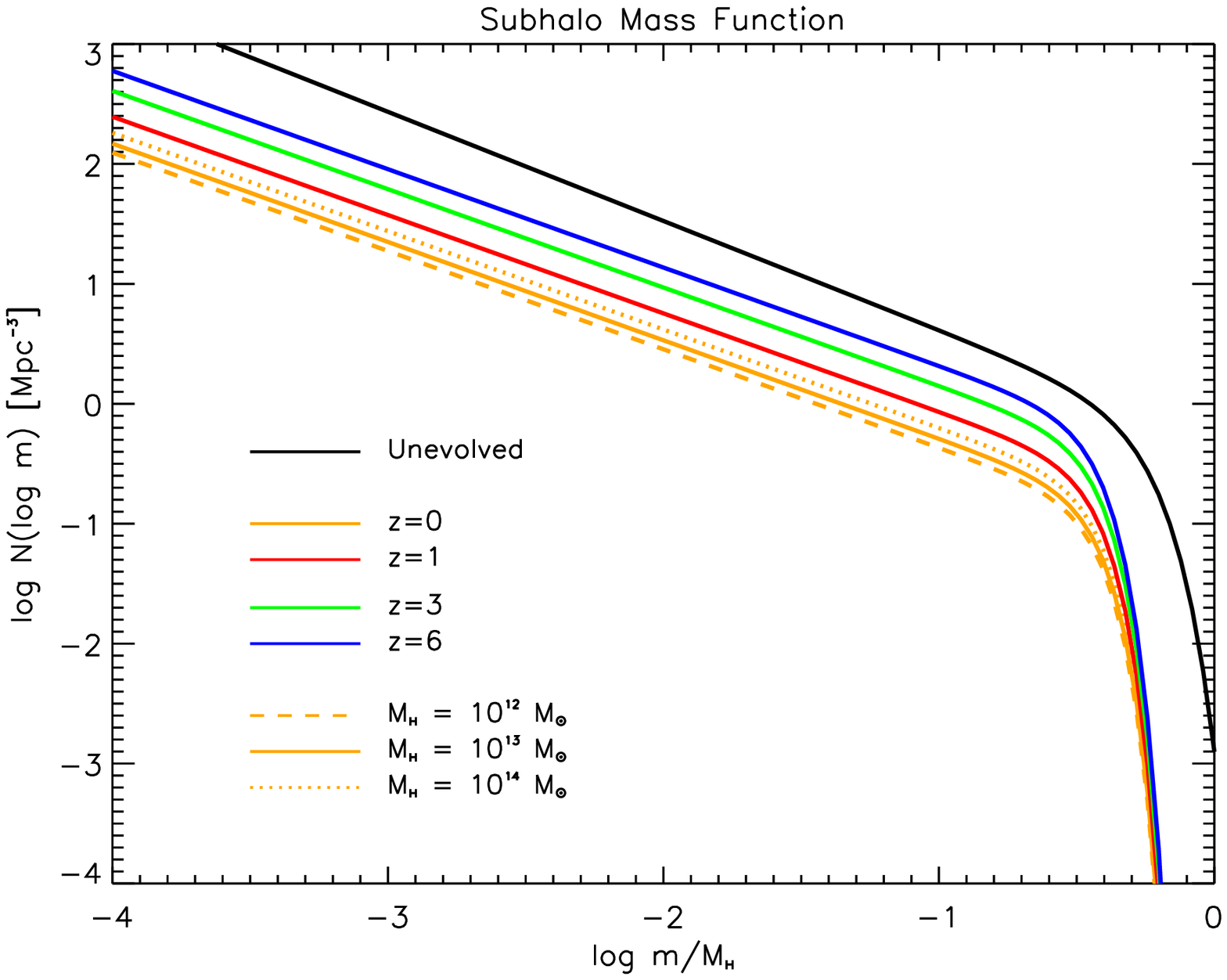}\caption{The subhalo mass function $N(\log m)$ vs.
the ratio between the satellite and the halo masses $m/M_{\rm H}$, computed
according to the prescriptions by van den Bosch \& Jiang (2014). The black
line refers to the unevolved mass function, and colored lines to the evolved
mass function at $z=0$ (orange), $1$ (red), $3$ (green), and $6$
(blue). At $z=0$ the solid line refers to a mass of the host of $M_{\rm
H}=10^{13}\, M_\odot$, dashed line to a $10^{12}\, M_\odot$, and dotted to
$10^{14}\, M_\odot$.}\label{fig|subhalo_MF}
\end{figure}

\clearpage
\begin{figure}
\figurenum{A2}
\plotone{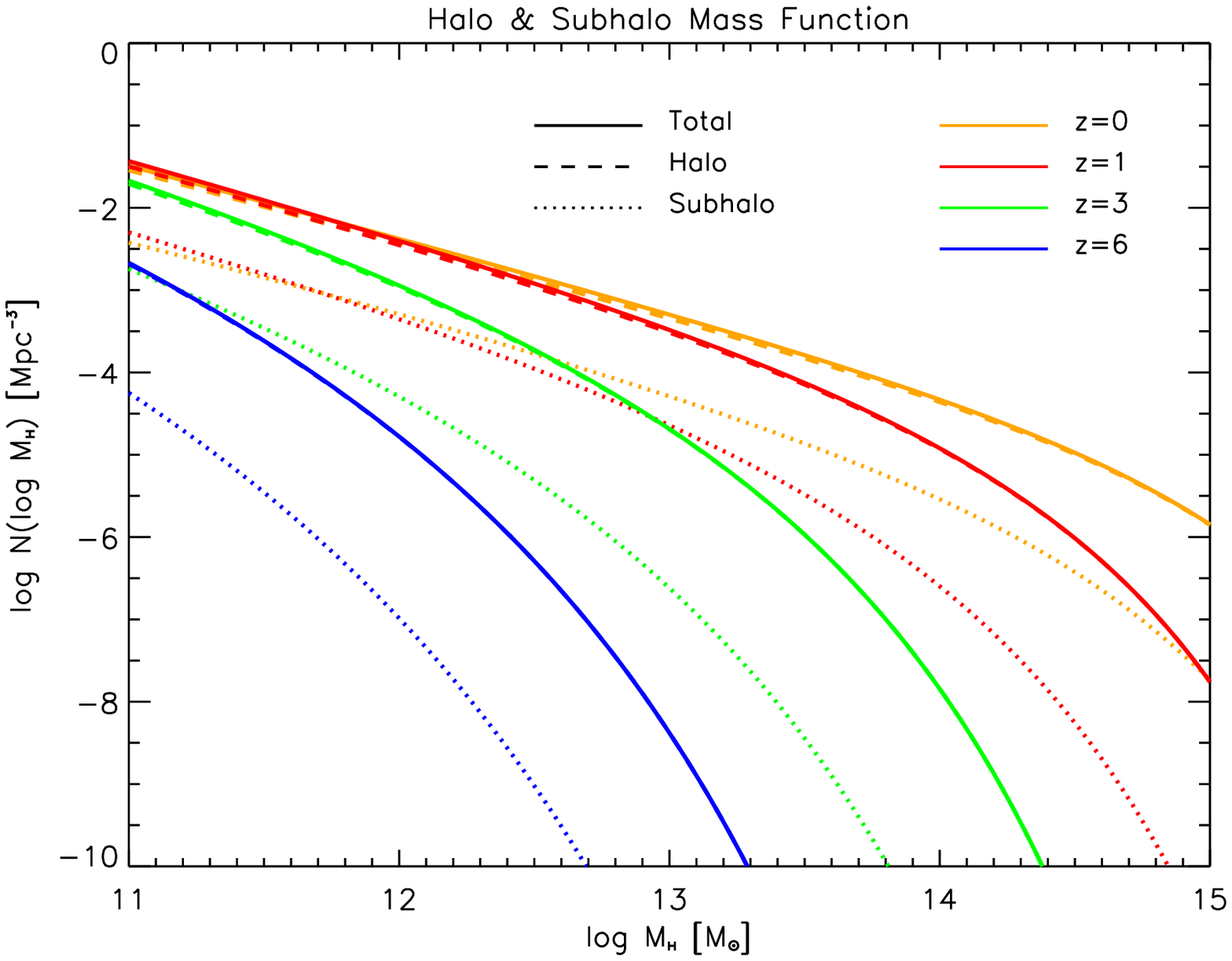}\caption{The overall
contribution of subhalos to the halo mass function, vs. the halo mass $M_{\rm
H}$. Solid lines show the halo mass function including subhalos, dashed lines
show the halo mass function without subhalos, and dotted lines show the
subhalo contribution. Results are plotted at redshift $z=0$ (orange), $1$
(red), $3$ (green), and $6$ (blue).}\label{fig|subhalo_MF_tot}
\end{figure}

\clearpage
\begin{figure}
\figurenum{A3}
\plotone{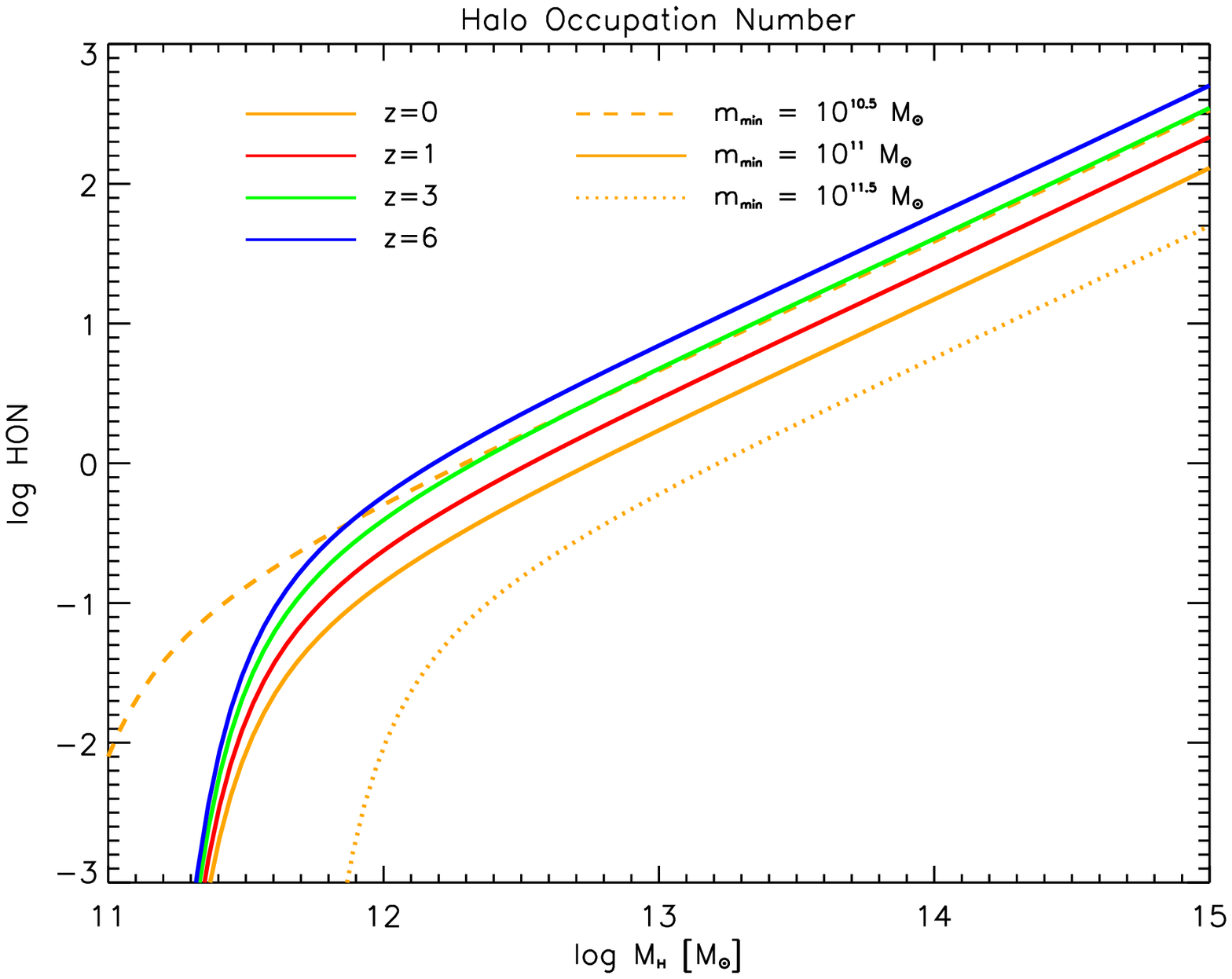}\caption{The halo occupation number vs. the host
halo mass $M_{\rm H}$. Results are plotted at redshift $z=0$ (orange), $1$
(red), $3$ (green), and $6$ (blue). At $z=0$ solid line refers to a
minimum satellite mass $m_{\rm min}=10^{11}\, M_\odot$, dashed to
$10^{10.5}\, M_\odot$, and dotted to $10^{11.5}\, M_\odot$.}\label{fig|subhalo_HON}
\end{figure}

\clearpage
\begin{figure}
\figurenum{B1}
\plotone{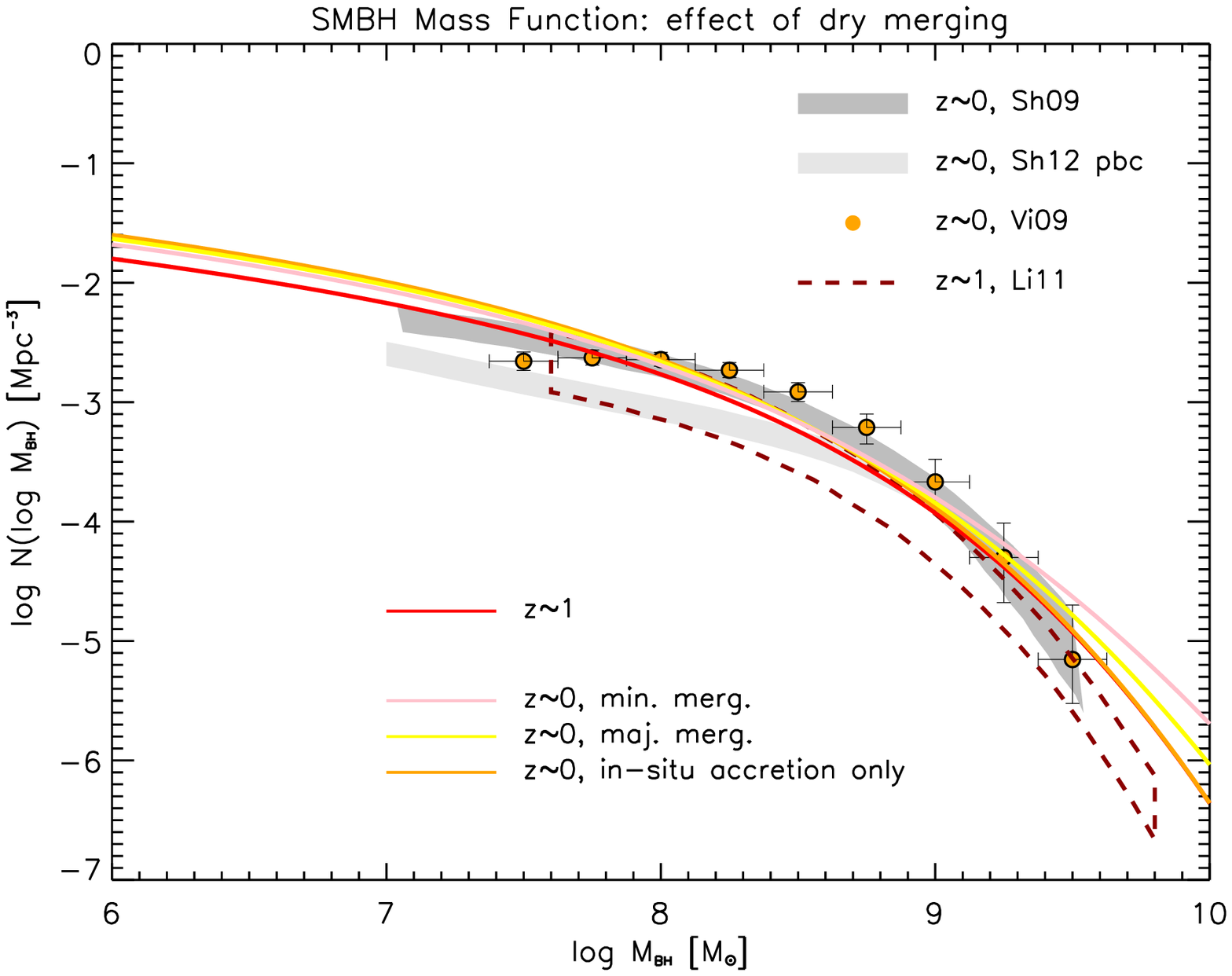}\caption{Effect of dry mergers on the
late $z\la 1$ evolution of the supermassive BH mass function. The red line
represents the mass function at redshift $z\sim 1$ (at higher redshift
dry merging effects are negligible), while the other colored lines illustrates
its evolution toward $z\sim 0$ due to merging and in-situ accretion.
Specifically, the BH merging rate is assumed to mirror the DM merging rates
as given by Stewart et al. (2009) for major mergers (yellow line) and for
minor mergers (pink line); the result for in-situ accretion only is plotted
for reference as an orange line. Data points and shaded areas as in
Fig.~\ref{fig|BH_MF}.}\label{fig|BH_merging}
\end{figure}

\clearpage
\begin{figure}
\figurenum{B2}
\plotone{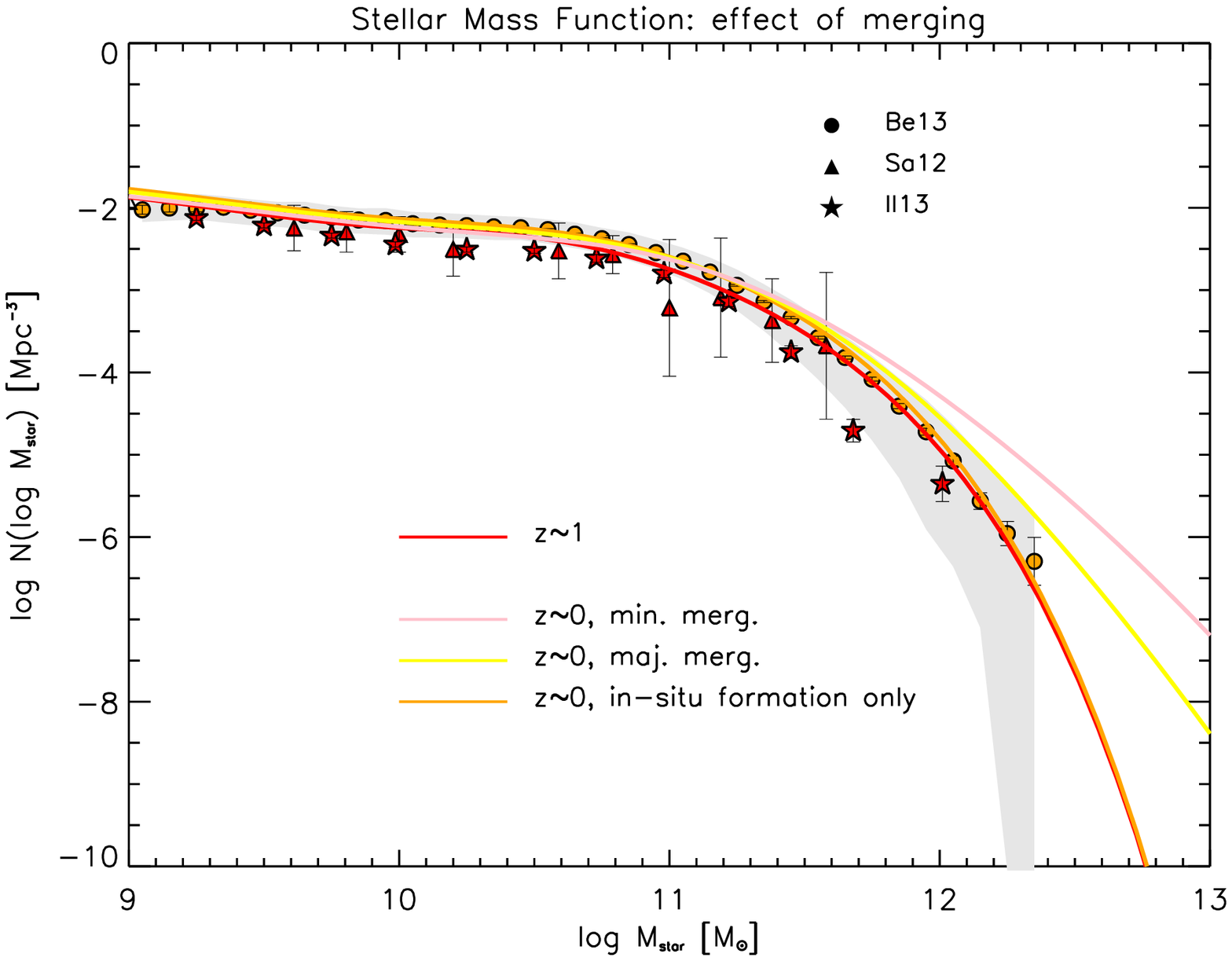}\caption{Effect of dry mergers on the late $z\la
1$ evolution of the galaxy stellar mass function as derived from the
continuity equation. The red line represents the mass function at redshift
$z\sim 1$ (at higher redshift dry merging effects are negligible), while the
other colored lines illustrate its evolution toward $z\sim 0$ due to merging
and in-situ formation. Specifically, the galaxy merging rate is computed
according to Stewart et al. (2009) for major mergers (yellow line) and for
minor mergers (pink line); the result for in-situ formation only is plotted
for reference as an orange line. Data points and shaded areas as in
Fig.~\ref{fig|STAR_MF}.}\label{fig|STAR_merging}
\end{figure}

\clearpage
\begin{figure}
\figurenum{C1}
\epsscale{1.}\plotone{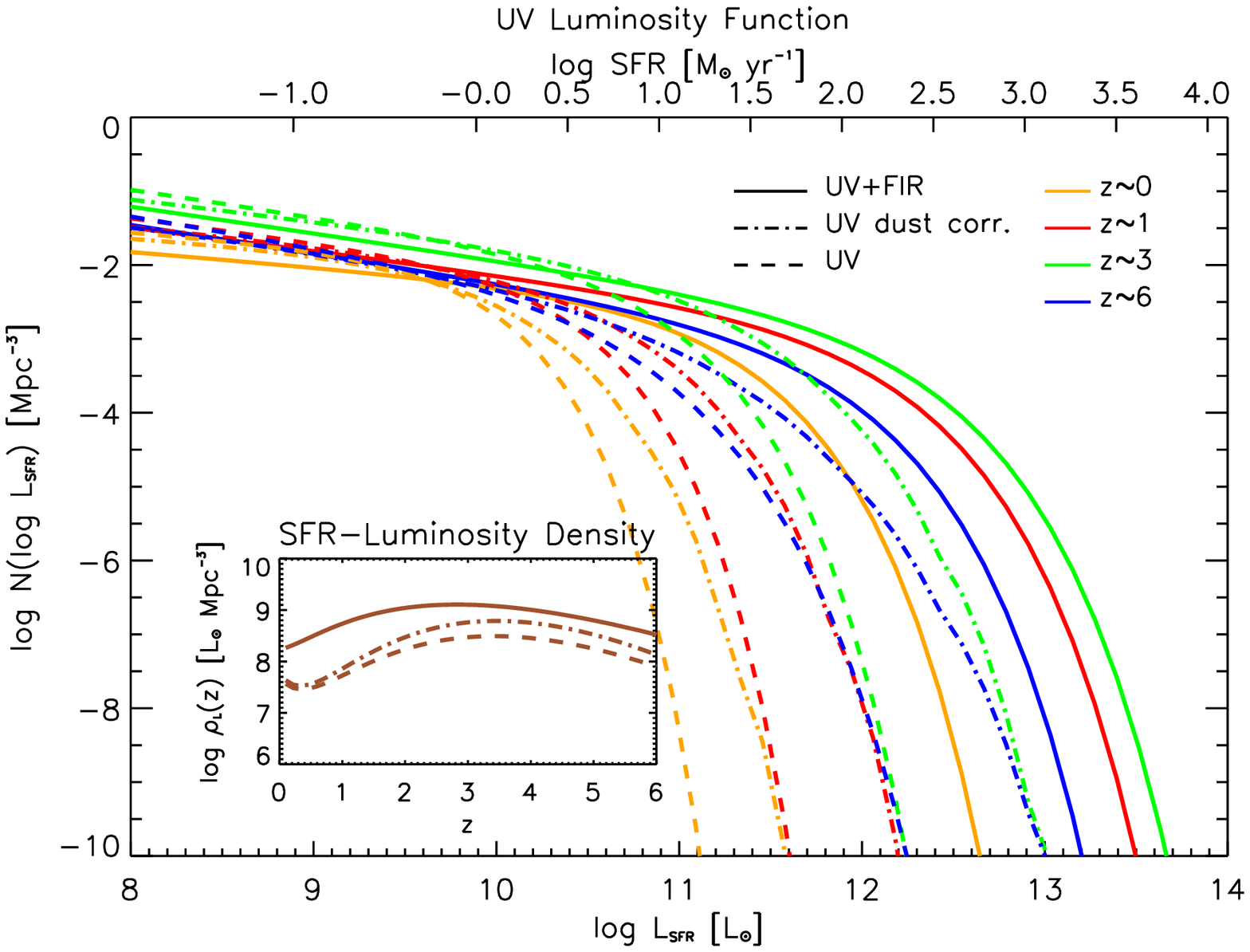}\caption{The SFR-luminosity function
$N(\log L_{\rm SFR})$ at redshift $z = 0$ (orange), $1$ (red), $3$ (green),
and $6$ (blue), vs. the luminosity $L_{\rm SFR}$ associated to the SFR (lower
axis) and vs. the SFR (upper axis). Solid lines is our rendition of the
luminosity function based on the UV data at the faint and FIR data at the
bright; this is the same plotted in Fig.~\ref{fig|STAR_LF}. Dot-dashed lines
is a rendition based only on dust-corrected UV data, and dashed lines from
dust-uncorrected UV data. The inset shows the corresponding SFR-luminosity
densities. Data points have been omitted for clarity.}\label{fig|STAR_UVLF}
\end{figure}

\clearpage
\begin{figure}
\figurenum{C2}
\epsscale{1.}\plotone{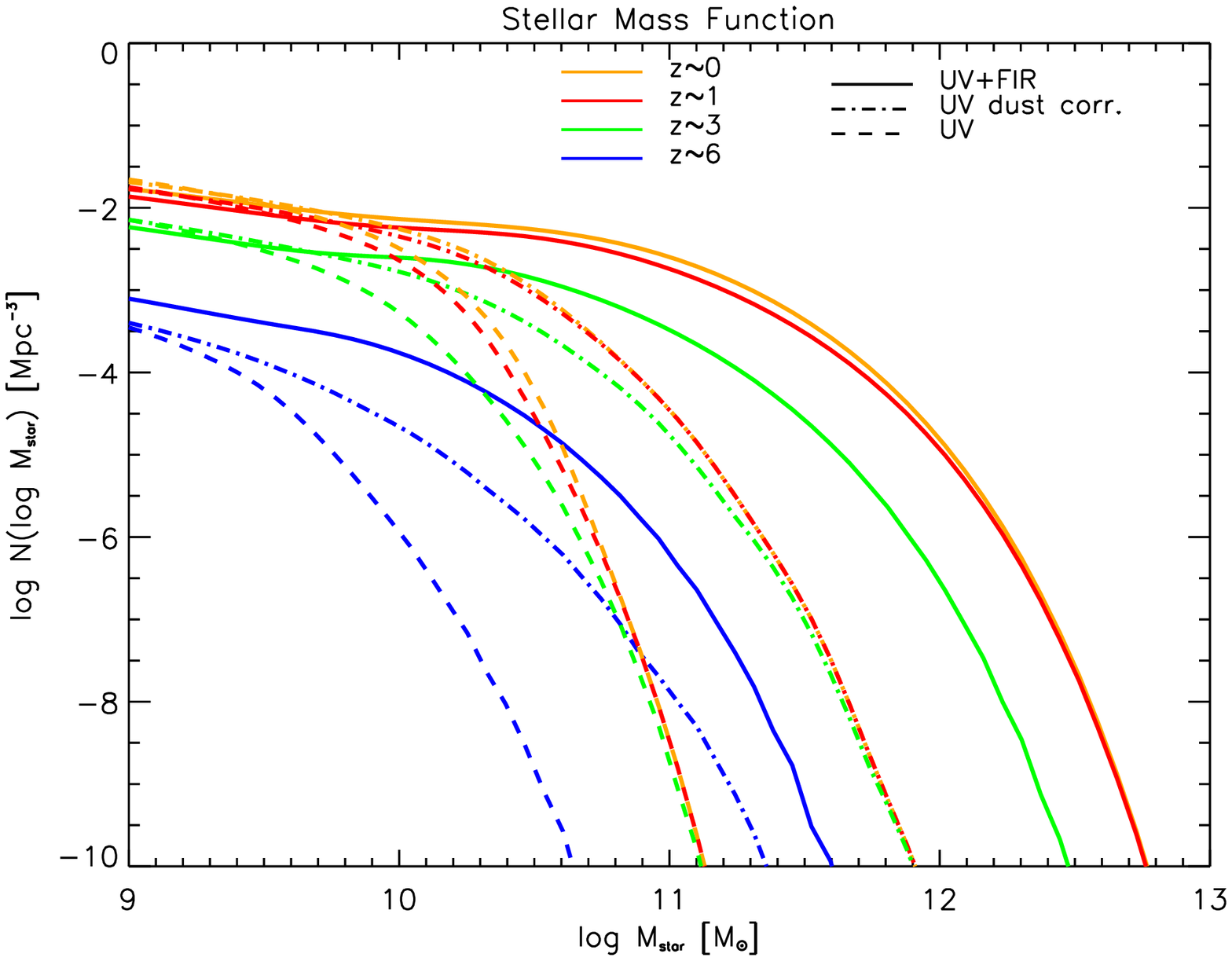}\caption{Effect of adopting the
SFR-luminosity functions obtained basing on FIR and UV data (solid lines),
only dust-corrected UV data (dot-dashed lines) and only dust-uncorrected UV
data (dashed lines) as input of the continuity equation to obtain the stellar
mass function. The solid lines are the same outputs shown in
Fig.~\ref{fig|STAR_MF} to be in very good agreement with data points (here
omitted for clarity).}\label{fig|STAR_UVMF}
\end{figure}

\clearpage
\begin{turnpage}
\begin{deluxetable}{lccccccccccccccccccccccccccccc}
\tabletypesize{\scriptsize}\tablewidth{0pt}\tablecaption{Input/Output
Generalized Schechter Functions.}\tablehead{\colhead{function} &
&\colhead{$\log \Phi_0$} & \colhead{$k_{\Phi 1}$} & \colhead{$k_{\Phi 2}$} &
\colhead{$k_{\Phi 3}$} & & \colhead{$\log X_0$} & \colhead{$k_{X1}$} &
\colhead{$k_{X2}$} & \colhead{$k_{X3}$} & & \colhead{$\alpha(z_0)$} &
\colhead{$k_{\alpha 1}$} & \colhead{$k_{\alpha 2}$} & \colhead{$k_{\alpha
3}$} & & \colhead{$\omega(z_0)$} & \colhead{$k_{\omega 1}$} &
\colhead{$k_{\omega 2}$} & \colhead{$k_{\omega 2}$}} \startdata
\cutinhead{$N(\log X,z) = \Phi(z)\, \left[X\over
X_c(z)\right]^{1-\alpha(z)}\,\exp{\left\{-\left[X\over
X_c(z)\right]^{\omega(z)}\right\}}$} AGN LF & &-3.80 & 0.45 & -1.00 & 0.00 &
& 10.90 & 1.10 & 6.94 & -11.55 &
&1.40 & -1.70 & 3.40 & -1.75 & &0.36 & 0.62 & -1.59 & 0.8\\
SFR LF & &-2.40 & -2.30 & 6.20 & -4.90 & & 10.90 & 3.20 & -1.40 &
-2.10 & &1.20 & 0.50 & -0.50 & 0.20 & & 0.70 & -0.15 & 0.16 & 0.00\\
BH MF & &-2.30 & -0.40 & -1.80 & -1.50 & & 8.07 & -0.80 & 7.30 & -9.20 &
& 1.35 & -0.10 & 0.40 & 0.30 & & 0.46 & 0.05 & 0.18 & -0.55\\
Stellar MF & &-2.10 & -0.80 & 1.65 & -3.10 & & 10.85 & 0.00 & 0.00 &
-1.90 & & 1.20 & 0.00 & -0.40 & 0.55 & & 0.65 & 0.00 & -0.40 & 0.55\\
HMF & &-3.97 & 0.00 & 0.00 & 1.50 & & 14.00 & -0.90 & -1.90 & -1.10 & &
1.80 & 0.50 & 0.10 & 0.70 & & 0.47 & 0.45 & -0.10 & -0.45\\
GHMF & &-3.35 & 0.50 & 0.1 & -1.50 & & 13.05 & -0.80 & 0.00 & -1.30 & &
1.88 & 0.30 & -0.40 & 1.30 & & 1.10 & -0.10 & 0.00 & -0.43\\
\enddata
\tablecomments{Typical tolerance on the parameters is less than $10\%$. See
Sect.~\ref{sec|AGN_LF} for details on the redshift evolution of the
parameters.}
\end{deluxetable}
\end{turnpage}

\clearpage
\begin{turnpage}
\begin{deluxetable}{lcccccccccccccccccccc}
\tabletypesize{\scriptsize} \tablewidth{0pt}\tablecaption{Fits to
Abundance Matching Results}\tablehead{\colhead{function} & &\colhead{$\log
N_0$} & \colhead{$k_{N 1}$} & \colhead{$k_{N 2}$} & \colhead{$k_{N 3}$} & &
\colhead{$\log M_{b 0}$} & \colhead{$k_{M1}$} & \colhead{$k_{M2}$} &
\colhead{$k_{M3}$} & & \colhead{$\alpha_0$} & \colhead{$k_{\alpha 1}$} &
\colhead{$k_{\alpha 2}$} & \colhead{$k_{\alpha 3}$} & & \colhead{$\omega_0$}
& \colhead{$k_{\omega 1}$} & \colhead{$k_{\omega 2}$} & \colhead{$k_{\omega
3}$}} \startdata \cutinhead{$Y=N\times \left[\left({X\over
M_b}\right)^{\alpha}+\left({X\over M_b}\right)^{\omega}\right]^{-1}$} $L_{\rm
AGN}-M_{\rm H}$ (1to1)  & & 12.25 & -1.90 & 1.45 & 1.10 & & 12.30 & 0.00 &
0.00 & 0.00 & & -1.65 & 1.20 & -4.50 & 0.60 & & -0.65 &
-3.50 & -0.40 & 0.80\\
$L_{\rm AGN}-M_{\rm H}$ (aver.) & & 12.33 & -2.00 & 2.70 & -1.30 & & 12.50 &
0.00 & 0.00 & 0.00 & & -1.60 & 0.00 & -3.30 & 1.30 & & -0.50 &
-2.10 & -1.60 & 2.60\\
$L_{\rm SFR}-M_{\rm H}$ (aver.) & & 11.25 & 0.00 & 0.00 & 0.00 & & 12.20 &
-1.20 & 0.00 & 0.00 & & -1.30 & -3.00 & -0.50 & 1.20 & & -0.50 & -1.50
& 0.00 & 1.50\\
$M_{{\rm BH}}-M_{\rm H}$ (1to1) & & 8.00 & -0.40 & 0.70 & -0.80 & & 11.90 &
0.00 & 0.00 & 0.00 & & -1.10 & -0.80 & -1.50 & 0.10 & & -1.10 &
-0.80 & -1.50 & 0.10\\
$M_{{\rm BH}}-M_{\rm H}$ (aver.) & & 8.20 & -0.20 & 0.80 & -1.50 & & 12.20 &
0.00 & 0.00 & 0.00 & & -1.40 & -1.30 & -0.30 & 0.10 & & -0.80 &
-0.40 & -1.10 & 0.10\\
$M_{\star}-M_{\rm H}$ (aver.) & & 10.40 & -0.80 & 0.80 & -0.20 & & 11.50 &
0.00 & 0.00 & 0.00 & & -2.20 & -1.90 & -1.60 & 4.70 & & -0.75 & -0.30
& -1.80 & 2.60\\
SFR$-M_{\star}$ (aver.) & & 1.90 & 0.00 & 0.00 & 0.00 & & 11.60 & -1.90 &
-2.50 & 1.70 & & -1.60 & 0.00 & 1.50 & -0.20 & & -0.50 & -1.70 & 3.50
& -2.00\\
\cutinhead{$Y=N\times \left[\left({X\over M_b}\right)^{\alpha}+\left({X\over
M_b}\right)^{\omega}\right]$} $M_{\rm BH}-M_\star$ (1to1) & & 7.33 & 0.10 &
-0.70 & 1.00 & & 10.60 & 0.00 & 0.00 & 0.00 & & 1.60 & 0.00 & -0.20 & 2.80 &
& 0.60 & 0.40 & -0.40 & 2.20\\
$M_{\rm BH}-M_\star$ (aver.) & & 7.15 & 0.00 & -0.60 & 0.00 & & 10.50 & 0.00
& 0.00 & 0.00 & & 1.30 & 0.30 & 0.00 & 0.90 & & 0.60 & 0.20 &
0.40 & 0.90\\
\enddata
\tablecomments{Typical tolerance on the parameters is less than $10\%$. See
Appendix D for details on the redshift evolution of the parameters.}
\end{deluxetable}
\end{turnpage}

\clearpage
\begin{deluxetable}{lcccccccc}
\tabletypesize{}\tablecaption{Main Results} \tablewidth{0pt}
\tablehead{\colhead{Results} && \colhead{Sections} &&
\colhead{Figures}}\startdata
AGN luminosity function && \ref{sec|AGN_LF} && \ref{fig|AGN_LF}\\
BH mass function && \ref{sec|BH_MF} && \ref{fig|BH_MF}\\
SFR luminosity function && \ref{sec|STAR_LF} && \ref{fig|STAR_LF}\\
Stellar mass function && \ref{sec|STAR_MF} && \ref{fig|STAR_MF}\\
Galaxy halo mass function && \ref{sec|abundance} && \ref{fig|GHMF}\\
$M_{\rm BH}-M_{\rm H}$ relationship && \ref{sec|abmatch_BHvshalo} && \ref{fig|Mbh_MH}\\
$L_{\rm AGN}-M_{\rm H}$ relationship &&\ref{sec|abmatch_BHvshalo} && \ref{fig|Lagn_MH}\\
AGN/BH bias && \ref{sec|abmatch_BHvshalo} && \ref{fig|AGN_bias}\\
$M_\star-M_{\rm H}$ relationship && \ref{sec|abmatch_STARvshalo} && \ref{fig|Mstar_MH},\ref{fig|Mstar_MH_z0comp}\\
$L_{\rm SFR}-M_{\rm H}$ relationship && \ref{sec|abmatch_STARvshalo} && \ref{fig|Lstar_MH}\\
Galaxy bias && \ref{sec|abmatch_STARvshalo} && \ref{fig|STAR_bias}\\
Cosmic sSFR vs. z && \ref{sec|abmatch_sSFR} && \ref{fig|sSFR_z}\\
sSFR vs. $M_\star$ && \ref{sec|abmatch_sSFR} && \ref{fig|sSFR}\\
$M_{{\rm BH}}-M_{\rm H}$ relationship && \ref{sec|abmatch_BHvsSTAR} && \ref{fig|Mbh_Mstar}\\
\enddata
\end{deluxetable}

\end{document}